\documentclass[12pt]{article}
\usepackage{graphicx} % Required for inserting images
\usepackage{natbib}
\usepackage{amsmath, amssymb}
\usepackage{bbm,bm}
\usepackage{algorithm} 
\usepackage{algpseudocode} 
\usepackage[table,xcdraw]{xcolor}
\usepackage[labelformat=simple]{subfig}
\usepackage{booktabs}
\usepackage{hyperref}
\hypersetup{
    colorlinks=true,
    linkcolor=blue,
    filecolor=magenta,      
    urlcolor=cyan,
    citecolor = red,
}

\usepackage{diagbox} 
\usepackage{tikz}
\usepackage{pgfplots}
\pgfplotsset{compat=1.18}

\usepackage{amsthm} % provides theorem/lemma environments
\newtheorem{theorem}{Theorem}[section]    % numbered within sections
\newtheorem{lemma}[theorem]{Lemma}
\newtheorem{corollary}[theorem]{Corollary}
\newtheorem{prop}[theorem]{Proposition}
\theoremstyle{definition}

\newcommand{\iid}{\stackrel{\rm iid}{\sim}}

\usepackage{lastpage}
\begin{document}

\title{Latent Modularity in Multi-View Data}

\author{A. Cremaschi\thanks{School of Science and Technology, IE University. \texttt{andrea.cremaschi@ie.edu}}
	\and M. De Iorio\thanks{Yong Loo Lin School of Medicine, National University of Singapore}
	\and G. L. Page\thanks{Department of Statistics, Brigham Young University}
	\and A. Jasra\thanks{School of Data Science, Chinese University of Hong Kong}
}

\date{} % or \date{\today}

	\maketitle

\begin{abstract}%   <- trailing '%' for backward compatibility of .sty file
In this article, we consider the problem of clustering multi-view data, that is, information associated to individuals that form heterogeneous data sources (the views). We adopt a Bayesian model and in the prior structure we assume that each individual belongs to a baseline cluster and conditionally allow each individual in each view to potentially belong to different clusters than the baseline. We call such a structure \emph{latent modularity}.
Then for each cluster, in each view we have a specific statistical model with an associated prior. We derive expressions for the marginal priors on the view-specific cluster labels and the associated partitions, giving several insights into our chosen prior structure. Using simple Markov chain Monte Carlo algorithms, we consider our model in a simulation study, along with a more detailed case study that requires several modeling innovations. 
\end{abstract}

%\begin{keywords}
  Multi-View Data, Latent Modularity, Clustering, Partitioning, Markov chain Monte Carlo
%\end{keywords}

\clearpage
\section{Introduction}

With advances in information acquisition technologies, the availability of multi-view data has grown significantly. These datasets encompass multiple modalities, each offering a complementary, yet unique, perspective on the same underlying system. However, they often feature heterogeneous data types, intricate inter-dependencies, varying degrees of missing data, and differing sample sizes. In fields such as biomedicine and public health, the increasing availability of large-scale, diverse datasets, linked through complex and only partially understood mechanisms (e.g., multi-omic patient data combined with clinical and demographic information), has created a pressing need for statistical methods capable of effectively handling both the scale and heterogeneity of such data.  As a result, multi-view learning has become increasingly prominent in statistics, machine learning, and data mining \citep{sun2013survey}. 

Techniques such as multi-view unsupervised and semi-supervised learning, including co-training and co-regularization, have drawn considerable attention. In particular, multi-view clustering (MVC) has seen rapid development due to its ability to integrate information from multiple sources, enhancing clustering performance. Clustering, which groups subjects into subpopulations based on similarities, has broad applications. Yet, challenges persist in balancing view-specific and shared information, ensuring robustness to noisy or incomplete views, and improving the scalability of MVC methods. As a brief aside, the term ``multi-view clustering'' has referred to techniques that produce one integrated clustering based on multiple views (\cite{ Chao_etal, fang_etal}). As alluded to however, we focus on the perspective contained in an emerging literature that considers the hierarchical structure inherent in multi-view data which results in a clustering per each view (\cite{franzolini2023conditional, Dombowsky_2025}). 
Borrowing terminology from the network literature, in the context of multi-view clustering we use \textit{latent modularity} to indicate the underlying shared (latent) modular structure of a set of subjects, inferred from multiple heterogeneous data, while still allowing for view-specific variations.
Latent modularity captures the idea that beneath potentially discordant views there exists a global modular organization (clusters or communities), which determines part of the observed clustering structure 
and explains the concordant structure across views, while accommodating view-specific deviations

Statistically, the main challenge in multi-view clustering is to take advantage of information from all views while accounting for differences in their representation. Early strategies generally fall into two extremes, each with evident drawbacks \citep{shen2009integrative}: (i) clustering each view independently, which neglects potentially valuable cross-view information; or (ii) enforcing a single joint clustering across all views, which unrealistically assumes that the data are partitioned identically in every view.

Bayesian approaches have been at the forefront of developing more flexible multi-view clustering models. \cite{kirk2012bayesian}, for example, introduce the multiple dataset integration framework, in which each dataset (view) is modeled with its own finite Dirichlet–multinomial mixture, while dependencies between cluster assignments across views are captured via explicit agreement parameters. \cite{lock2013bayesian} propose Bayesian Consensus Clustering, which assumes both an overall consensus partition of the observations and separate partitions for each data view. The view-specific clusters may deviate from the global structure, but a hierarchical prior links them probabilistically, enforcing partial but not complete agreement.
More recent work has tried to address the trade-off between shared and view-specific latent structures. \cite{shapiro2024bayesian}, for instance, propose the Bayesian Multi-View Clustering (BMVC) model for more complex scenarios. Unlike earlier methods that assume identical entities or strict alignment, BMVC handles partially overlapping or hierarchically related sets and many-to-many cluster correspondences. It uses a graphical model with dependence weights to adaptively tune coupling between views, strengthening links when structures agree and down-weighting them when they diverge.

A recent and interesting contribution is the Conditional Partial Exchangeability (CPE) paradigm \citep{franzolini2023conditional}, which provides a probabilistic framework for dependent random partitions of the same objects across distinct domains, potentially originating from different support spaces. CPE induces dependencies between cluster assignments across features while preserving within-subject dependence, and it captures the marginal contribution of each view. In particular, two subjects have an increased probability of being clustered together if they were previously co-clustered (e.g., at an earlier time point). However, a limitation of CPE is its reliance on at least a partial ordering of the views.
Most recently, \cite{Dombowsky_2025} introduce the product-centered Dirichlet process (PCDP) as the basis of the CLIC model (Clustering with Inter-View Correlation). This construction allows continuous tuning of the dependency between view-specific partitions via a single parameter, thereby bridging the extremes of independent clustering and fully shared clustering. Although PCDP provides a principled way to model correlations between partitions across views, the single dependence parameter may oversimplify cross-view relationships. The clustering dependence and number of clusters are also highly sensitive to the prior hyperparameters and the model may require careful prior tuning or empirical calibration to avoid degenerate results (either fully independent or identical partitions). Finally, PCDP  leads to computationally intensive posterior inference, making it impractical in high-dimensional settings.

The contributions of this paper are as follows.
\begin{itemize}
\item{We derive a new Bayesian model for multi-view data. This consists of allowing a baseline cluster for each subject and then conditionally a view-specific cluster, which can be different from the baseline.}
\item{We prove expressions for the marginal prior on the clustering and associated partitions. This provides several insights of the prior and the prior parameters on statistical inference.}
\item{We develop a simple Metropolis-within-Gibbs Markov chain Monte Carlo (MCMC) algorithm to infer the model, which is parallelizable.}
\item{We implement our model on a simulation study and then an in-depth case study. The latter requires modeling innovations. In particular, we  propose a novel technique to modeling zero-inflated panel count data.} 
\end{itemize}

The proposed model provides a flexible Bayesian framework for multi-view clustering by jointly capturing both shared and view-specific structures. The model enables inference on both view-specific and global clustering structures, while allowing for varying degrees of dependence across views. This formulation accommodates heterogeneous relationships among data domains, enabling some views to exhibit strong alignment while others remain relatively independent. By modeling dependence through a common latent process, the framework can adapt to a continuum of clustering relationships, from fully shared to completely independent partitions, thus offering a unified approach to integrative clustering across multiple data sources.
Posterior inference is achieved through a computationally efficient Gibbs sampling algorithm, ensuring scalability to  high-dimensional and multi-view datasets.

This article is structured as follows. In Section \ref{sec:stat_model} we give details of our overarching statistical model.
It should be remarked that to apply the model in practice, one needs to specify models on different views, and this is investigated later in the paper.
In Section \ref{sec:theory} we provide theoretical results with a discussion of their statistical relevance.
In Section \ref{sec:sim_study} we provide a simulation study that investigates several practical implications of our modeling choices.
Section \ref{sec:gusto} provides a comprehensive case study for real data, which requires several modeling choices associated with the different views of the data. The article is concluded in Section \ref{sec:conc}. There is a significant Appendix section which contains details on simulation algorithms, simulation studies, and the proofs of our mathematical results. 

%cluster then assigne 

%In many studies it is common to record or measure multiple variables on each experimental or observational unit which enables viewing subjects from numerous perspectives.  These type multivariate data are often referred to as multi-view data in the clustering community and clustering techniques that are able to employ all perspectives to group units has garnered considerable interest.  A variety of multiview clustering techniques have been developed in the  machine learning literature as evidenced by the existence of a number of survey articles \cite{fang_etal, Chao_etal} .These type of data e are clustering has recently attracted quite a bit of interest from the Bayesian model-based clustering community.  For example, Need to cite \cite{Dombowsky_2025}

\section{Statistical Model}\label{sec:stat_model}

\subsection{Basic Model}

We will use the notation for any $m\in\mathbb{N}$ that $[m]:=\{1,\dots,m\}$.
Let $Y_{ji}$ be a multi-view dataset with $i\in[n]$ subjects observed over $j\in[J]$ views.
We indicate with $\bm c_0 = \left(c_{01}, \dots, c_{0n}\right)$ the vector of clustering allocations at a baseline level, and by $\bm c_j = \left(c_{j1}, \dots, c_{jn}\right)$ the clustering allocation in
the $j^{\textrm{th}}$ view, $j\in[J]$. Note that $\bm c_0 $ corresponds to a latent partition $\rho_0$ of the $n$ subjects, capturing a global clustering structure, while $\bm c_j$ is associated with a view-specific partition $\rho_j$; a formal definition is given later on, as $\rho_j$ is not needed to define the model.
We specify the following model, called model (1), for $(i,j)\in[n]\times[J]$:
\begin{align}
    Y_{ji} \mid \bm \theta_j^*, \bm c_j &\overset{\text{ind}}{\sim} F_j(\cdot |\bm \theta^*_{jc_{ji}}) \nonumber\\
    \bm \theta^*_{jm} \mid M & \overset{\text{ind}}{\sim} G_j(\cdot), \quad m\in[M] \nonumber \\
    c_{ji} \mid \bm \omega_i, M & \overset{\text{ind}}{\sim} \text{Mult}_M(1;\omega_{i1},\ldots,\omega_{iM}) \label{eq:model_1}\\
    (\omega_{i1},\ldots,\omega_{iM}) \mid c_{0i}, \alpha, M & \overset{\text{ind}}{\sim}  \text{Dir}_M\left(\alpha + 
\mathbb{I}_{\{1\}}(c_{0i}), \dots, \alpha + \mathbb{I}_{\{M\}}(c_{0i})\right) \label{eq:model_2}\\
    c_{0i} \mid M, \bm \omega_0 & \overset{\text{ind}}{\sim}
    \text{Mult}_M(1; \omega_{01}, \dots, \omega_{0M}) \label{eq:model_3}\\
    (\omega_{01}, \dots, \omega_{0M})\mid \alpha_0 , M  &\sim
    \text{Dir}_M\left(\alpha_0,\ldots, \alpha_0\right) \label{eq:model_4}\\
    M &\sim q_M(\cdot) \nonumber
\end{align}
where, 
we recall the indicator of some set $\mathsf{D}$, $\mathbb{I}_{\mathsf{D}}(x)$ which is one if $x\in D$ and zero otherwise, and
within each of the $J$ views, 
$F_j(\cdot |\bm \theta^*_{jc_{ji}})$ is the distribution (with density $f_j(\cdot \mid \bm \theta^*_{jc_{ji}})$) representing the data parametrized by $\bm \theta^*_{jc_{ji}}$, $\bm \theta_j^* = \left(\bm \theta_{j1}^*, \dots, \bm \theta_{jM}^*\right)$, and $G_j$ the distribution for the parameters of $F_j$.  We will specify particular models
for $F_j$ and $G_j$ in sections \ref{sec:sim_study} and  \ref{sec:gusto} which are appropriate for the application under study.
We indicate by $\text{Dir}_M\left(1; \alpha_1, \dots, \alpha_M\right)$ the Dirichlet distribution with support on the $(M-1)$-dimensional simplex and associated vector of shape parameters $\alpha_1, \dots, \alpha_M$, and by $\text{Mult}_M\left(1; \omega_1, \dots, \omega_M\right)$ the Multinomial distribution with support over the classes $[M]$ and associated vector of probabilities $\omega_1, \dots, \omega_M$. We assume a random number of components with prior distribution $q_M$, often chosen to be a shifted Poisson distribution $M \sim \text{Poi}_1\left(\Lambda\right)$. 

\subsubsection{Discussion}

We remark that the parameters of the Dirichlet distribution 
in \eqref{eq:model_2} across the $J$ views are specified conditionally on the allocation variable $c_{0i}$ and are subject-specific. This allows information on subject cluster allocation to be transferred across views.  We further note that the structure in  \eqref{eq:model_1}-\eqref{eq:model_2} could be generalized.
For instance, for $i\in[n]$, setting $\bm{\pi}_i=(\pi_{i1},\dots,\pi_{iM})$
a probability vector one could use the model for $(i,j,m)\in[n]\times[J]\times[M]$
$$
\mathbb{P}(c_{ji}=m|\bm{\pi}_i,C_{0i},M) = 
\pi_{ic_{0i}}
$$
with $\bm{\pi}_i|c_{0i},M\sim
\overset{\text{ind}}{\sim}  \text{Dir}_M\left(\alpha + 
\mathbb{I}_{\{1\}}(c_{0i}), \dots, \alpha + \mathbb{I}_{\{M\}}(c_{0i})\right).
$
 The prior distribution on $\bm c_0$, and therefore on $\rho_0$ in \eqref{eq:model_3}-\eqref{eq:model_4} can be replaced by an informative prior when expert knowledge is available. For instance, it is straightforward to incorporate in model (1) a construction such as the 
Centered Partition Processes \citep{paganin2020centered}, the anchor prior \citep{dahl2025dependent} and the informed random partition prior \citep{paganin2023informed}. Finally, model (1) is conditionally partially exchangeable as defined in \cite{franzolini2023conditional}, as it corresponds polytree-structured dependence among the views, when the root $\rho_0$ is not observed.

\subsection{Unnormalized weights representation}

Model (1) can be re-written in terms of unnormalized weights used in the construction of the Dirichlet-distributed random vectors and in an equivalent manner. In other words, we introduce an equivalent
formulation of model (1). %, in expressions  \eqref{eq:model_1}-\eqref{eq:model_4} above.
For each $M$, let $\bm s_0 = \left(s_{01}, \dots, s_{0M}\right)$ and $\bm s_i = \left(s_{i1}, \dots, s_{iM}\right)$ for $i\in[n]$ be the unnormalized weights, corresponding to $\boldsymbol{\omega}_0$ and $\boldsymbol{\omega}_j$, respectively. We obtain model (2), for $(i,j,m)\in[n]\times[J]\times[M]$:
\begin{align}%\label{eq:model_full_2}
%    Y_{ji} \mid \bm \theta_j^*, \bm c_j &\overset{\text{ind}}{\sim} f_j(y_{ji} |\bm \theta^*_{jc_{ji}}) \nonumber\\
%    \bm \theta^*_{jm} \mid M &\sim G_j, \quad m=1, \dots, M \nonumber \\
    \mathbb{P}\left(c_{ji} = m \mid \bm s_i, M\right) &\propto s_{im} \nonumber\\
    s_{im} \mid c_{0i}, \alpha &\sim  \text{Gamma}\left(\alpha + \mathbb{I}_{\{m\}}(c_{0i}), 1\right)  \nonumber\\
    \mathbb{P}\left(c_{0i} = m \mid \bm s_0, M\right) &\propto  
    s_{0m} \nonumber\\
    s_{0m}\mid \alpha_0 , M  &\sim
   \text{Gamma}\left(\alpha_0,1\right)  \nonumber%\\
%    M &\sim q_M \nonumber
\end{align}
where $\text{Gamma}\left(\alpha, \beta\right)$ is the Gamma distribution with shape $\alpha>0$ and rate $\beta>0$. We introduce the variables $\bm s_0$ and $\bm s_i$ for $i\in[n]$ by exploiting the constructive definition of the Dirichlet distribution via normalization of Gamma random variables. 
We will use the notation  $t_0 = \sum\limits_{m=1}^Ms_{0m}$ and $t_i = \sum\limits_{m=1}^M s_{im}$ and note that  $\omega_{0m} = \frac{s_{0m}}{t_0}$ and $\omega_{im} = \frac{s_{im}}{t_i}$, for $m\in[M]
$ and $i \in[n]$.  We consider Model (2) in addition to Model (1) as it facilitates computation \citep{argiento2022infinity}, details of which are provided in Appendix.

%Model~\eqref{eq:model_full_1} can be recovered from model~\eqref{eq:model_full_2} by introducing the random variables $T_0 = \sum\limits_{m=1}^MS_{0m}$ and $T_i = \sum\limits_{m=1}^MS_{im}$ such that $\omega_{0m} = \frac{S_{0m}}{T_0}$ and $\omega_{im} = \frac{S_{im}}{T_i}$, for $m = 1, \dots, M$ and $i = 1, \dots, n$.

\section{Theoretical Properties}\label{sec:theory}

\subsection{Multi-view co-clustering probabilities}

To build intuition associated with the co-clustering behavior at the view/feature level, we provide some technical results that are consequences of the hierarchical structure of the partition modeling found in model (1).  In particular, in Proposition \ref{prop:conditional_joint_feature_labels} we derive the probability of co-clustering at the feature level given the baseline cluster information. In Corollary \ref{cor:limiting_probabilities} we also explore how these probabilities behave as $\alpha$ approaches the boundaries of its support. This provides additional insight to simulations detailed in Section \ref{sec:sim_study}. To this end we provide the joint probability distribution of the $n$-dimensional component label vectors for each of the $J$ features in the following proposition. Before giving the statement we give some definitions which are used in the case that $n=J=2$. Set
\begin{eqnarray}
n_{\mathsf{A}} & = & \sum_{i=1}^2\mathbb{I}_{\mathsf{A}}(c_{0i},c_{1i},c_{2i}) \label{eq:na}\\
n_{\mathsf{B}} & = & \sum_{i=1}^2 \mathbb{I}_{\mathsf{B}}(c_{0i},c_{1i},c_{2i})\label{eq:nb} \\
n_{\mathsf{C}} & = & \sum_{i=1}^2\left\{\mathbb{I}_{\mathsf{C}_1}(c_{0i},c_{1i},c_{2i})+\mathbb{I}_{\mathsf{C}_2}(c_{0i},c_{1i},c_{2i})+\mathbb{I}_{\mathsf{C}_3}(c_{0i},c_{1i},c_{2i})\right\}\label{eq:nc}
\end{eqnarray}
where
\begin{eqnarray}
\mathsf{A} &= &\{(i,j,k)\in\{1,2\}:i=j=k\} \label{eq:seta}\\
\mathsf{B} &= & \{(i,j,k)\in\{1,2\}:i\neq j \neq k\} \label{eq:setb}\\
\mathsf{C}_1 &=& \{(i,j,k)\in\{1,2\}:i=j,j\neq k\}\label{eq:setc1}\\
\mathsf{C}_2 &=& \{(i,j,k)\in\{1,2\}:j\neq k,i=k\}\label{eq:setc2}\\
\mathsf{C}_3 &=& \{(i,j,k)\in\{1,2\}:i\neq j,j=k\}.\label{eq:setc3}
\end{eqnarray}
In words, $n_{\mathsf{A}}$ is the number of units whose feature cluster labels are equal and also equal to the baseline cluster label, $n_{\mathsf{B}}$ is number of units that has no feature labels equal nor are any equal to the baseline cluster label, and $n_{\mathsf{C}}$ are the number of units that either have non-equal feature cluster labels,  but one feature cluster is equal to the baseline or equal feature cluster labels which are different from the baseline.   We have the following result whose proof can be found in Appendix \ref{app:prf_prop1}.

\begin{prop}\label{prop:conditional_joint_feature_labels}
Let $(m,\alpha,\alpha_0)\in\mathbb{N}\times(\mathbb{R}^+)^2$ be given.  Under Model (1) the conditional distribution of component labels $\bm{c}_1, \ldots, \bm{c}_J$ given  $\bm{c}_0$ is
\begin{align}\label{eq:joint_prob}
\mathbb{P}(\bm{c}_1, \ldots, \bm{c}_J ~|~ \bm{c}_0, \alpha, m) & = 
Z(m,\alpha,n)
\prod_{i=1}^n\prod_{s=1}^m \Gamma\left(\alpha + 
\mathbb{I}_{\{s\}}(c_{0i}) + \sum_{j=1}^J
\mathbb{I}_{\{s\}}(c_{ji})\right)\\
Z(m,\alpha,n)& =\left[\frac{ m \Gamma(\alpha m)}{ \Gamma(\alpha)^m\Gamma(\alpha m + J + 1)}\right]^{n}\nonumber
\end{align}
If  $n=J=2$, then \eqref{eq:joint_prob} simplifies to
\begin{align} \label{eq:cond_prob_n2}
\mathbb{P}(c_{11}, c_{12}, c_{21}, c_{22} | c_{01}, c_{02}, \alpha, m) 
& = \frac{ [(\alpha+2)(\alpha+1)]^{ n_{\mathsf{A}} } [\alpha^2]^{ n_{\mathsf{B}} }[(\alpha+1)\alpha]^{ n_{\mathsf{C}} }}{[(\alpha m +2)(\alpha m + 1)]^2}
\end{align}
\end{prop}

Interestingly, and as expected, the magnitude of the probability in \eqref{eq:cond_prob_n2} depends heavily on the ``majority vote''.  That is, as the number of view/feature cluster labels that are equal to their corresponding baseline cluster labels increases, then so does the probability in \eqref{eq:cond_prob_n2}. What is somewhat unexpected is that the probability increases when the baseline cluster labels are equal in addition to the baseline cluster labels being equal (at least for $J=2$). 

What is of more interest in the current study is determining the probability that two units co-cluster at a feature level, given that they belong to the same cluster (or do not) at the baseline level. These probabilities are provided in \eqref{eq:prob_cond_equal} and \eqref{eq:prob_cond_notequal} of the Appendix. As expected, the probabilities depend heavily on $\alpha$. As $\alpha$ increases, more and more prior mass associated with $\bm{\omega}_i$ concentrates on a small number of dimensions. This results in the Corollary \eqref{cor:limiting_probabilities} which is proved in Appendix \ref{app:prf_cor1}.

\begin{corollary}\label{cor:limiting_probabilities}
Let $(m,\alpha,\alpha_0)\in\mathbb{N}\times(\mathbb{R}^+)^2$ be given.  Then for $n=J=2$ we have
%Fix $M>1$ and set $n=2$, $J=2$.  %With out lost of generality, for $c_{01} = c_{02}$, set $c_{01} = c_{02} = 1$ and for $c_{01} \ne c_{02}$, set $c_{01} =1$ and $c_{02}=2$. 
%Then under the same assumptions in Proposition \eqref{prop:conditional_joint_feature_labels}, we have
%\begin{align*}
%Pr(c_{11} = c_{12}, c_{21} = c_{22} ~|~ c_{01}=1,c_{02}=1, \alpha, M) & = \frac{\alpha^4M^2 + \alpha^36M + \alpha^2(3M+10) + \alpha 12 + 4}{[(\alpha M +2)(\alpha M + 1)]^2}, \\[1em]
%Pr(c_{11} = c_{12}, c_{21} = c_{22} ~|~ c_{01}=1,c_{02}=2, \alpha, M) & = \frac{\alpha^4(M-4M+8)^2 + \alpha^3(2M+6) + \alpha^2(M+6) + \alpha}{[(\alpha M +2)(\alpha M + 1)]^2}, \\[1em]
%Pr(c_{11} \ne c_{12}, c_{21} = c_{22} ~|~ c_{01}=1,c_{02}=1, \alpha, M) & = \frac{\alpha^4M^2 + \alpha^36M + \alpha^2(3M+10) + \alpha 12 + 4}{[(\alpha M +2)(\alpha M + 1)]^2}, \\[1em]
%Pr(c_{11} \ne c_{12}, c_{21} = c_{22} ~|~ c_{01}=1,c_{02}=2, \alpha, M) & = \frac{\alpha^4M^2 + \alpha^36M + \alpha^2(3M+10) + \alpha 12 + 4}{[(\alpha M +2)(\alpha M + 1)]^2}, \\[1em]
%Pr(c_{11} \ne c_{12}, c_{21} \ne c_{22} ~|~ c_{01}=1,c_{02}=2, \alpha, M) & = \frac{\alpha^4M^2 + \alpha^36M + \alpha^2(3M+10) + \alpha 12 + 4}{[(\alpha M +2)(\alpha M + 1)]^2}. 
%\end{align*}
\begin{eqnarray*}
\lim_{\alpha \rightarrow 0} \mathbb{P}(c_{11} = c_{12}, c_{21} = c_{22} ~|~ c_{01},c_{02}, \alpha,  m) 
& = & \left\{\begin{array}{ll}
1 & \text{\emph{if}}~c_{01} = c_{02}\\
0 & \text{\emph{if}}~c_{01} \neq c_{02}
\end{array}\right.
\\
\lim_{\alpha \rightarrow 0} \mathbb{P}(c_{11} \ne c_{12}, c_{21} = c_{22} ~|~ c_{01} = c_{02}, \alpha,  m) & = & 0 \\
\lim_{\alpha \rightarrow 0} \mathbb{P}(c_{11} \ne c_{12}, c_{21} \ne c_{22} ~|~ c_{01} \ne c_{02}, \alpha,  m) & = & 1 \\
\lim_{\alpha \rightarrow \infty} \mathbb{P}(c_{11} = c_{12}, c_{21} = c_{22} ~|~ c_{01},c_{02}, \alpha,  m)  & = & 1/m^2~\forall~(c_{01},c_{02})\in\{1,2\}^2 \\
\lim_{\alpha \rightarrow \infty} \mathbb{P}(c_{11} \ne c_{12}, c_{21} = c_{22} ~|~ c_{01}, c_{02}, \alpha,  m) & = & m(m-1)/m^2 ~\forall~(c_{01},c_{02})\in\{1,2\}^2 \\
\lim_{\alpha \rightarrow \infty} \mathbb{P}(c_{11} \ne c_{12}, c_{21} \ne c_{22} ~|~ c_{01}, c_{02}, \alpha,  m) & = & (m-1)^2/m^2 ~\forall~(c_{01},c_{02})\in\{1,2\}^2
\end{eqnarray*}
\end{corollary}

As expected, if the two units are grouped at the baseline level then the probability of co-clustering at the feature level tends to one as $\alpha$ tends to zero (all prior mass concentrates on component identified by $c_{01}$ and $c_{02}$). This behavior also explains the other probability calculations as $\alpha \rightarrow 0$. The remaining probability calculations in Corollary \eqref{cor:limiting_probabilities} behave as expected given that as $\alpha$ approaches $\infty$ prior mass is uniformly distributed across the $m$ components.

\begin{figure}
\begin{center}
\includegraphics[scale=0.55]{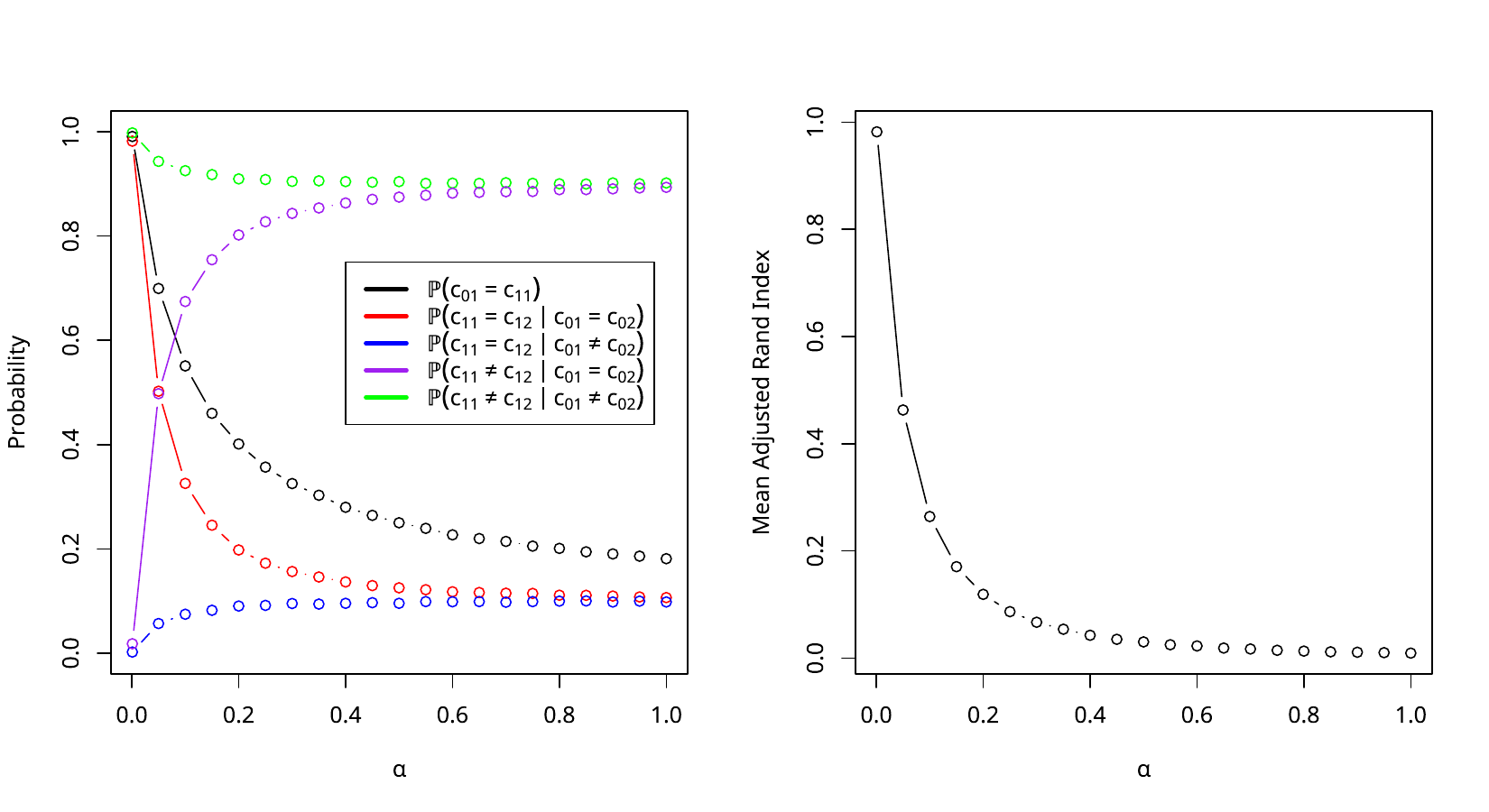}
\end{center}
\caption{The left figure displays Monte Carlo estimates of the probability of two unit co-clustering given baseline cluster configuration and marginalized over the baseline cluster.   The right figure displays the adjusted rand index between a single feature clustering configuration and the baseline cluster configuration.}
\label{fig:prior_simulation}
\end{figure}

To visualize the co-clustering probabilities studied in Proposition \eqref{prop:conditional_joint_feature_labels} and Corollary \eqref{cor:limiting_probabilities}, we conduct a small Monte Carlo study that samples from equations \eqref{eq:model_1} - \eqref{eq:model_4} after having fixed $m=10$, $J=2$, and $n=10$. Using $10^4$ Monte Carlo samples of $(\bm{c}_1, \bm{c}_2, \bm{c}_3)$ and $\bm{c}_0$, as a function of $\alpha$, we estimate $\mathbb{P}(c_{01} = c_{11})$,  $\mathbb{P}(c_{11} = c_{12} ~|~ c_{01} = c_{02})$, $\mathbb{P}(c_{11} = c_{12} ~|~ c_{01} \ne c_{02})$, and $\mathbb{E}[ARI(\bm{c}_1, \bm{c}_0)]$. Here $ARI(\cdot, \cdot)$ denotes the adjusted rand index between two cluster configurations with values close to 1 indicating that the two clusterings are equal and values close to 0, dissimilar \citep{hubert1985comparing}. Results are presented in Figure \ref{fig:prior_simulation}.

\subsection{Law of the multi-view partition}
In order to provide the law of the multi-view partition, we consider the baseline partition $\rho_0$ (implied by $\bm{c}_0$)  and the multi-view level partitions $\rho_j$ (implied by $\bm{c}_j$).
More precisely let $(n,k)\in\mathbb{N}\times[n]$ be given and  $\mathtt{A}_{k}(n)$ be the set of all possible
mutually disjoint subsets of $[n]$ with exactly $k$ subsets.  For example,  if $n=3$, $k=2$ then we would have
$$
\mathtt{A}_{2}(3) = \left\{
(\{1,2\},\{3\}),(\{1,3\},\{2\}),(\{2,3\},\{1\})
\right\}
$$
Then for $j\in[J]\}$,  $\rho_j\in\mathtt{P}(n):=\bigcup_{k=1}^n \mathtt{A}_{k}(n)$.
Note that $\rho_j$ can be determined by $\bm c_j$, albeit in a many-to-one fashion.
For example, if $n=3$, $m=3$ then either $c_j=(1,1,3)$ or $c_j=(2,2,3)$ would give the identical partition
$\rho_j=(\{1,2\},\{3\})$.
For a partition $\rho_j$, let $k_j$ be the number of clusters.
Let $(k_j,s)\in[m]^2$ be given and define the sets:
\begin{eqnarray}
\overline{\mathtt{C}}_s(m) & := & \{(c_{1},\dots,c_{n})\in [m]^n:\sum_{i=1}^n\mathbb{I}_{\{s\}}(c_{i})>0\}\label{eq:cs1}\\
\mathtt{C}_{k_j}(m) & := & \{(c_{1},\dots,c_{n})\in[m]^n:\sum_{s=1}^m\mathbb{I}_{\overline{\mathtt{C}}_s(m)}(c_{1},\dots,c_{n})=k_j\}\label{eq:cs2}
\end{eqnarray}
The set $\overline{\mathtt{C}}_s(m)$ is the collection of clusterings which have at least one allocation to the label $s$.
The set $\mathtt{C}_{k_j}(m)$ is the collection of clusterings which have exactly $k_j$ different clusters.
Let $\bm c_j$ and $\mathbf{c}_0$ be given and
denote $n_{j0} = \text{Card}\left(\{i \in [n] : c_{ji} = c_{0i}\}\right)$ the number of elements with labels in agreement between $\bm c_j$ and $\bm c_0$. 
%Let $C_{k_j,m}^n(n_{j0})$, be the number of the %partitions of $n$ elements into $k_j$ clusters with %label values in $[m]$ for which exactly $n_{j0}$ are %in agreement with $\bm c_0$.  
For $\mathbf{c}_0\in \mathtt{C}_{k_0}(m)$ we write $n_j$ for the number of units in cluster $j$.
We have the following result whose proof is in Appendix \ref{app:theo_prf}. 

\begin{theorem}\label{theo:part}
Let $(\alpha,\alpha_0)\in(\mathbb{R}^+)^2$ be given.
Under model (1) the joint probability mass function of $(\rho_1, \dots, \rho_J, \rho_0)$ is
\begin{align*}
  p\left(\rho_1, \dots, \rho_J, \rho_0\right)   & =
  \sum_{m=1}^{\infty}
\mathbb{I}_{[m]}(k_0)
\frac{\Gamma(m\alpha_0)}{\Gamma(\alpha_0)^{k_0}\Gamma(n+m\alpha_0)}
\sum_{\mathbf{c}_0\in \mathtt{C}_{k_0}(m)}
\prod_{j=1}^J \mathbb{I}_{[m]}(k_j) 
\Big\{\\
& 
\sum_{\mathbf{c}_j\in\mathtt{C}_{k_j}(m)}
 \frac{\left(\alpha + 1\right)^{n_{j0}} \alpha^{n - n_{j0}}}{\left(m\alpha + 1\right)^n}
\prod_{j=1}^{k_0}\Gamma(\alpha_0 + n_j)\Big\}
 q_M(m)
\end{align*}
\end{theorem}

\begin{figure}
\begin{center}
\includegraphics[scale=0.75]{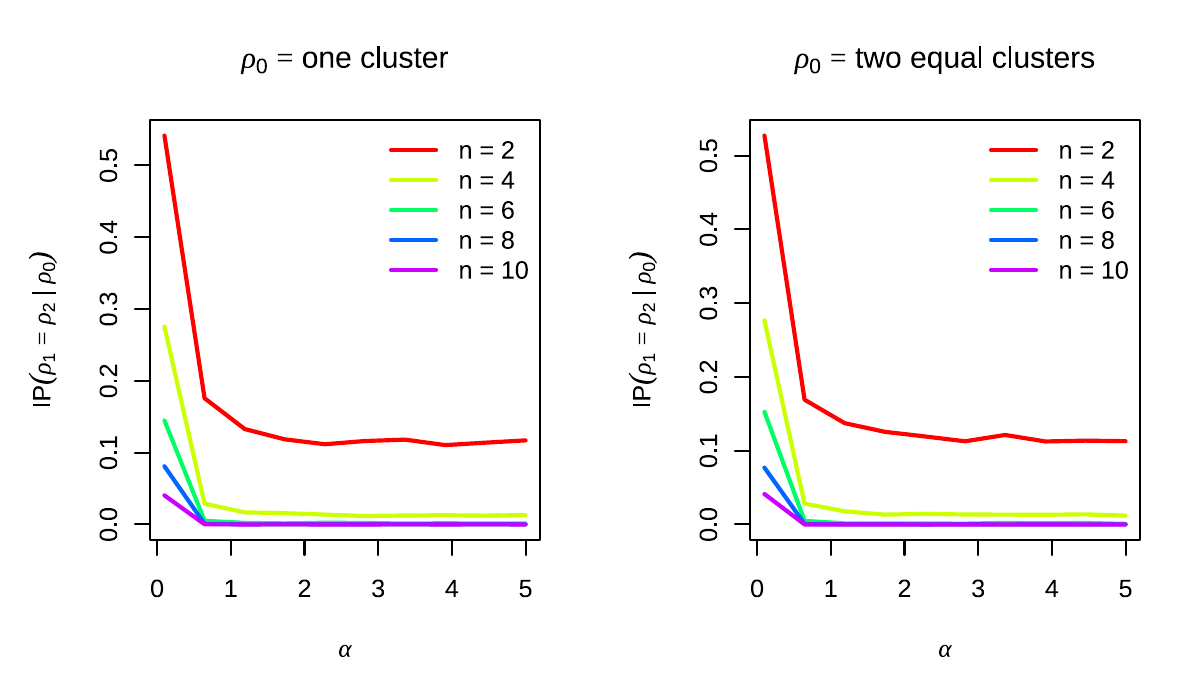}
\end{center}
\caption{Monte Carlo estimates of the probability of observing two equal partitions across two views, conditionally on a fixed baseline allocation vector $\bm \rho_0$. The two figures refer to baseline partitions with one or two equal-sized clusters, respectively. The probabilities are computed for different values of $\alpha$ (on the x-axis) and sample size $n$ (colors).}
\label{fig:apriori_prob_equal}
\end{figure}

Using Theorem \ref{theo:part} and Lemma \ref{lem:1} (in the appendix), it is straightforward to establish that:
\begin{align*}
p\left(\rho_1, \dots, \rho_J| \rho_0\right)
& = 
\sum_{m=1}^{\infty}
\mathbb{I}_{[m]}(k_0)
\frac{\Gamma(m\alpha_0)}{\Gamma(\alpha_0)^{k_0}\Gamma(n+m\alpha_0)}
\sum_{\mathbf{c}_0\in \mathtt{C}_{k_0}(m)}
\prod_{j=1}^J \mathbb{I}_{[m]}(k_j) 
\Big\{\\
& 
\sum_{\mathbf{c}_j\in\mathtt{C}_{k_j}(m)}
 \frac{\left(\alpha + 1\right)^{n_{j0}} \alpha^{n - n_{j0}}}{\left(m\alpha + 1\right)^n}
\prod_{j=1}^{k_0}\Gamma(\alpha_0 + n_j)\Big\}
 q_M(m) \Bigg/ \\
 & \left\{\sum_{m=1}^{\infty}
 \mathbb{I}_{[m]}(k_0)
\frac{1}{ \Gamma(\alpha_0)^{k_0}}
\frac{\Gamma(m\alpha_0)}{\Gamma(n+m\alpha_0)}
\sum_{\mathbf{c}_0\in \mathtt{C}_{k_0}(m)} 
\prod_{l=1}^{k_0}\Gamma(\alpha_0 + n_l)
q_M(m)\right\}
\end{align*}
In general, it is difficult to interpret what the implications of this probability mass function are. However, to gain some insight, one can sample the distribution and some results are presented in
Figure \ref{fig:apriori_prob_equal}. This displays the Monte Carlo estimates of the a-priori probability of observing two equal partitions in a setting with $J = 2$ views, conditionally on $\rho_0$, for different values of $\alpha$ and $n$. We note that, with increasing $\alpha$, the probability of observing two equal partitions across views decreases rapidly toward zero; the sample size also contributes to this behavior. 
This seems to be consistent with the discussion associated to the clustering vectors in the previous section.

\section{Simulation Study}\label{sec:sim_study}
In this section, we present  the results obtained from a simulation study, aimed at investigating the model performance under different data-generating mechanisms. Specifically, we consider a scenario where the view level partitions $\bm c_j$s are fixed in the simulation process, and interest lies in the study of the posterior distribution of $\bm c_0$. This analysis addresses the question of interpretability of the posterior distribution of $\bm c_0$, in light of the true view-level partitions. Moreover, a sensitivity analysis in a simulated setting is presented in Appendix~\ref{app:sensitivity}.

We simulate data from mixtures of univariate Gaussian distributions, such that for $(i,j)\in[n]\times[J]$:
\begin{align*}
    Y_{ji} \mid \bm \mu_j, \bm \sigma_j^2, \bm c_j &\sim \text{N}\left(y_{ji} \mid \mu_{jc_{ji}}, \sigma^2_{jc_{ji}}\right) \nonumber\\
    \mu_{jm}, \sigma_{jm}^2 \mid M &\sim \text{N}\left(\mu_{jm} \mid m_0, \sigma_{jm}^2/k_0\right) \text{Inv-Gamma}\left(\sigma_{jm}^2 \mid a_0, b_0\right) , \quad m\in[M] \nonumber
\end{align*}

In principle, the model should be able to recover both the view-specific partitions $\bm c_1, \dots, \bm c_J$, as well as a global  $\bm c_0$ which should account for the fact that subjects might belong or not to the same cluster across the $J$ views. This principle is analogous to  a ``majority vote'' estimate, where the clusters in $\bm c_0$ are assigned based on how many times a unit is in the same cluster across the $J$ views. 

\begin{figure}
\centering
\includegraphics[scale=0.75]{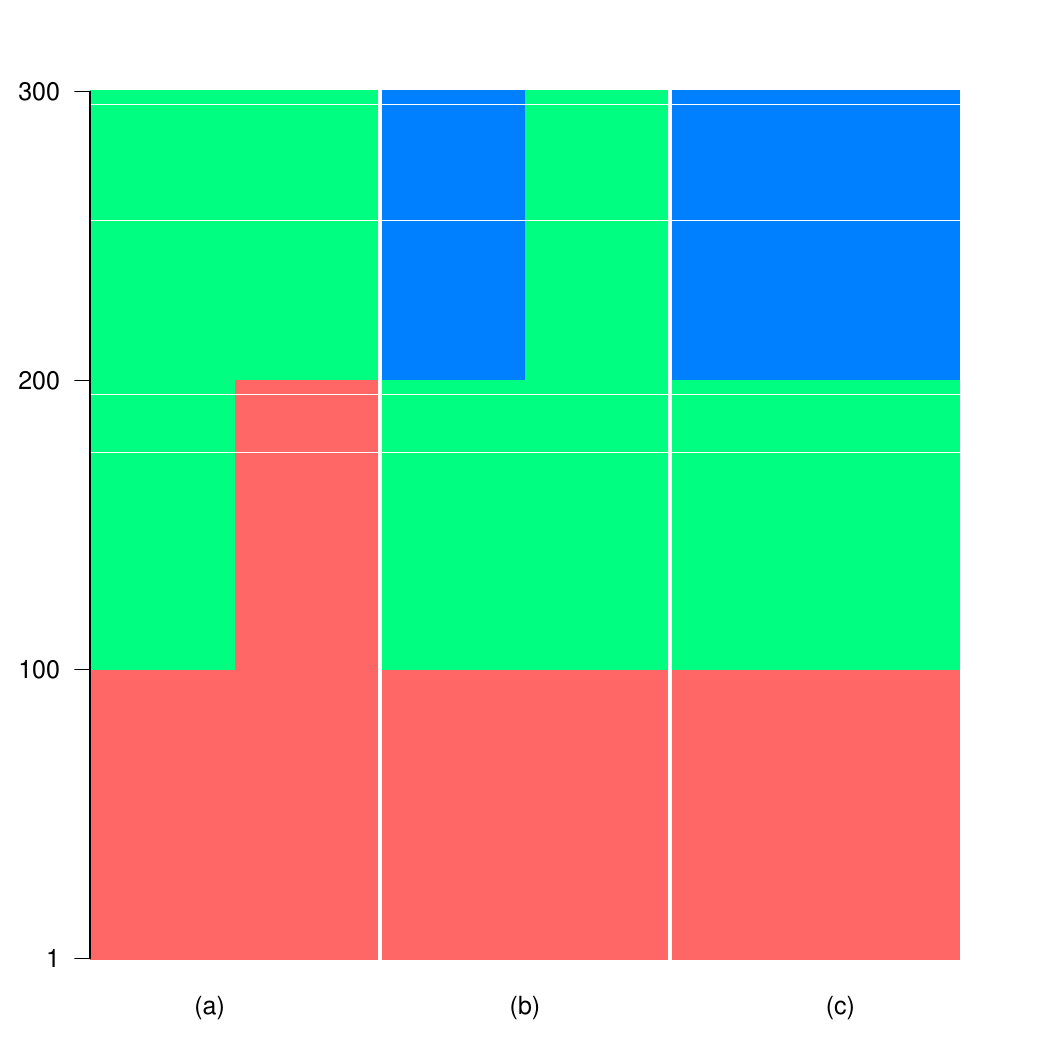}
\caption{Simulation Study. Pattern of clustering assignment underlying the four simulation settings. The heatmap rows indicate the observations units, while the three simulation settings are separated by vertical spaces.}
\label{fig:Simul2_truecj}
\end{figure}

\begin{figure}
\centering
\subfloat[Scenario (a)]{\includegraphics[width=0.35\linewidth]{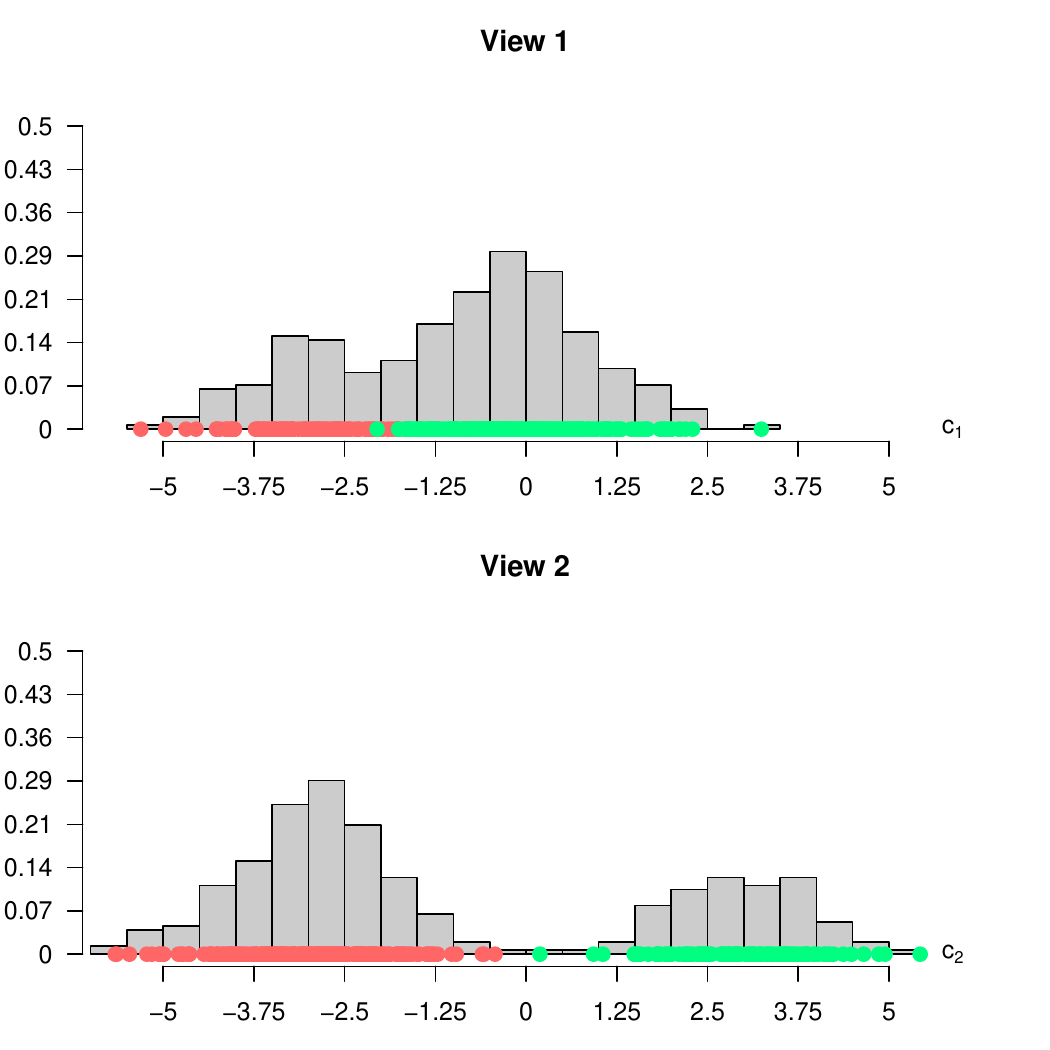}}
\subfloat[Scenario (b)]{\includegraphics[width=0.35\linewidth]{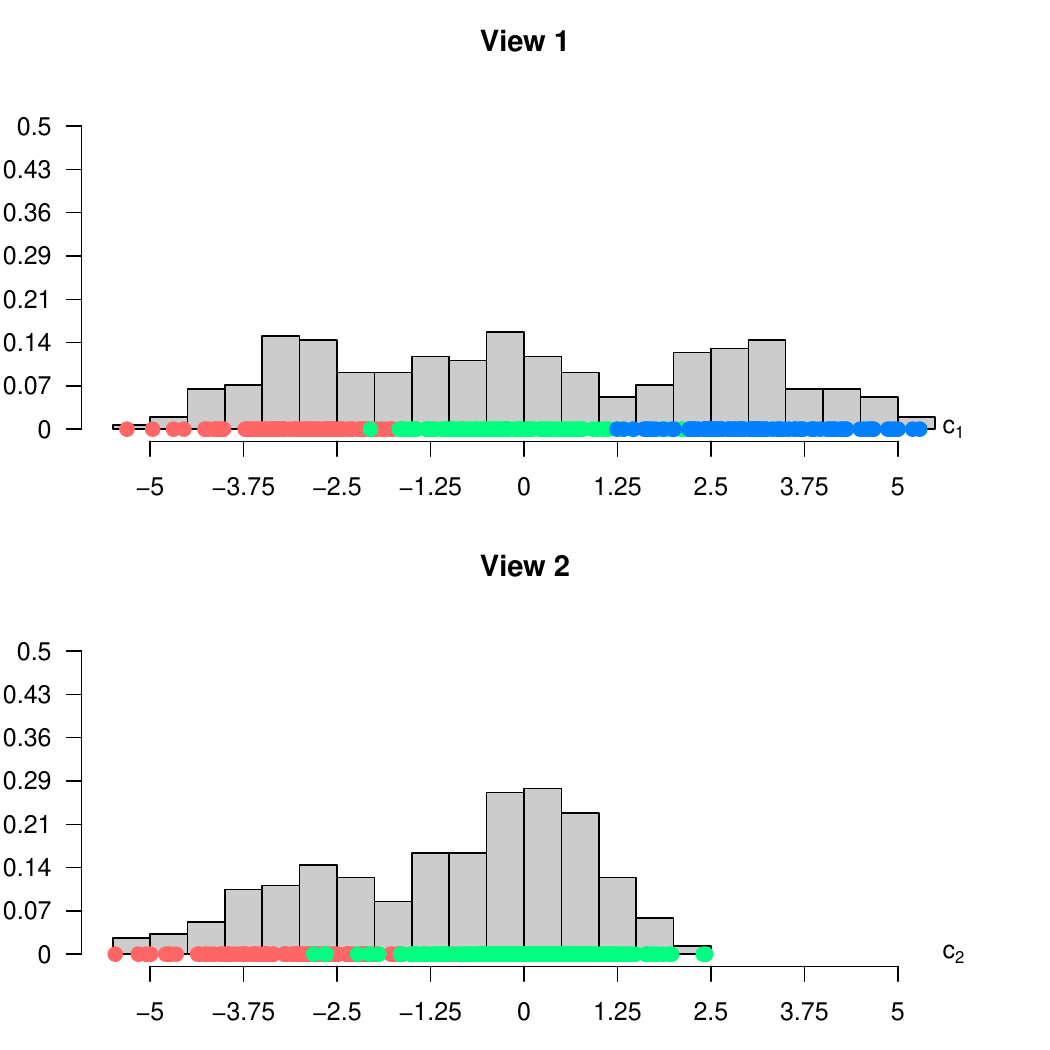}}
\subfloat[Scenario (c)]{\includegraphics[width=0.35\linewidth]{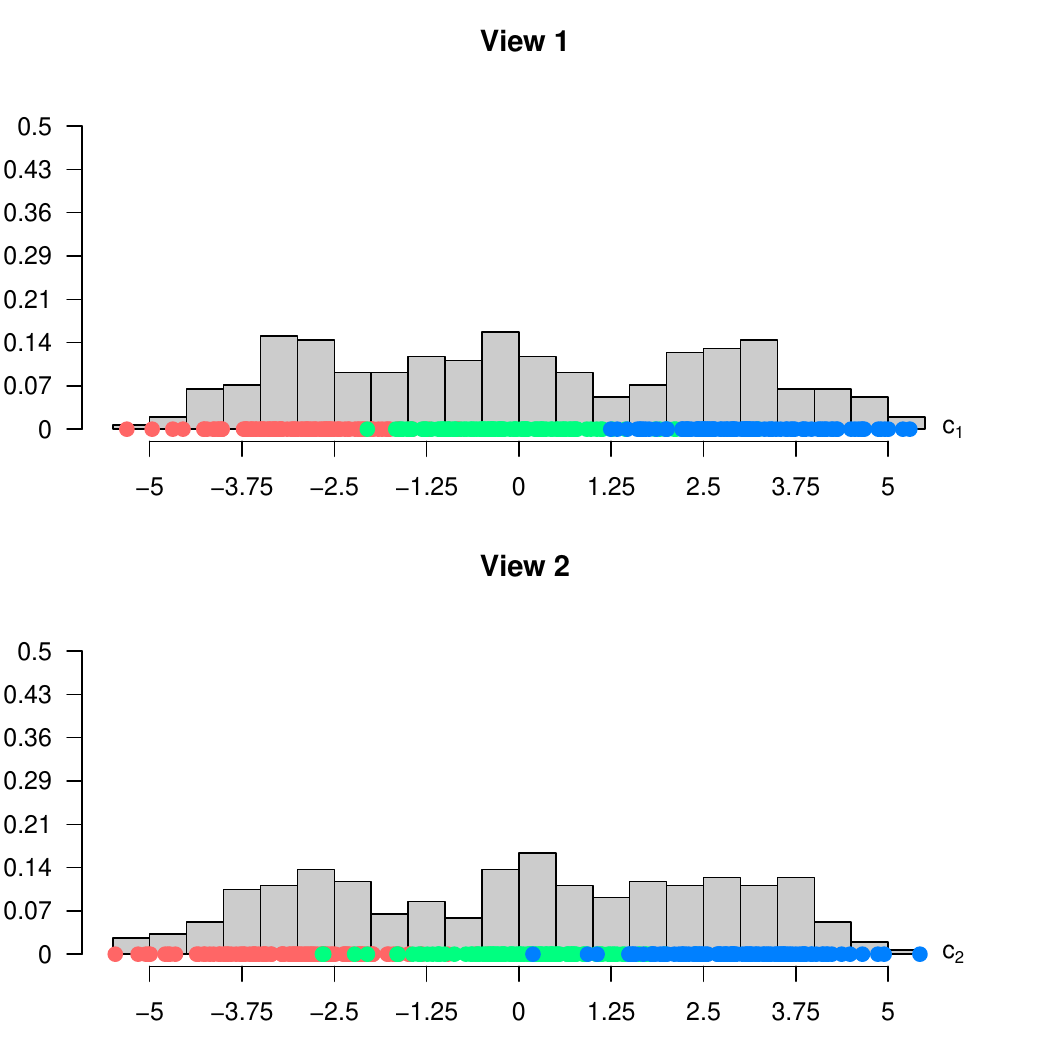}}
\caption{Simulation Study. Histograms of the simulated dataset across the $J=2$ views. Each panel refers to a simulation setting (a) - (c). Dots represent the sampled data, colored by the view's true partitions.}
\label{fig:Simul2_hist}
\end{figure}

We simulate data for the two views ($n = 300$, $J=2$) under three simulation settings, featuring an increasing proportion of overlapping clusters. Specifically, we have that the two views share (a) one third (100 units), (b) two thirds (200 units) or (c) all (300 units) of the labels across clusters. However, (a) and (b) formally share the same clustering structure with 200 units grouped in the same way, with the difference that in (a) both views only have only two clusters, while in (b) three clusters are present in the first view. This means that only 100 units in both (a) and (b) are effectively in different clusters across the two views.
A graphical representation of the true partitions $\bm c_1$ and $\bm c_2$ is reported in Figure \ref{fig:Simul2_truecj}. The $n=300$ observations in each view of the synthetic dataset are simulated from a mixture of univariate normals with parameters $\bm \theta_{jm} = \left(\mu_{jm}, \sigma^2_{jm}\right)$, for $(m,j)\in[M]\times[J]$ where $M=3$ and $J=2$, $\mu_{jm} \in \{-3,0,3\}$, and $\sigma^2_{jm} = 1$ for all $j$ and $m$.  The resulting generated data sets are shown in Figure \ref{fig:Simul2_hist} where each point's color indicates the cluster to which the observation belongs as shown in Figure \ref{fig:Simul2_truecj}. Note once again that the color refers to the cluster in each view and not to the unique values of $\bm \theta_{jm}$ associated with it. For example, in view 1 of data set (a) $\mu_{11} = -3$, $\mu_{12} = 0$ and for view 2 $\mu_{21} = -3$, $\mu_{22} = 3$.

\begin{figure}
\centering
\subfloat[(a) $\bm c_0$]{\includegraphics[width=0.35\linewidth]{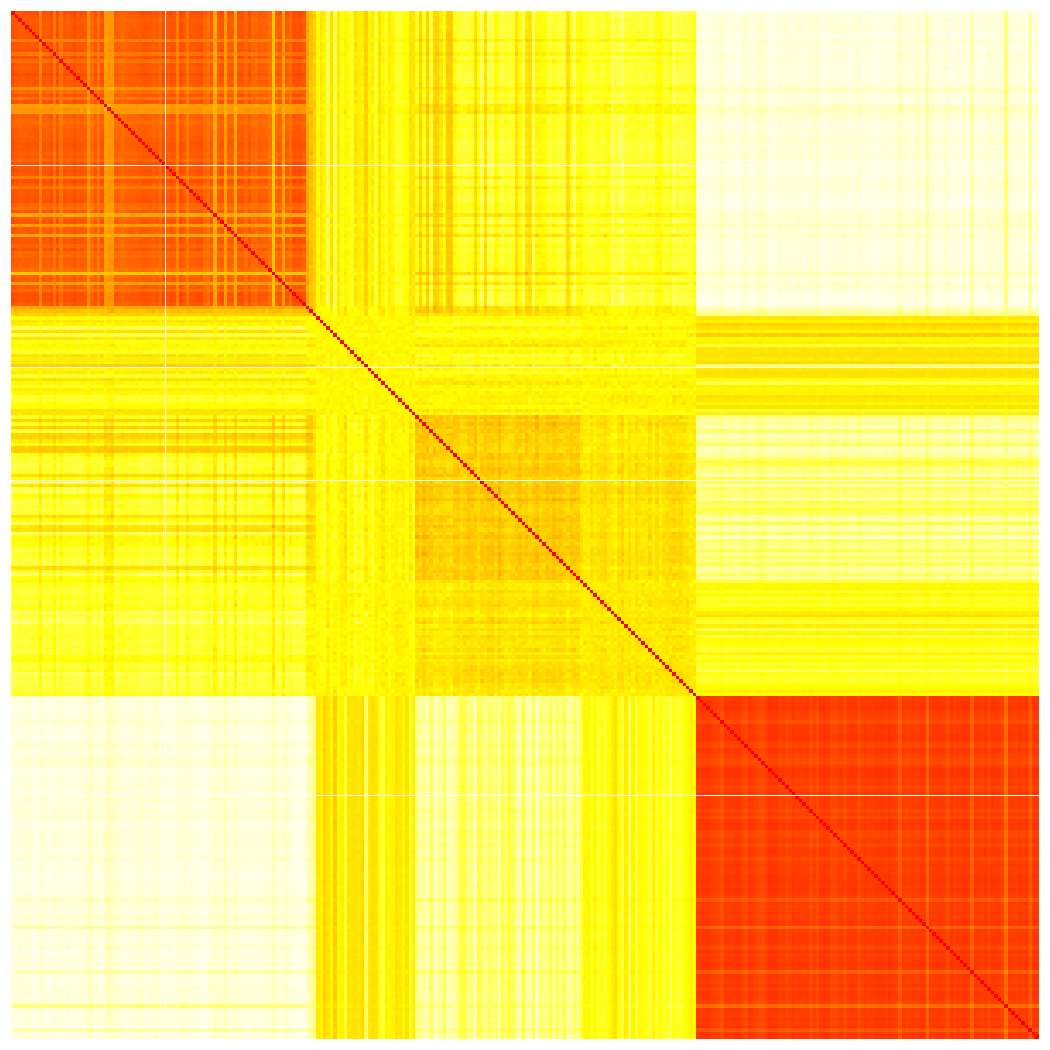}}
\subfloat[(a) $\bm c_1$]{\includegraphics[width=0.35\linewidth]{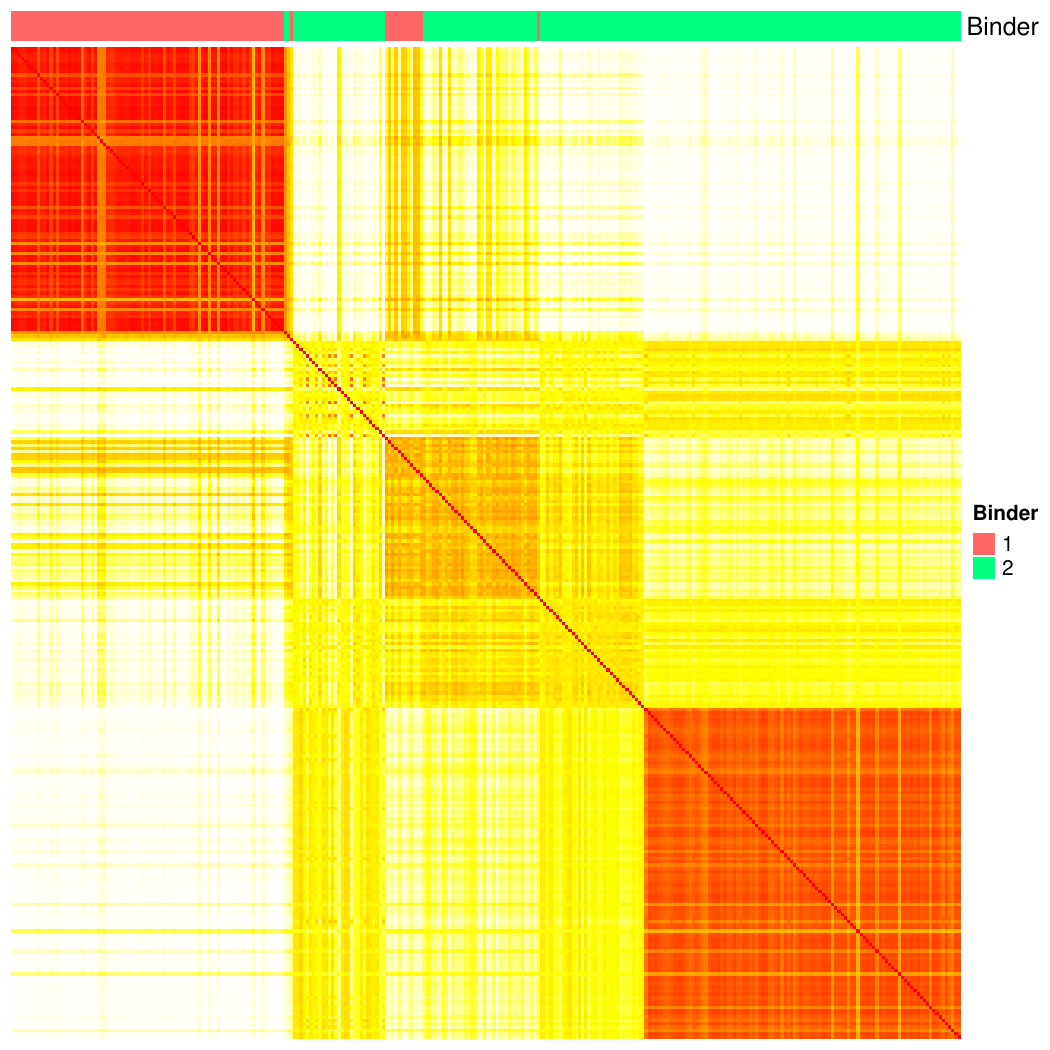}}
\subfloat[(a) $\bm c_2$]{\includegraphics[width=0.35\linewidth]{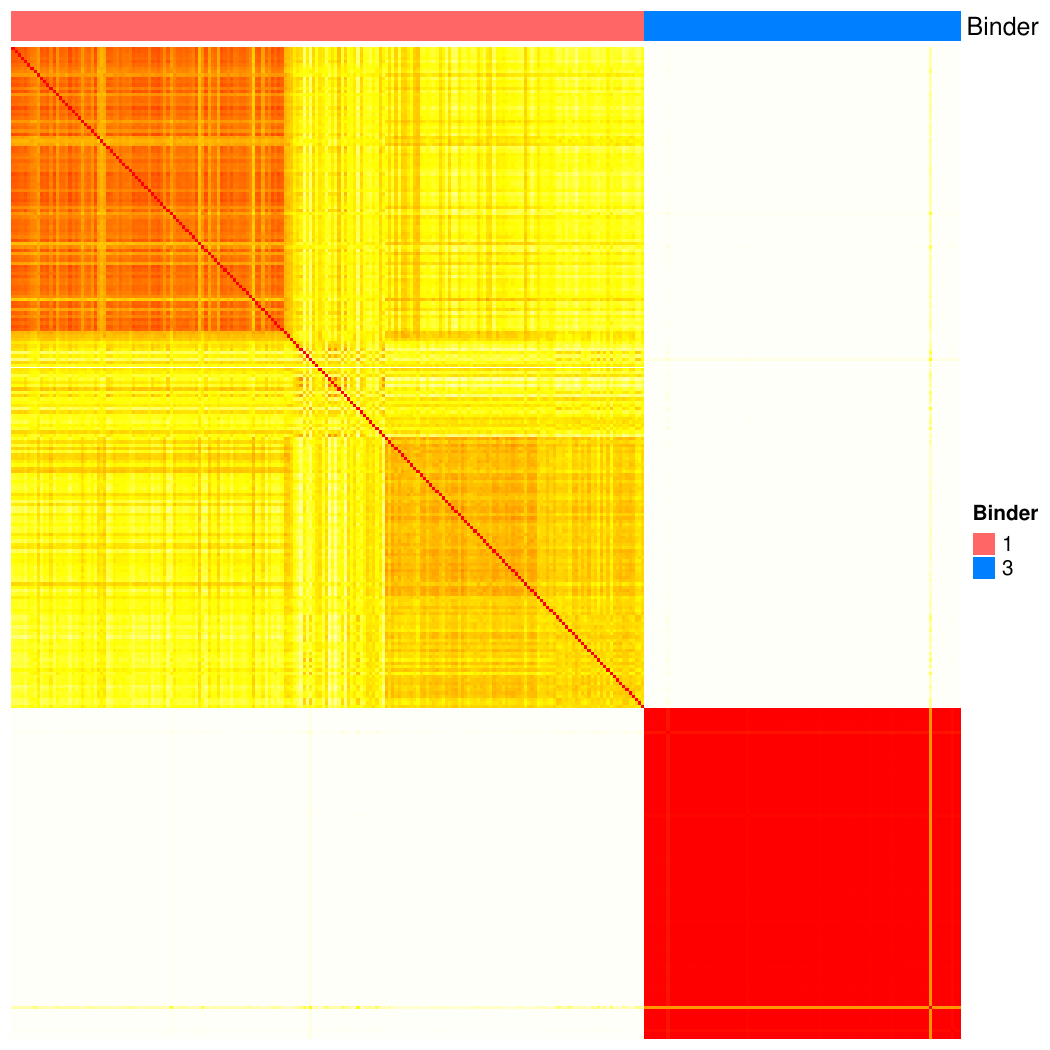}}
\caption{Simulation Study. Posterior co-clustering probabilities for the simulation scenario (a), sorted according to the Binder estimate of $\bm c_0$.}
\label{fig:Simul2_c_a_est}
\end{figure}

\begin{figure}
\centering
\subfloat[(b) $\bm c_0$]{\includegraphics[width=0.35\linewidth]{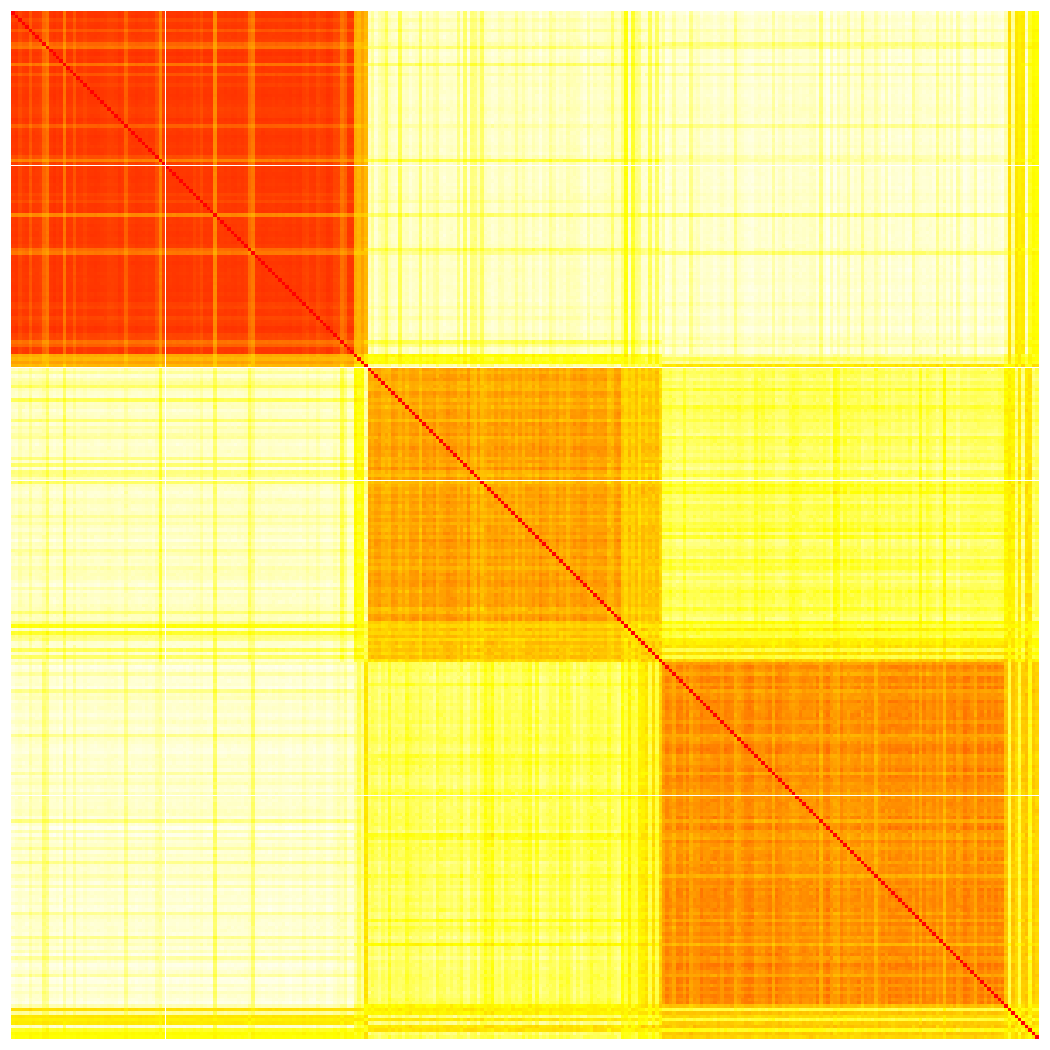}}
\subfloat[(b) $\bm c_1$]{\includegraphics[width=0.35\linewidth]{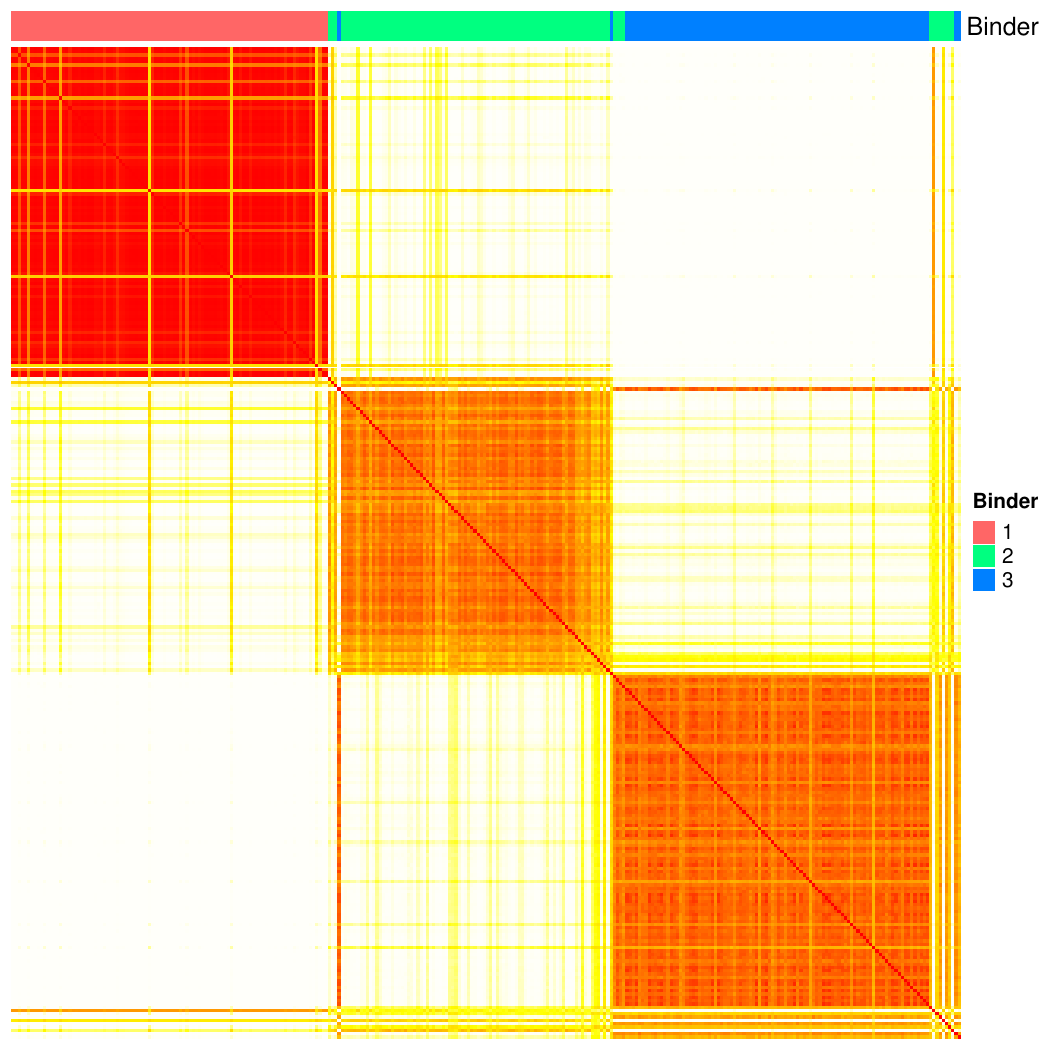}}
\subfloat[(b) $\bm c_2$]{\includegraphics[width=0.35\linewidth]{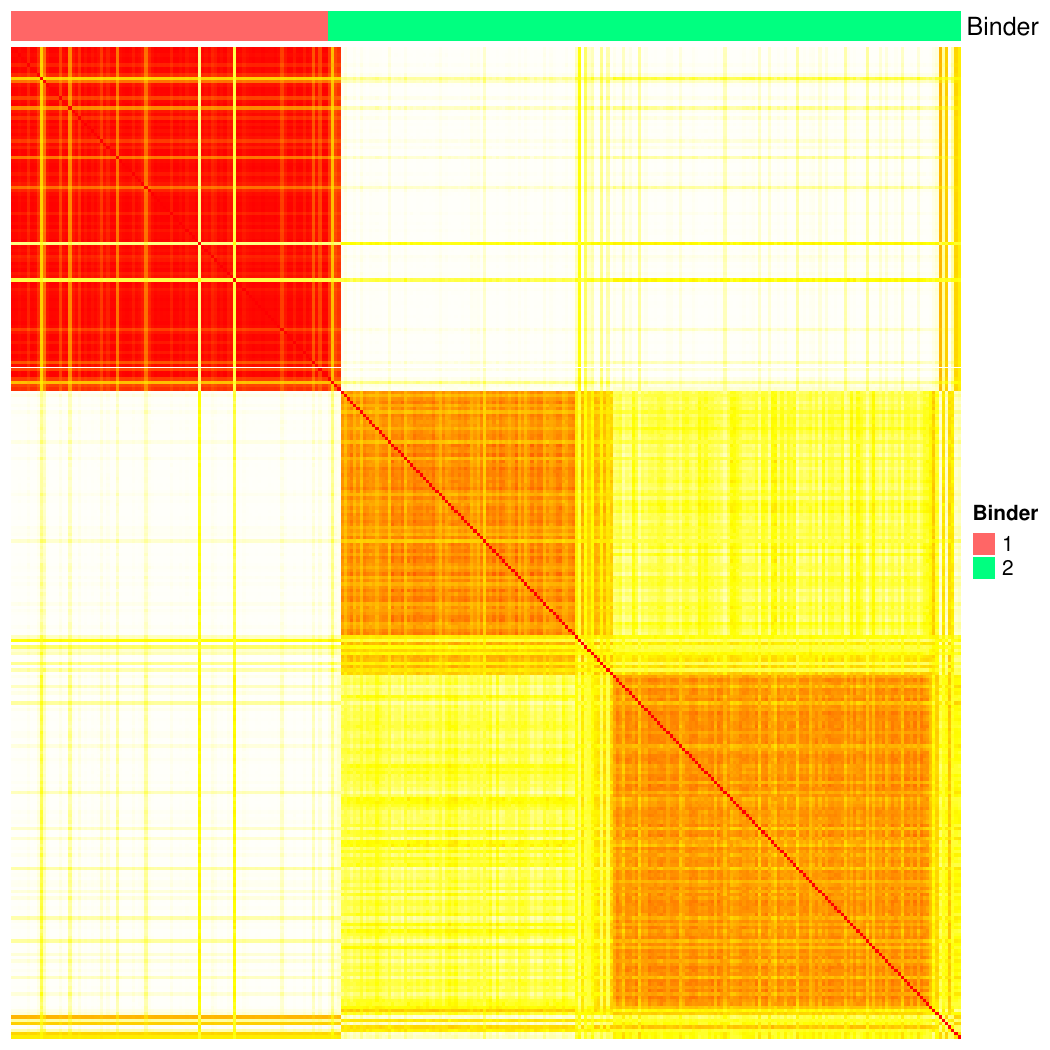}}
\caption{Simulation Study. Posterior co-clustering probabilities for the simulation scenario (b), sorted according to the Binder estimate of $\bm c_0$.}
\label{fig:Simul2_c_b_est}
\end{figure}

\begin{figure}
\centering
\subfloat[(c) $\bm c_0$]{\includegraphics[width=0.35\linewidth]{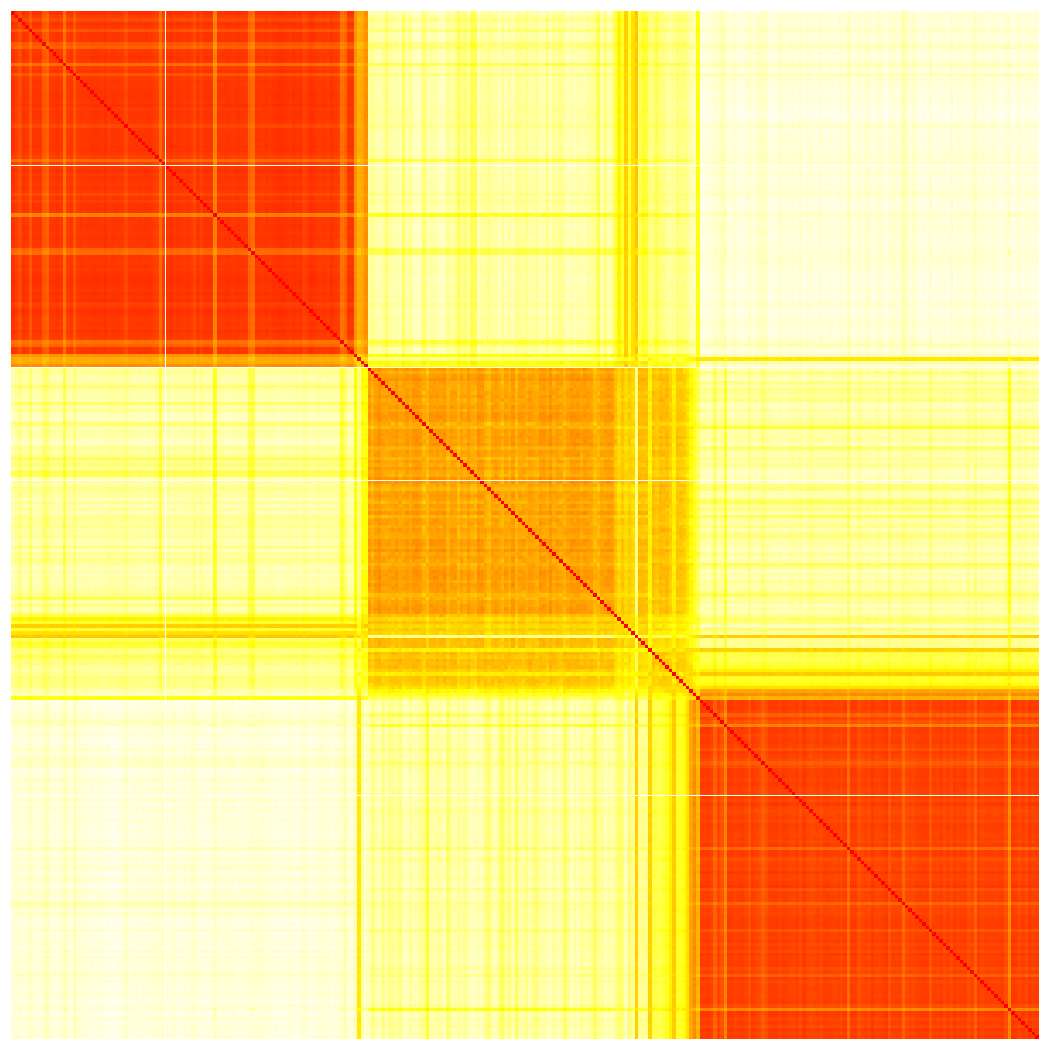}}
\subfloat[(c) $\bm c_1$]{\includegraphics[width=0.35\linewidth]{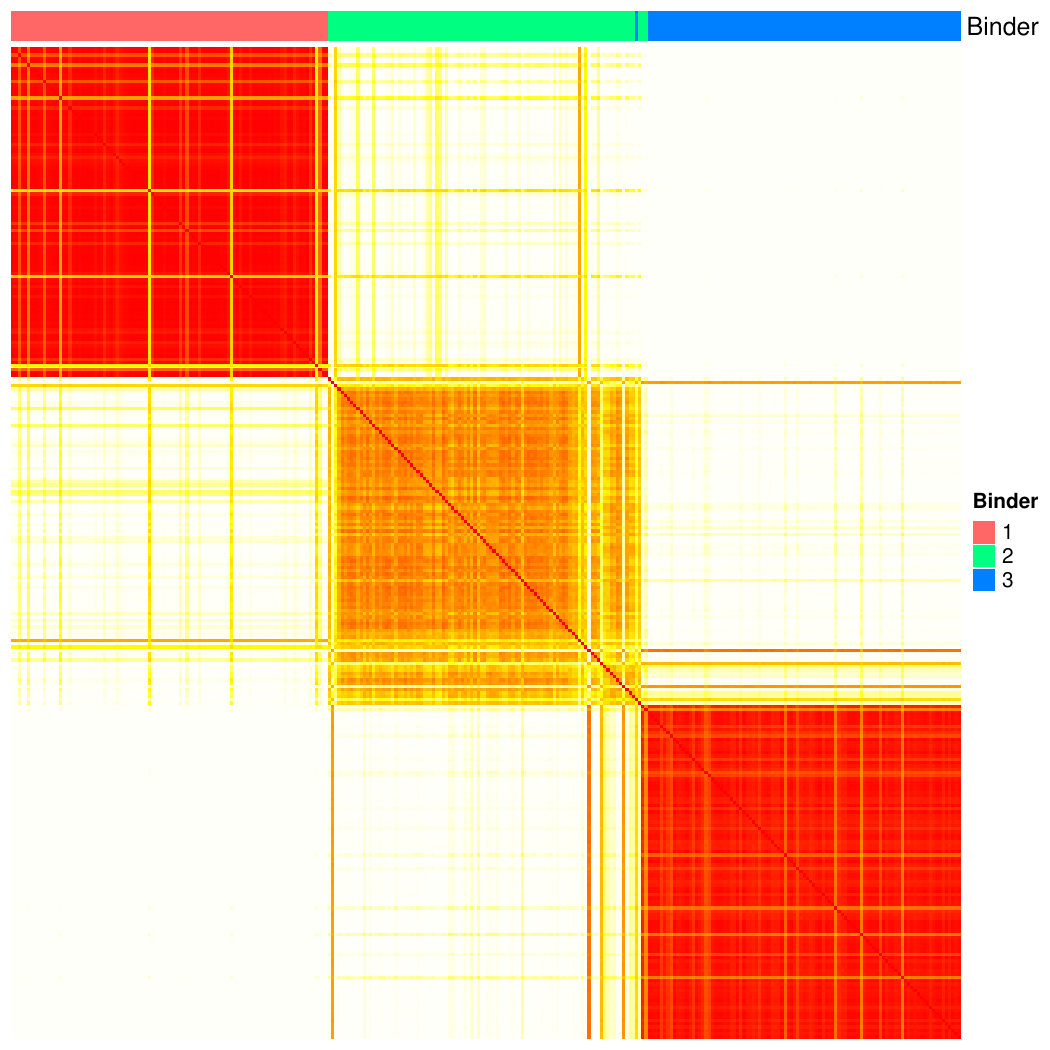}}
\subfloat[(c) $\bm c_2$]{\includegraphics[width=0.35\linewidth]{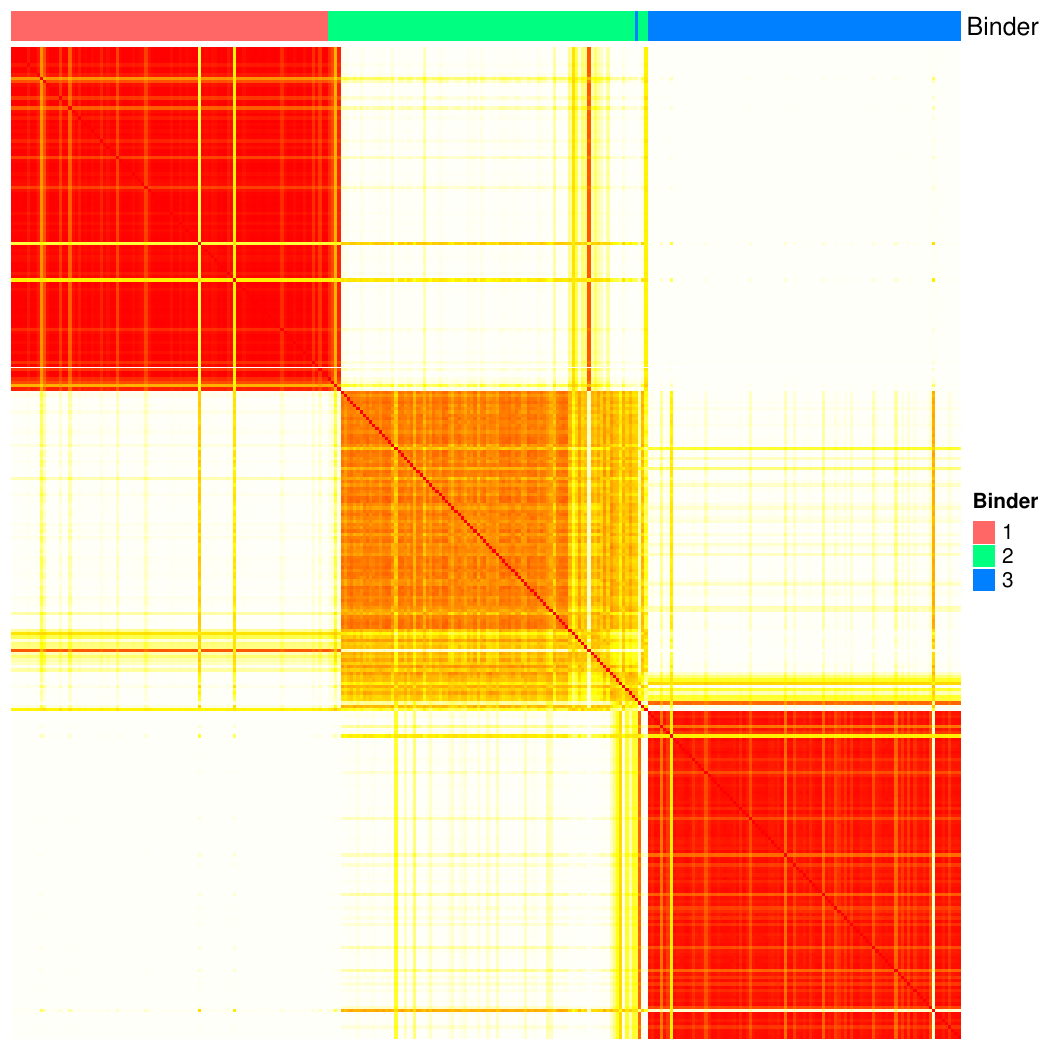}}
\caption{Simulation Study. Posterior co-clustering probabilities for the simulation scenario (c), sorted according to the Binder estimate of $\bm c_0$.}
\label{fig:Simul2_c_d_est}
\end{figure}

We run the MCMC algorithm for 3500 iterations, described in details in Appendix~\ref{app:MCMC}, discarding the first 1000 as burn-in and using the remaining 2500 for posterior inference. In all scenarios, we fix $G_j(\mu, \sigma^2) = \text{N}\left(\mu \mid 0, \sigma^2\right) \text{Inv-Gamma}\left(\sigma^2 \mid 3, 2\right)$ for $j = 1, 2$, and $\alpha_0 = \alpha = 0.1$. We show in Figures \ref{fig:Simul2_c_a_est}-\ref{fig:Simul2_c_d_est} the heatmaps of the posterior co-clustering probabilities, for the baseline partition $\bm c_0$, as well as the two views $\bm c_1$ and $\bm c_2$. We can observe how the posterior estimates for $\bm c_0$ correspond to a consensus between the two other views, increasing the co-clustering uncertainty when the partition for the two views do not agree. However, when there is agreement across views, the posterior estimates of $\bm c_0$ show much less uncertainty. This behavior is also reflected in the posterior co-clustering probabilities across the two views $\bm c_1$ and $\bm c_2$.  For example, both views in data set (a) are composed of two clusters which are clearly seen in the right two co-clustering probability matrices of Figure \ref{fig:Simul2_c_a_est}.  However, view 1 is composed of clusters that are less separated and this uncertainty is evident in the middle co-clustering probability matrix in Figure \ref{fig:Simul2_c_a_est}.  The ``majority rule'' behavior inherent in our modeling approach is on full display in the left co-clustering probability matrix of Figure \ref{fig:Simul2_c_a_est}.  Notice that the units that are co-clustered with high probability in view 1 and view 2 are co-clustered with high probability in the baseline as well.  Where as, units for which there is one view with co-clustering probability is small (i.e., clustering is more uncertain), this uncertainty is propagated to the baseline clustering.

\clearpage
\section{Application to the GUSTO cohort}\label{sec:gusto}

\subsection{Data Description}
Studying the clustering of obesity, asthma, and hypertension in children is critical, as these interrelated conditions not only impair health and quality of life in early years but also track into adulthood, amplifying the long-term burden of cardiometabolic and respiratory disease.
Comorbidities between obesity, asthma, and hypertension in children are increasingly recognized as a major public health concern \citep{reyes2022obesity, di2023pediatric, ma2025interaction}. Obesity has emerged as the central driver of this triad, with excess adiposity contributing to systemic inflammation, altered lung mechanics, and metabolic dysregulation, thereby increasing the risk of both asthma and elevated blood pressure \citep{pulgaron2013childhood}. Epidemiological studies consistently show that obese children have a significantly higher prevalence of asthma, with each unit increase in BMI associated with an incremental rise in asthma risk, and are up to seven times more likely to develop hypertension compared to their normal-weight peers \citep{ma2025interaction}. Moreover, children with coexisting obesity and asthma are more prone to poorly controlled symptoms, severe exacerbations, and higher healthcare utilization, while the presence of hypertension in this group compounds long-term cardiovascular risk \citep{lang2021contribution, averill2024management}. These overlapping conditions not only share biological mechanisms, such as chronic low-grade inflammation, insulin resistance, and dysregulated autonomic activity \citep{pulgaron2013childhood, di2023pediatric}, but are also affected by social determinants, disproportionately affecting children in lower socioeconomic and minority populations \citep{herrera2024holistic, zhang2024global}. Collectively, the clustering of obesity, asthma, and hypertension in childhood highlights the need for integrated prevention and management strategies targeting lifestyle, environmental exposures, and equitable access to care.

In this section we apply the proposed framework to data from the GUSTO cohort study \citep{soh2014cohort}. The Growing Up in Singapore Towards healthy Outcomes (GUSTO) study is a longitudinal cohort study, started in 2009 and still ongoing, following Singaporean mothers and their children, collecting a plethora of clinical and biomedical data ranging from mental health, to growth measurements, as well as perinatal information such as gestational diabetes. The cohort study aims to provide a comprehensive view of the life development of mother-child pairs, spanning a time frame that goes from prenatal months to adolescence. 
In Singapore, a 2001 study found asthma in 27.4\% of children aged 12–15 years \citep{wang2004prevalence}, while childhood obesity has risen tenfold over the past four decades, reaching 12\% in 2016 among school-aged children \textbf{(from singhealth website)}. Among these, 70\% have at least one cardiovascular risk factor and 39\% have two or more \citep{friedemann2012cardiovascular}. In the GUSTO cohort study nearly half of children with signs of pre-hypertension, are associated to maternal blood pressure and early-life adiposity \citep{yuan2021trajectories}.

In this work, we focus on the GUSTO children ($n = 771$), and measurements related to obesity, hypertension and asthma. To study these, we use data on growth trajectories, blood pressure levels and frequency of wheezing episodes, respectively. These subsets of the GUSTO data represent the three views in the proposed latent modularity  approach. Additionally, we include in the model the following fixed-effects covariates: \textit{Gestational Age}, \textit{Age at Delivery}, \textit{pre-Pregnancy BMI}, \textit{Sex} of the infant, \textit{Ethnicity} (Chinese, Malay, Indian), \textit{Education} level (below University, University and above), self-reported \textit{Diabetes} diagnose. After creating suitable dummy variables, we obtain $q_X = 8$ covariates. The continuous covariates (i.e., \textit{Gestational Age}, \textit{Age at Delivery}, \textit{pre-Pregnancy BMI}) are standardized before the analysis. Below, we provide a detailed description of the data used in each of the three views, as well as the view-specific sub-models. Here we the define the view-specific likelihood as well as the prior distributions for view-specific parameter vectors.  It is important to notice how the support spaces of the different views are widely different. As such, a novel contribution of our work is the development of a statistical method for the analysis of zero-inflated panel count data corresponding to view $j=3$. 

\subsection{Modeling}

In this section, to connect with the global model in Section \ref{sec:stat_model}, we will now describe $J=3$ datasets and the associated models $f_j$,  the data model, and
$G_j$ the prior model on the unknown parameters of the model $f_j$.  These specifications feature original modeling contributions.

\subsubsection{Z-BMI}
Body mass index (BMI) is measured at $T^Z = 22$ unequally spaced time points from birth to 10 years of age. The raw measurements are standardized, following age- and sex-specific WHO criteria \citep{onyango2004managing, dinsdale2011simple}. We model the resulting Z-BMI trajectories via a spline regression, taking into account that subjects may only have measurements at a subset of the $T^Z$ observation times.  This view is represented by the index
$j=1$ and this is reflected in the notation below.

Let $\bm Z_i=(Z_{t^Z_{i1}},\dots,Z_{t^Z_{iT^Z_i}})$ be the longitudinal Z-BMI score for subject $i$ over the time points $\bm t^Z_i = \left(t^Z_{i1}, \dots, t^Z_{iT^Z_i} \right)$, which may not be the same across subjects. We model these trajectories via B-splines regression:
\begin{align*}
    \bm Z_i = \bm B_i \bm \beta^*_{c_{1i}} + \bm \eta^Z \bm X_i \mathbf{1}_{T^Z_i} + \bm \epsilon_i, \quad i\in[n]
\end{align*}
where $\bm \beta^*_{c_{1i}}$ is the unique value associated with cluster $c_{1i}$ in the Z-BMI data view, to which the $i$-th observation belongs, and $\mathbf{1}_{T^Z_i}$ is a vector of ones of length $T^Z_i$. In the likelihood above, $\bm B_i$ is matrix of  B-spline basis functions for the $i$-th subject of dimension $T^Z_i \times d_B$, where $d_B$ is the number of spline functions, calculated as $d_B = \bar{m} + 2 + d$, with $\bar{m}$ the number of internal knots and $d$ the degree of the spline. Here, we have $\bar{m}=2$ and $d=3$, and $\bm \beta$ is of dimension $d_B = 7$. The columns of matrix $\bm{X}_i$ contain the $q_X=8$ subject-specific time-invariant covariate values described earlier and $\bm{\eta}^Z$ the corresponding vector of regression coefficients. Furthermore, we assume time-specific variance parameters $\bm \Sigma_Z = \text{diag}\left(\sigma^2_{Z,1}, \dots, \sigma^2_{Z,T^Z}\right)$, so that $\bm \epsilon_i \sim N\left(\bm 0, \left[\bm \Sigma_Z\right]_{\bm t^Z_i}\right)$, for $i\in[n]$. We complete this sub-model by specifying the following prior distributions:
\begin{align*}
    \bm \beta^*_1, \dots, \bm \beta^*_M \mid M &\sim \text{N}_{n_B}\left(\bm \mu_{\bm \beta}, \bm \Sigma_{\bm \beta}\right)\\
    \sigma^2_{Z,1}, \dots, \sigma^2_{Z,T^Z} &\iid \text{Inv-Gamma}\left(\alpha_{\sigma^2_Z}, \beta_{\sigma^2_Z}\right)\\
    \bm \eta^Z &\sim \text{N}_{q_X}\left(\bm \mu_{\eta^Z}, \bm \Sigma_{\eta^Z}\right)
\end{align*}

\subsubsection{Hypertension}
Hypertensive and normal state are defined in terms of blood pressure measurements, available in  the GUSTO study \citep{lim2015maternal}.
A two-state Markov model in continuous time is adopted to model the hypertension data. For each subject, this binary response is observed at time points $t \in \bm t^H_i = \{t^H_{i1}, \dots, t^H_{iT^H_i}\}$ with $H_{it} = 0$ indicating normal blood pressure and $H_{it}=1$ indicating high blood pressure (hypertension). Just as in the previous data view, the subject-specific observation times may not be the same across subjects. The main assumption is that 
\begin{align*}
    \mathbb{P}(H_{it}\mid H_{i1},\dots,H_{it-1}) = \mathbb{P}(H_{it}\mid H_{it-1})
\end{align*}
This view is represented by the index $j=2$ which is reflected in the notation below.

The transition intensities $\lambda_{rs}$, for $r,s \in \{ 0,1\}$ and $r=1-s$, of the multi-state model indicate the instantaneous risk of moving from state $r$ to state $s$, are defined as follows
\begin{align*}
    \lambda_{rs} = \displaystyle\lim_{\Delta_t\to 0} \frac{\mathbb{P}(H_{t+\Delta_t}=s\mid H_t=r)}{\Delta_t}
\end{align*} 
The probability of changing state over a time period $\epsilon$ is given by the following stochastic matrix:
$$ P^H(\epsilon; \bm \lambda) =\begin{bmatrix}
p_{0\rightarrow 0}(\epsilon; \bm \lambda) & p_{0\rightarrow 1}(\epsilon; \bm \lambda) \\
p_{1\rightarrow 0}(\epsilon; \bm \lambda) & p_{1\rightarrow 1}(\epsilon; \bm \lambda) 
\end{bmatrix} = 
\begin{bmatrix}
1-p_{0\rightarrow 1}(\epsilon; \bm \lambda) & p_{0\rightarrow 1}(\epsilon; \bm \lambda) \\
p_{1\rightarrow 0}(\epsilon; \bm \lambda) & 1-p_{1\rightarrow 0}(\epsilon; \bm \lambda) 
\end{bmatrix}
$$
By solving the Chapman-Kolmogorov equations, the above probabilities can be expressed in closed form \citep{ross1995stochastic}:
\begin{align*}
    p_{0\rightarrow 1}(\epsilon; \bm \lambda) &= \frac{\lambda_{01}}{\lambda_{01}+\lambda_{10}}(1-\exp(-(\lambda_{01}+\lambda_{10})\epsilon))\\
    p_{1\rightarrow 0}(\epsilon; \bm \lambda) &= \frac{\lambda_{10}}{\lambda_{01}+\lambda_{10}}(1-\exp(-(\lambda_{01}+\lambda_{10})\epsilon))
\end{align*}

We let the transition intensities be subject-specific $\bm \lambda_i = (\lambda_{i01},\lambda_{i10})$, and we include them in the model for clustering the Hypertension view. Moreover, homogeneous covariate effects can be included in the transition intensities. We specify the following hierarchical sub-model for the Hypertension view:
\begin{align*}
    p\left(\bm H_i \mid \bm \lambda^*_{c_{2i}},H_{i1} \right) &= \prod_{t \in \bm t^H_i} p_{H_{it-1}\rightarrow H_{it}} \left(\epsilon_{it}; \bm \lambda^*_{c_{2i}} \right) \\
    \lambda_{i,rs} &= \lambda^*_{c_{2i},rs} \exp(\bm \eta^H_{rs} \bm X_i ), \quad \bm \lambda^*_m = \left(\lambda^*_{m,01}, \lambda^*_{m,10} \right) \\
    \bm \lambda^*_1, \dots, \bm \lambda^*_M \mid M & \iid \text{log-Normal}\left(\bm \mu_{\bm \lambda}, \text{diag}\left(\bm \sigma^2_H \right) \right), \quad \bm \sigma^2_H = \left(\sigma^2_{H,01}, \sigma^2_{H,10} \right) \\
    \sigma^2_{H,rs} &\sim \text{Inv-Gamma}\left(\alpha_{\sigma^2_H}, \beta_{\sigma^2_H}\right) \\
    \bm \eta^H_{rs} &\sim \text{N}_{q_X}\left(\bm \mu_{\bm \eta^H_{rs}}, \bm \Sigma_{\bm \eta^H_{rs}}\right)
\end{align*}
where $\lambda_{irs}$ is the transition intensity for subject $i$ with state changing from $r$ to $s$. We indicate by $\bm c_2 = \left(c_{21}, \dots, c_{2n} \right)$ the vector of clustering allocations of the Hypertension data, and by $\bm \eta^H_{rs}$ the vector of fixed effects for covariates $\bm X$.

\subsubsection{Wheezing}

%Panel Count Data -- 
Asthma-related data are recorded on the same subjects from three months of age up to six years. Specifically, for subject $i$ at time $t$, we observe the number of new wheezing episodes recorded since the previous time point. We denote the vectors of panel counts as $\bm W_i = \left(W_{i1}, \dots, W_{iT^W_i} \right)$, for $i\in[n]$, where $T^W_i$ is the number of time points observed for subject $i$. Just as before, the time points $\bm t^W_i = \{t^W_{i1}, \dots, t^W_{iT^W_i}\}$ at which the wheezing episodes are measured may not be all the same across individuals.  This view represents the index $j=3$ which is reflected in the notation below.

We propose a zero-inflated Poisson process to model these data as follows. For subject $i\in[n]$, at time $t\in[T^W_i]$ we have:
\begin{align*}
    \mathbb{P}\left(W_{it}=0 \mid p_i, \mu_{it}\right) &= p_i + (1-p_i) e^{-\mu_{it}} \\
        \mathbb{P}\left(W_{it} = k \mid p_i, \mu_{it}\right) &= (1-p_i) \frac{\mu_{it}^k e^{-\mu_{it}}}{k!}, \quad k = 1, 2, \ldots  
\end{align*}
where $p_i$ is the probability of zero inflation and $\mu_{it}$ is the mean of the Poisson process.

To simplify inference, we introduce the auxiliary random binary vectors $\bm b_i = \left(b_{it^W_{i1}}, \dots, b_{it^W_{iT^W_i}} \right)$ indicating, conditionally on $W_{it}=0$, whether this comes from the point mass or from the Poisson part of the zero-inflated distribution
\begin{equation*}
    \mathbb{P}\left(b_{it} = z \mid W_{it} = 0 \right) =  
    \begin{cases}
        \displaystyle \frac{p_i e^{-\mu_{it}}}{p_i + (1 - p_i) e^{-\mu_{it}}} & \text{if } z = 0 \quad \text{(point mass)} \\
        \displaystyle \frac{(1 - p_i) e^{-\mu_{it}}}{p_i + (1 - p_i) e^{-\mu_{it}}} & \text{if } z = 1 \quad \text{(Poisson)}
    \end{cases}
\end{equation*}
and conditionally on $W_{it} > 0$
\begin{equation*}
    \mathbb{P}\left(b_{it} = z \mid W_{it} > 0 \right) =  
    \begin{cases}
        \displaystyle 0 & \text{if } z = 0 \\
        \displaystyle 1 & \text{if } z = 1
    \end{cases}
\end{equation*}

For each subject $i$, define three subset of time indices for the vector of observations $\bm W_i$: 
\begin{align*}
    A_i &= \{ t \in \bm t^W_i : W_{it}=0, b_{it}=0 \} \\
    B_i &= \{ t \in \bm t^W_i : W_{it}=0, b_{it}=1 \} \\
    C_i &= \{ t \in \bm t^W_i : W_{it}>0 \}
\end{align*}
Denote by $n_{A_i}=\text{Card}(A_i), n_{B_i}=\text{Card}(B_i), n_{C_i}=\text{Card}(C_i)$.  It is straightforward to see that $T^W_i = n_{A_i} +n_{B_i}+ n_{C_i}$. The joint distribution of $\bm W_i$ and $\bm b_i$ has the form:
$$
    p\left(\bm W_i, \bm b_i \mid  \bm \mu_i, p_i\right) = p_i^{n_{A_i}} \left(1-p_i\right)^{n_{B_i}} \left(1-p_i\right)^{n_{C_i}} \prod_{t \in A_i \cup B_i} e^{-\mu_{it}} \prod_{t \in C_i} \frac{\mu_{it}^{W_{it}} e^{-\mu_{it}}}{W_{it}!}
$$

We model the mean of the Poisson process via monotone I-splines \citep{ramsay1988monotone}:
$$
\mu_{it} = \left(\sum_{l=1}^L r_{il} \left(I_l(t_{it}) - I_l(t_{it-1}) \right)\right) \exp\left(\bm \eta^W \bm X_i\right)
$$
where $I$ is the I-spline basis function matrix with $L$ functions, which is determined as $L=m+d+1$ with $m$ the number interior knots over the time window and $d$ the degree. We fix $m = 5$ and $d = 3$. We indicate by $r_{il}$ the coefficients of the I-spline functions. Denote covariates as $\bm X_i$ and coefficient as $\bm \eta^W$. 

To allow for a conjugate update of $r_{il}$, we exploit the infinite divisibility property of the Poisson distribution. We introduce $L$ independent Poisson-distributed random variables $Y_{itl}$ with rate parameters $\mu_{itl} = r_{il} \left( I_l(t_{it}) - I_l(t_{it-1}\right)$, for $l\in[L]$, such that $Y_{it} = \sum_{l=1}^L Y_{itl}$ and $\sum_{l=1}^L \mu_{itl} = \mu_{it}$, for $t \in B_i \cup C_i$, i.e. for those counts that are generated from a Poisson distribution. Then, we have:
$$
    Y_{itl} \mid r_{il} \sim \text{Poi}\left(\mu_{itl}\right), \quad \mu_{itl} = \left( r_{il} \left( I_l(t_{it}) - I_l(t_{it-1}\right) \right) \exp\left(\bm \eta^W \bm X_i\right)
$$
Note that we can write $W_{it} = b_{it} Y_{it}$, yielding the following joint distribution for $\bm W_i$ and $\bm b_i$:
\begin{align*}
    p\left(\bm W_i, \bm b_i \mid  \bm \mu_i, p_i\right) &= p_i^{n_{A_i}} \left(1-p_i\right)^{n_{B_i}} \left(1-p_i\right)^{n_{C_i}} \prod_{t \in A_i \cup B_i} p\left(W_{it}=0 \mid \mu_{it}\right) \prod_{t \in C_i} p\left(W_{it} \mid \mu_{it}\right) \\
    &= p_i^{n_{A_i}} (1-p_i)^{n_{B_i}+n_{C_i}} \prod_{t \in A_i \cup B_i} \prod_{l=1}^L p\left(W_{itl}=0 \mid \mu_{itl}\right)  \prod_{t \in C_i} \prod_{l=1}^L p\left(W_{itl} \mid \mu_{itl}\right) \\
    &= p_i^{n_{A_i}} (1-p_i)^{T^W_i - n_{A_i}}  \exp\left(-\sum_{t \in A_i \cup B_i} \sum_{l=1}^L\mu_{itl}\right) \left(\prod_{t \in C_i} \prod_{l=1}^L \frac{\mu_{itl}^{W_{itl}}}{W_{itl}!}\right) \times \\ & \exp\left(-\sum_{t \in C_i} \sum_{l=1}^L\mu_{itl}\right)
\end{align*}

Conditionally on the view-level partition $\bm c_3 = \left(c_{31}, \dots, c_{3n}\right)$, the final zero-inflated Poisson process model for our count panel data is:
\begin{align*}
    &W_{it} = b_{it} Y_{it} \\
    &b_{it} \mid p^*_{c_{3i}}, c_{3i} \sim \text{Ber}\left(1 - p^*_{c_{3i}}\right)\\
    &Y_{it} \mid r_{i1},\dots,r_{iL} \sim \text{Poi}\left(\mu_{it}\right) \\
    &\mu_{it} = \left(\sum_{l=1}^L r_{il}\left( I_l(t_{it}) - I_l(t_{it-1}) \right)\right) \exp\left(\bm \eta^W \bm X_i\right)\\
    &p^*_i\mid M \iid \text{Beta}\left(\alpha_p,\beta_p\right),\quad i\in[M]\\
    &r_{il} \mid \zeta^*_{c_{3i}}, c^H_i \sim \text{Exp}\left(\zeta^*_{c_{3i}}\right)\\
    &\zeta^*_i \mid M \iid \text{Gamma}\left(\alpha_{\zeta},\beta_{\zeta}\right),\quad i\in[M] \\
    &\bm \eta^W \sim \text{N}_{q_X}\left(\bm \mu_{\bm \eta^W}, \Sigma_{\bm \eta^W}\right)
\end{align*}

Finally we note that our model differs from the one proposed by  \cite{juarez2017joint} who models different count processes via zero-inflated Poisson processes with frailty term shared.  In a Hurdle model setting, instead of monotonic I-splines, they propose a transformation of B-splines to adapt to the parameters of interest (probability of zero inflation as well as intensity of the Poisson part). The zero-inflated and count components are linked via shared random effects.

\subsection{Model settings}
We fit the proposed model to the GUSTO data. Hyperparameters are specified in the following way: for multivariate Normal prior distributions, we fix the mean vectors to $\bm 0$ and the covariance matrices to the identity matrix $\mathbb{I}_q$ of appropriate dimension $q$, with exception of $\bm \Sigma_{\bm\beta} = 50 \mathbb{I}_{n_B}$; for inverse Gamma distributions, we choose shape and rate parameters equal to 3 and 2, respectively, corresponding to unitary prior expectation and variance. The shape parameters of the Beta distribution are set equal to 8 and 2, respectively, to reflect the excess of zeros in the asthma data set. The parameters governing the underlying clustering structures are set to $\alpha_0 = \alpha = 0.1$.

We develop a tailored MCMC algorithm, whose details are reported in Appendix~\ref{app:MCMC}. After 100 iterations used to initiate the adaptive steps of the algorithm, the MCMC chain is run for 50000 iterations. Of these, the first 40000 are discarded as burn-in period and the remaining 10000 are thinned to produce chains of size 5000 to use for posterior inference.

\subsection{Posterior Inference}
\begin{figure}
\centering
\scalebox{0.8}{\input{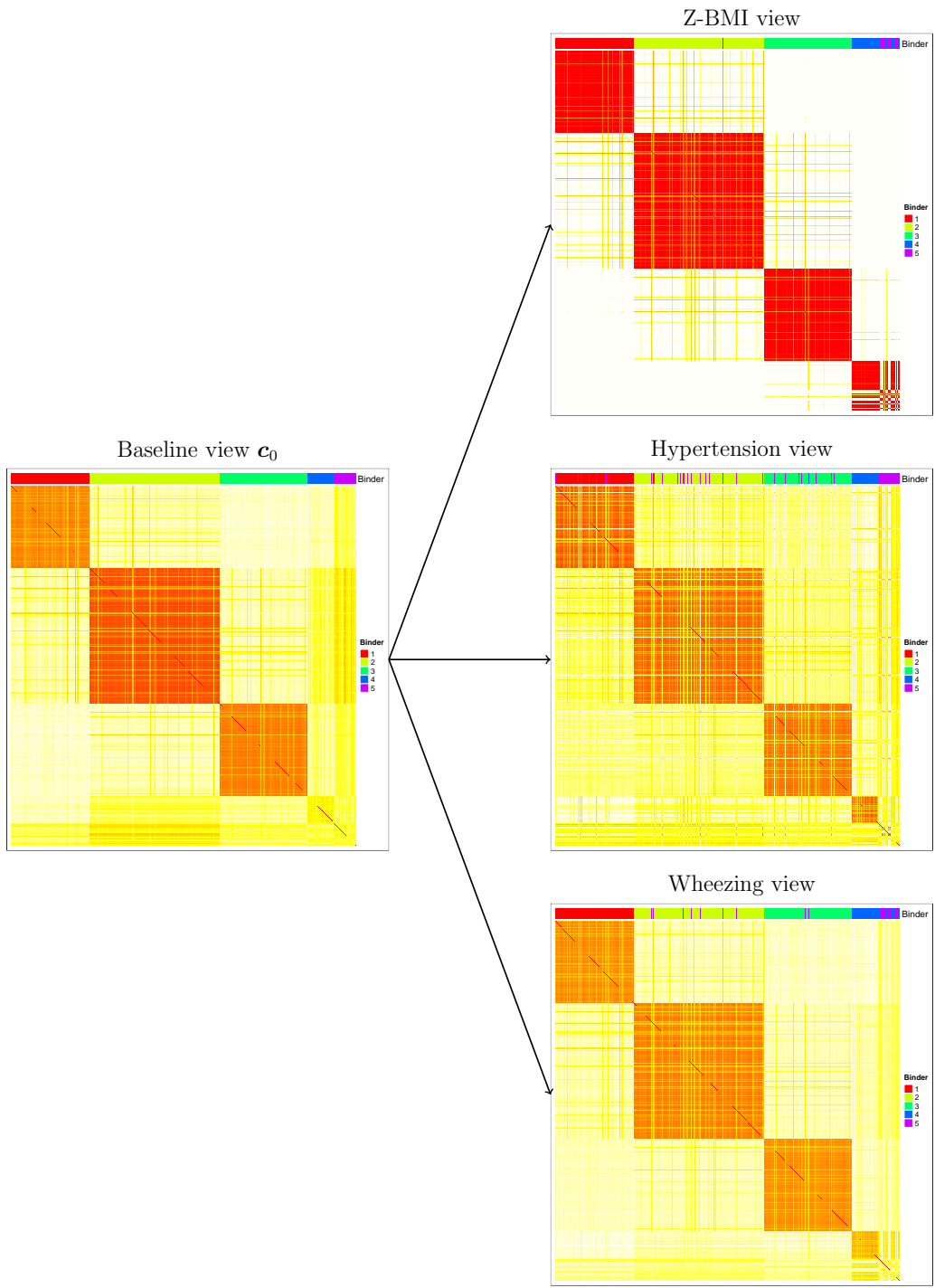}}
\caption{GUSTO data. Heatmaps of the posterior co-clustering probabilities within the baseline $\bm c_0$ (left node) and the three view (right nodes). The entries are sorted according to the Binder estimate of the baseline partition $\bm c_0$. The views show different degree of uncertainty, reflected in the posterior co-clustering probabilities for $\bm c_0$.}
\label{fig:GUSTO_Post_Heatmaps}
\end{figure}

Figure \ref{fig:GUSTO_Post_Heatmaps} displays the posterior co-clustering probabilities for each of the three views (through $\bm{c}_1$, $\bm{c}_2$ and $\bm{c}_3$ along with the baseline through $\bm{c}_0$.  It appears that the co-clustering of the Z-BMI view displays the least amount of uncertainty.  As before, the "majority vote" property of the hierarchical partition model is evident in Figure \ref{fig:GUSTO_Post_Heatmaps} as the units that are co-clustered in the three views have high co-clustering probability in the baseline view and units that are clustered in one or two views have smaller co-clustering probabilities in the baseline view.   
Figures \ref{fig:GUSTO_flow_c0} and \ref{fig:GUSTO_flow_cj} in Appendix~\ref{app:figures} help visualize how the partitions in each view differ. From Figure \ref{fig:GUSTO_flow_c0} it is clear that the Z-BMI partition and the baseline partition estimates are essentially the same, with the Hypertension and baseline being the most different.  This is seen further in the pairwise view comparisons of Figure \ref{fig:GUSTO_flow_cj}.  Here it is evident that the Z-BMI partition estimate is most different from that of Hypertension.  

\begin{figure}[ht]
\centering
\includegraphics[width=0.9\linewidth]{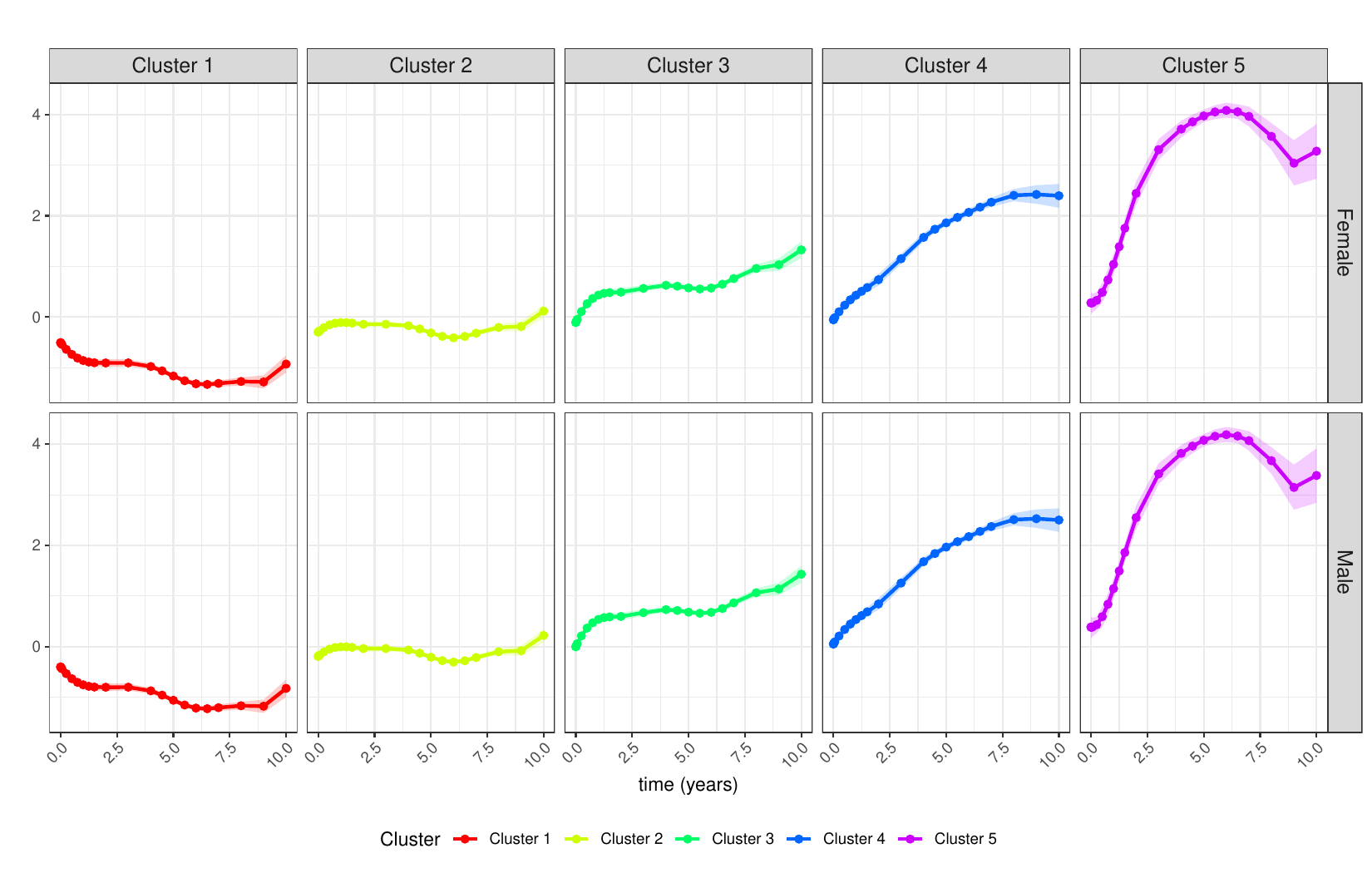}
\caption{GUSTO data. Sex-specific predictive Z-BMI trajectories within each cluster across the three views. The estimates are obtained conditionally on the Binder partition estimated in the first run of the MCMC.}
\label{fig:GUSTO_predictive_BMI}
\end{figure}

\begin{figure}[ht]
\centering
\includegraphics[width=0.9\linewidth]{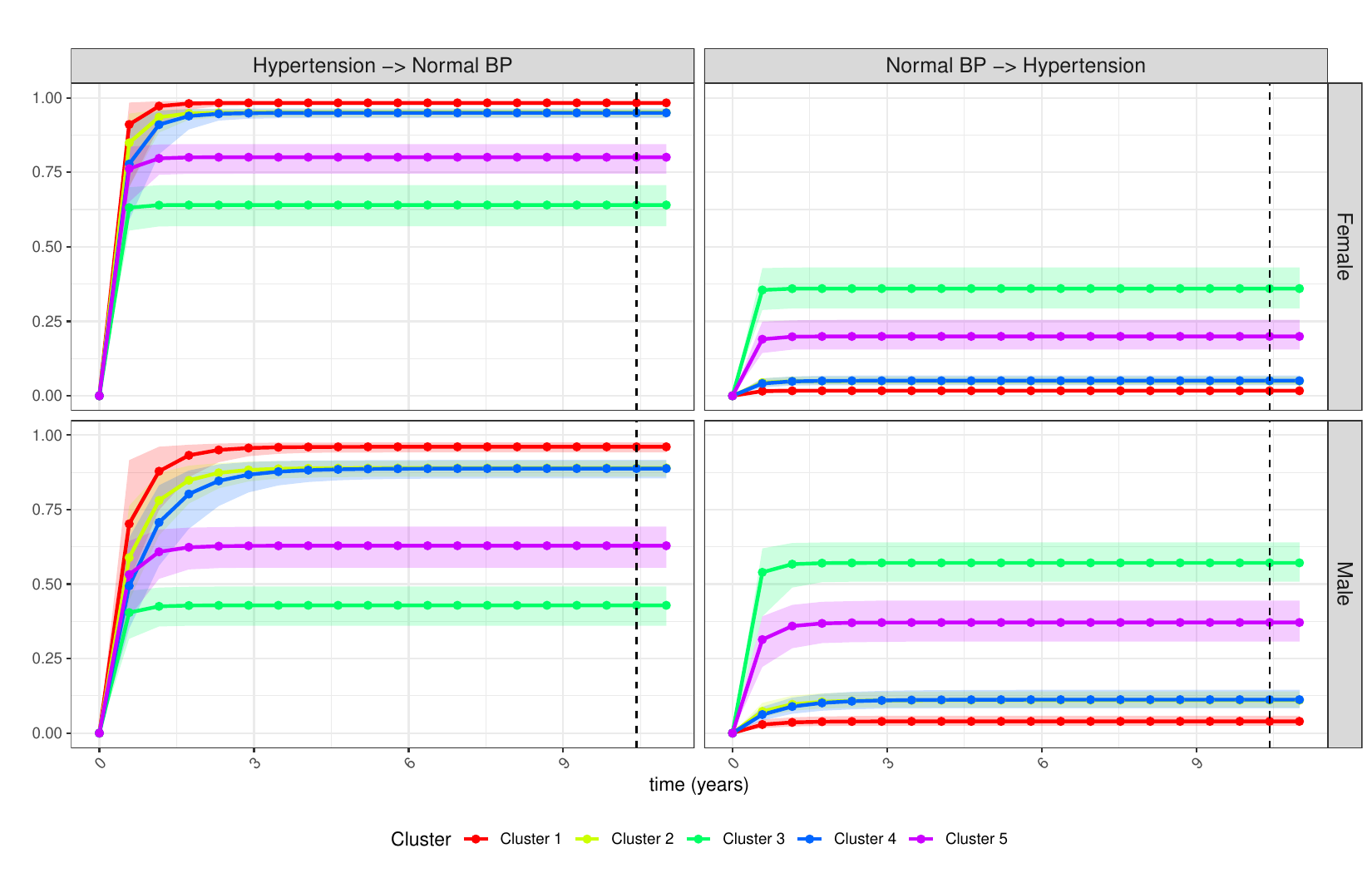}
\caption{GUSTO data. Sex-specific predictive transition probabilities between regular and elevated blood pressure states within each cluster across the three views. The figures include the predictive probability of changing state one year ahead of the last time point of observation. The estimates are obtained conditionally on the Binder partition estimated in the first run of the MCMC.}
\label{fig:GUSTO_predictive_HYP}
\end{figure}

\begin{figure}[ht]
\centering
\includegraphics[width=0.9\linewidth]{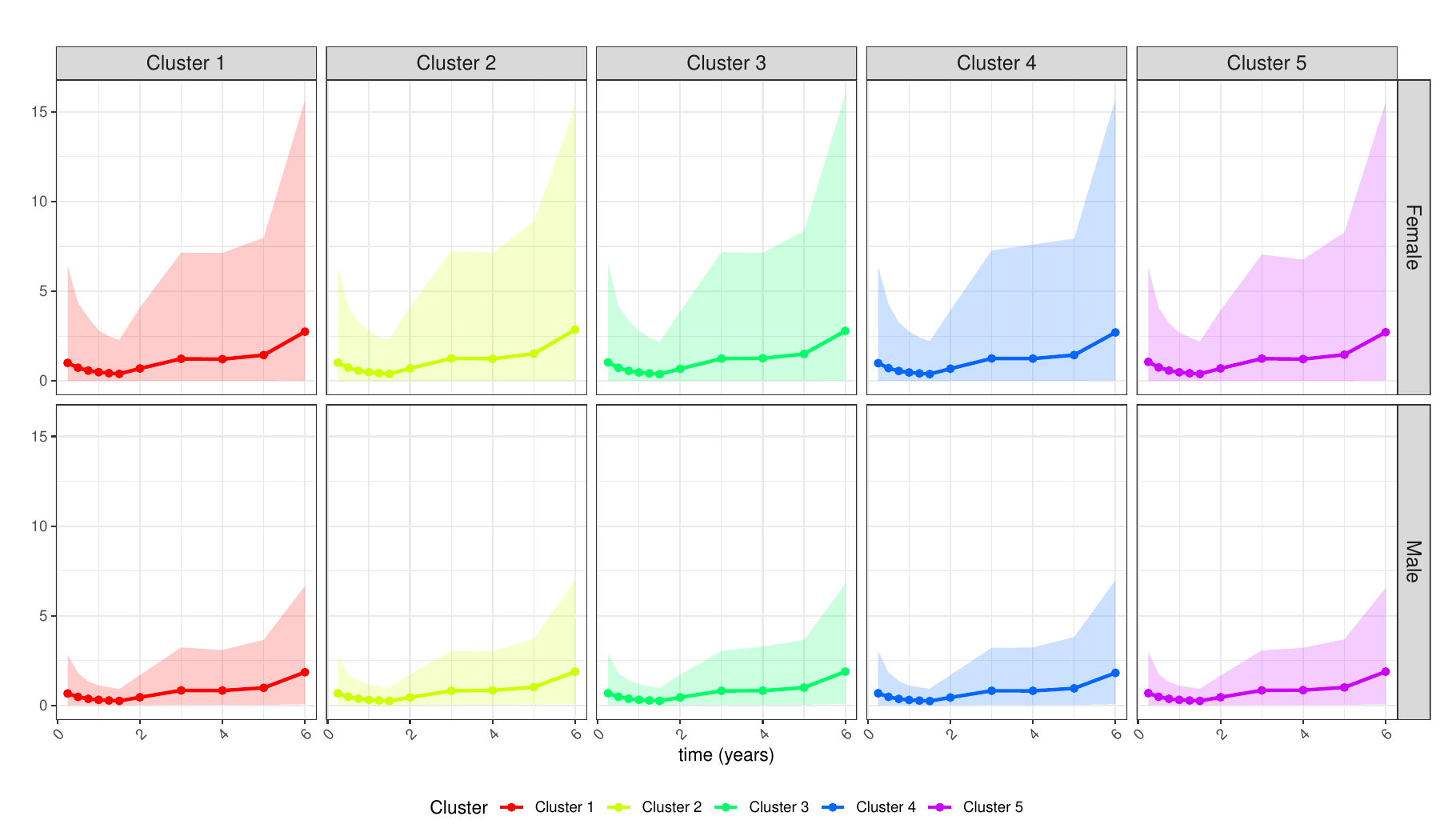}
\caption{GUSTO data. Sex-specific predictive mean intensities of the number of wheezing episodes within each cluster across the three views. The estimates are obtained conditionally on the Binder partition estimated in the first run of the MCMC.}
\label{fig:GUSTO_predictive_ASM}
\end{figure}

We show in Figures \ref{fig:GUSTO_predictive_BMI}-\ref{fig:GUSTO_predictive_ASM} predictive distributions within each cluster of quantities of interest across the three views. Cluster-specific objects were estimated conditioned on the Binder partition estimates.  Figure \ref{fig:GUSTO_predictive_BMI} shows predictive Z-BMI trajectories; \ref{fig:GUSTO_predictive_HYP} transition probabilities between normal and elevated blood pressure states; and \ref{fig:GUSTO_predictive_ASM} the mean intensities of the non-zero inflated wheezing count distribution. Estimates are obtained by first fixing the Binder partition obtained with the first MCMC run, and then run an additional MCMC chain, conditionally on these. Except for the covariate \textit{Sex}, all other covariates are set to baseline if categorical or zero if continuous). 
Health outcomes are known to differ between sexes across a wide range of conditions, and these disparities are influenced by a complex interplay of biological and other factors, something that can be seen already in early life.
Across all clusters, we observe small changes between male and female subjects, with more negative transitions (from normal to elevated blood pressure) in males, and higher intensities of wheezing episodes in females.

\begin{table}[htbp]
\centering
\caption{GUSTO data. Relative transition frequencies (\%) for the Hypertension data (“0” = regular BP, “1” = elevated BP), within the five clusters identified in the estimated partition (Binder estimate).}
\label{tab:transition_clusters_combined}
\resizebox{\textwidth}{!}{%
\begin{tabular}{l|cc|cc|cc|cc|cc|}
\cmidrule(lr){2-11}
 & \multicolumn{2}{c|}{Cluster 1 ($n = 175$)} 
 & \multicolumn{2}{c|}{Cluster 2 ($n = 261$)} 
 & \multicolumn{2}{c|}{Cluster 3 ($n = 66$)} 
 & \multicolumn{2}{c|}{Cluster 4 ($n = 185$)} 
 & \multicolumn{2}{c|}{Cluster 5 ($n = 84$)} \\
\midrule
\multicolumn{11}{l}{\textbf{(a) Transition counts}} \\
\midrule
From $\rightarrow$ To & 0 & 1 & 0 & 1 & 0 & 1 & 0 & 1 & 0 & 1 \\
\cmidrule(lr){2-3}\cmidrule(lr){4-5}\cmidrule(lr){6-7}\cmidrule(lr){8-9}\cmidrule(lr){10-11}
0 & 90.53 & 9.47 & 86.40 & 13.60 & 89.47 & 10.53 & 88.77 & 11.23 & 89.38 & 10.62 \\
1 & 61.90 & 38.10 & 61.16 & 38.84 & 45.65 & 54.35 & 58.33 & 41.67 & 72.55 & 27.45 \\
\bottomrule
\end{tabular}%
}
\end{table}

\begin{figure}
\centering
\subfloat[Z-BMI]{\includegraphics[width=0.35\linewidth]{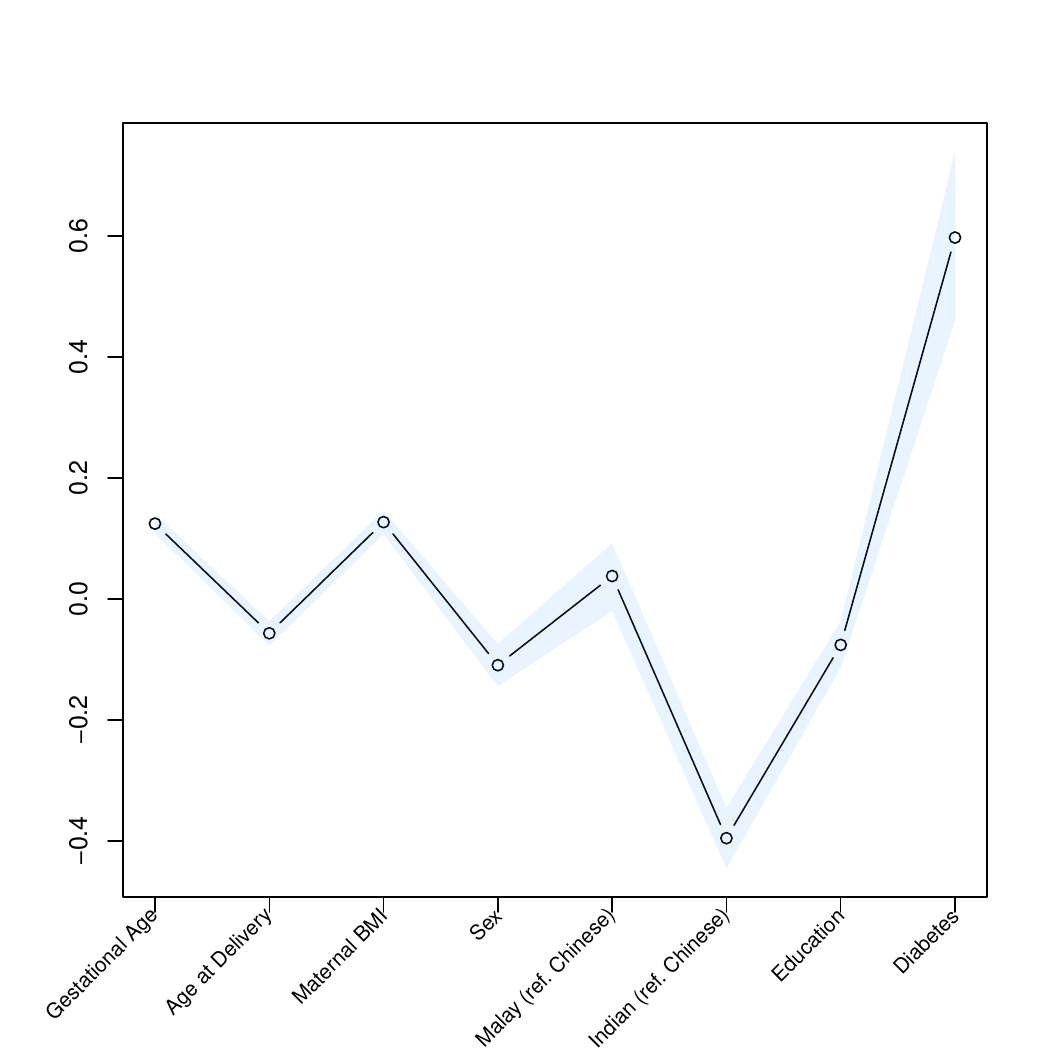}}
\subfloat[Hypertension]{\includegraphics[width=0.35\linewidth]{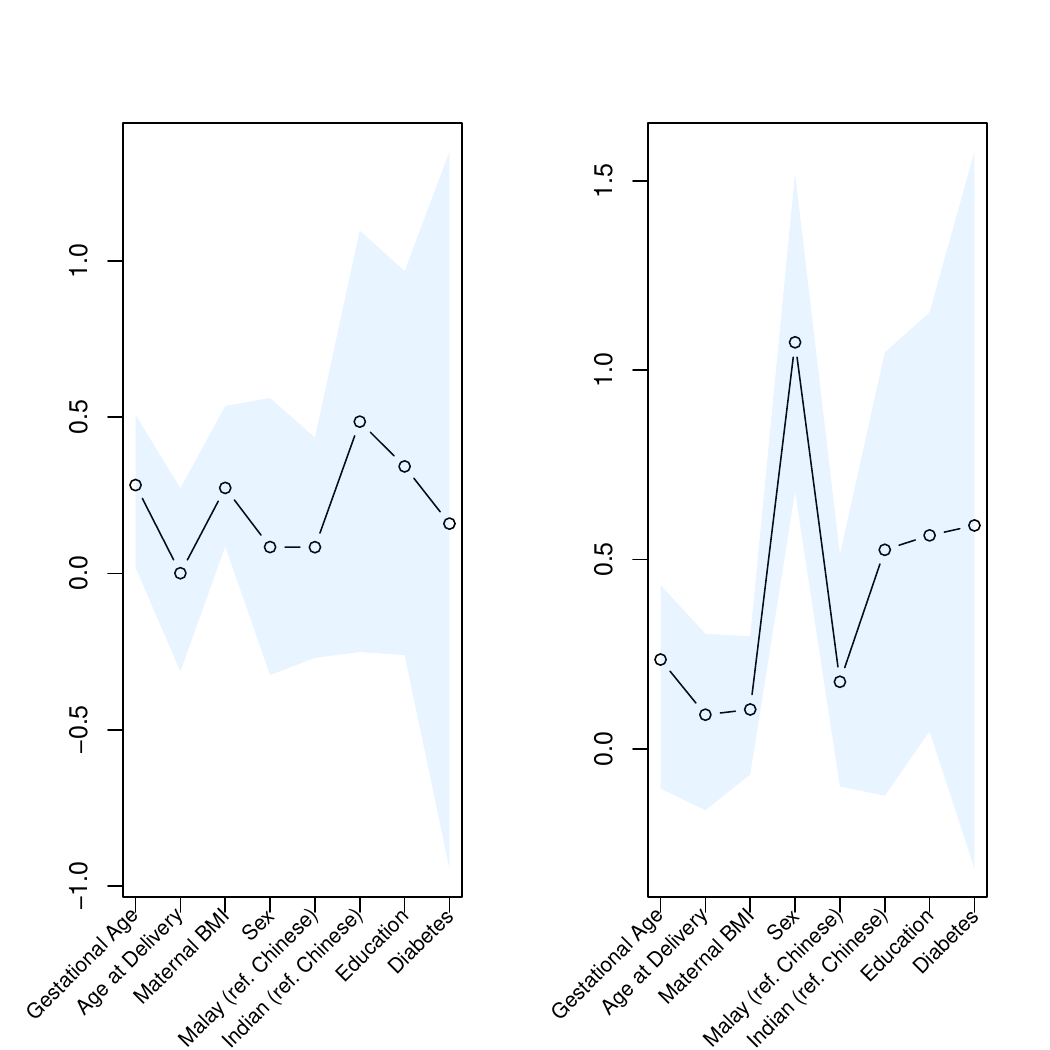}}
\subfloat[Wheezing]{\includegraphics[width=0.35\linewidth]{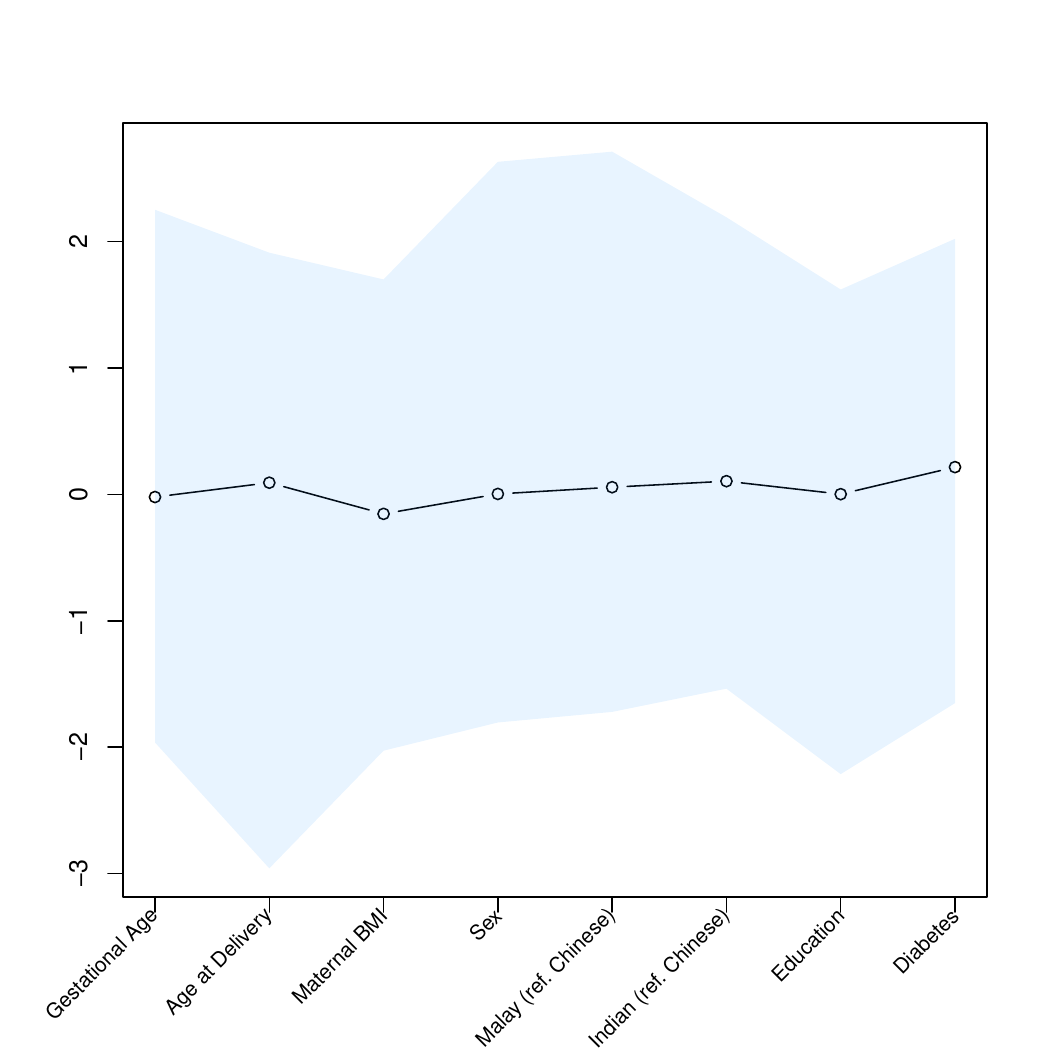}}
\caption{GUSTO data. Effect of covariates in the three different views.}
\label{fig:GUSTO_covs}
\end{figure}

Our latent modularity model identifies five distinct clusters reflecting heterogeneous cardiometabolic and respiratory profiles among the GUSTO children. Each cluster captures specific patterns of growth, blood pressure regulation, and wheezing intensity, highlighting complex comorbidities between obesity, hypertension, and asthma in early life.

Cluster 1 (red) comprises children with low Z-BMI trajectories and consistently normal blood pressure. The wheezing episodes are rare. Clinically, this group represents the metabolically and respiratory-healthy reference population.
Cluster 2 (yellow) includes children with average Z-BMI and stable normotensive states, with negligible respiratory morbidity.
Cluster 3 (green) is characterized by higher-than-average Z-BMI and an elevated likelihood of transitioning to hypertensive states. This cluster reflects early signs of metabolic dysregulation. The coexistence of overweight and blood pressure elevation suggests systemic inflammatory activation and altered vascular responsiveness.
Clusters 4 (blue) and 5 (purple) exhibit markedly elevated Z-BMI trajectories, but differ in vascular and respiratory profiles. Cluster 4 shows moderate hypertension risk and stable respiratory function, whereas Cluster 5 presents both high BMI and frequent hypertensive transitions, indicating an early phenotype of metabolic syndrome. These children may represent the highest-risk group for future cardiometabolic complication.
Across all clusters, the intensity of wheezing episodes remains largely unchanged; however, there is a consistent \textit{sex effect}, with females showing slightly higher wheezing intensity. This aligns with evidence that airway caliber and hormonal influences modulate respiratory reactivity even in childhood. In contrast, males display higher rates of progression from normotensive to elevated blood pressure states, consistent with early androgen-related differences in vascular tone.

In reference to the effect of the covariate, ethnicity and maternal metabolic factors play dominant roles. Indian ethnicity and maternal diabetes are the strongest predictors of higher Z-BMI and hypertensive trajectories, consistent with known genetic and epigenetic susceptibility to insulin resistance and reduced nephron endowment. Malay children also show increased risk, though less pronounced. Education level has a modest protective effect, reflecting socio-economic gradients in child health outcomes.

In general, the model uncovers latent interconnections between the growth, cardiovascular and respiratory pathways, suggesting that early-life exposures, including maternal metabolic status and postnatal growth patterns, shape clustered trajectories of cardiometabolic and respiratory risk.

\section{Conclusions}\label{sec:conc}

In this article, we have presented a new Bayesian model for multi-view clustering based upon the principle of latent modularity. We considered several theoretical aspects of our prior structure and in particular the implications upon clustering and partitions. We then derived a simple Gibbs sampler and implemented it on several simulated and real examples, the latter of which has demonstrated the flexibility of our modeling framework for practical examples.

In the GUSTO application, the latent modularity analysis provides new insights into how early-life metabolic and respiratory trajectories cluster within the pediatric population. The identification of subgroups characterized by concurrent obesity and hypertension suggests that metabolic risk stratification can begin as early as childhood, allowing for timely lifestyle and dietary interventions. The observation that wheezing intensity is largely stable across clusters but modulated by sex supports the idea that airway reactivity and vascular function may evolve independently yet share common inflammatory pathways. Ethnic disparities and maternal metabolic influences highlight the importance of prenatal and early-life preventive care, particularly for children of Indian and Malay descent and those exposed to maternal diabetes.

Several extensions to this work are possible. For instance, one can modify the structure of the prior on the view-specific cluster labels, so as to depend on the global (base) cluster. In addition, one can consider a more detailed understanding and decomposition of the result in our main theorem.

\bibliographystyle{plainnat}
\bibliography{ref}

\newpage

\appendix

\baselineskip 25pt
\begin{center}
{\LARGE{Appendix of \\ \bf Latent Modularity in Multi-View Data}} 

  % \title{\bf L}
  % \author{A. Cremaschi\thanks{
  %   The authors gratefully acknowledge \textit{please remember to list all relevant funding sources in the unblinded version}}\hspace{.2cm}\\
  %   IE University, Spain\\
  %   and \\
  %   M. De Iorio \\
  %   National University of Singapore, Singapore \\
  %   and \\
  %   A. Jasra \\
  %   School of Data Science (SDS), CUHK, Shenzhen \\
  %   and \\
  %   G. Page \\
  %   Department of Statistics, Brigham Young University, United States}
  % \maketitle

\end{center}

%\setcounter{equation}{0}
%\setcounter{definition}{0}
%\setcounter{page}{1}
%\setcounter{table}{1}
%\setcounter{figure}{0}
%\setcounter{section}{0}
%\numberwithin{table}{section}
%\renewcommand{\theequation}{A.\arabic{equation}}
%\renewcommand{\thesubsection}{A.\arabic{section}.\arabic{subsection}}
%\renewcommand{\thesection}{A.\arabic{section}}
%\renewcommand{\thepage}{A.\arabic{page}}
%\renewcommand{\thetable}{A.\arabic{table}}
%\renewcommand{\thefigure}{A.\arabic{figure}}
%\renewcommand{\thedefinition}{A.\arabic{definition}}

% \appendix
\section{Proofs}

\subsection{Proof of Proposition \ref{prop:conditional_joint_feature_labels}}\label{app:prf_prop1}
\begin{proof}
In the following proof we use the short-hand notation
$(\bm{c}_{1},\ldots,  \bm{c}_{J})=\bm{c}_{1:J}$,  
$(\bm{\omega}_1,\ldots, \bm{\omega}_n)=\bm{\omega}_{1:n}$ and
$d\bm{\omega}_1,\ldots, d\bm{\omega}_n=d\bm{\omega}_{1:n}$.
We write the normalisation constant of the density function associated to a $\text{Dir}_m(\alpha_{1:m})$ distribution
as $Z_m(\alpha_{1:m})$ and if all $\alpha_1=\cdots=\alpha_m$ we write $Z_M(\alpha_1)$.
\begin{align*}
 \mathbb{P}(\bm{c}_{1:J}|\bm{c}_{0}, \alpha, m) & = 
\int  \mathbb{P}(\bm{c}_{1:J}|\bm{\omega}_{1:n},m)  \mathbb{P}(
\bm{\omega}_{1:n}|\bm{c}_0, \alpha, m)d\bm{\omega}_{1:n}  \\
&  = \int \left[\prod_{j=1}^J \mathbb{P}(\bm{c}_{j}|\bm{\omega}_{1:n},, m)\right] 
\left[\prod_{i=1}^n\mathbb{P}(\bm{\omega}_i|\bm{c}_0, \alpha, m)\right]  d\bm{\omega}_{1:n,}  \\
& =  \int 
\prod_{i=1}^n\left\{
\left[\prod_{j=1}^J \mathbb{P}(\bm{c}_{ji}|\bm{\omega}_i, m)\right] \left[ 
 \prod_{s=1}^m
Z_m\left(\alpha + \mathbb{I}_{\{s\}}(c_{0i})\right)
\omega_{is}^{\alpha + \mathbb{I}_{\{s\}}(c_{0i}) -1}\right]\right\}  d\bm{\omega}_{1:n}  \\
%& = \int \left[\prod_{j=1}^J \prod_{i=1}^n \left\{ \prod_{m=1}^M\omega_{im}^{\mathbb{I}_{\{m\}}(c_{ji})}\right\}\right] \left[\prod_{i=1}^n\left\{ \frac{\Gamma(\alpha M + 1)}{\Gamma(\alpha)^{M-1} \Gamma(\alpha + 1)} \prod_{m=1}^M\omega_{im}^{\alpha + \mathbb{I}_{\{m\}}(c_{0i}) -1}\right\}\right]  d\bm{\omega}_{1:n} = \\
& =  
Z_m\left(\alpha+1,\alpha,\dots,\alpha\right)^n
 \int \left[\prod_{i=1}^n \prod_{s=1}^m\omega_{is}^{\sum_{j=1}^J
\mathbb{I}_{\{s\}}(c_{ji})}\omega_{is}^{\alpha + 
\mathbb{I}_{\{s\}}(c_{0i})-1}\right]  d\bm{\omega}_{1:n} \\
& =   
Z_m\left(\alpha+1,\alpha,\dots,\alpha\right)^n
 \prod_{i=1}^n
\left[\int \prod_{s=1}^m\omega_{is}^{\sum_{j=1}^J
\mathbb{I}_{\{s\}}(c_{ji})+ \alpha + \mathbb{I}_{\{s\}}(c_{0i}) -1}  d\bm{\omega}_i \right]\\
&  = \left[\frac{\alpha m\Gamma(\alpha m)}{\alpha\Gamma(\alpha)^{m}}\right]^n \prod_{i=1}^n 
\left[\frac{\prod_{s=1}^m\Gamma\left(\alpha + \mathbb{I}_{\{s\}}(c_{0i}) + \sum_{j=1}^J
\mathbb{I}_{\{s\}}(c_{ji})\right) }{\Gamma\left(\sum_{s=1}^m \left\{\alpha + \mathbb{I}_{\{s\}}(c_{0i}) + \sum_{j=1}^J
\mathbb{I}_{\{s\}}(c_{ji})\right\}\right)}\right] \\
& =  
Z(m,\alpha,n)
\prod_{i=1}^n \prod_{s=1}^m\Gamma\left(\alpha +  \mathbb{I}_{\{s\}}(c_{0i}) + \sum_{j=1}^J\mathbb{I}_{\{s\}}(c_{ji})\right).
\end{align*}
This completes the proof for the general case.  We now consider $n=J=2$, 
and to make the notation more compact let 
$$
I_{\mathsf{C}}(c_{0i},c_{1i},c_{2i}) := \mathbb{I}_{\mathsf{C}_1}(c_{0i},c_{1i},c_{2i}) + \mathbb{I}_{\mathsf{C}_2}(c_{0i},c_{1i},c_{2i}) + \mathbb{I}_{\mathsf{C}_3}(c_{0i},c_{1i},c_{2i})
$$
where $\mathsf{C}_1,\mathsf{C}_2,\mathsf{C}_3$ are defined in \eqref{eq:setc1}-\eqref{eq:setc3}.
Recall also $\mathsf{A}$ and $\mathsf{B}$ in \eqref{eq:seta} and \eqref{eq:setb},  which will be used in the below calculations.
Then we have that
\begin{eqnarray*}
  \mathbb{P}(\mathbf{c}_{1:2}|\mathbf{c}_{0},\alpha, m) & = &
Z(m,\alpha,2)
\prod_{i=1}^2 \prod_{s=1}^m\Gamma\left(\alpha +  \mathbb{I}_{\{s\}}(c_{0i}) + \sum_{j=1}^2\mathbb{I}_{\{s\}}(c_{ji})\right)\\
    & = & \left[\frac{1}{\alpha\Gamma(\alpha)^m(\alpha m + 2)(\alpha m + 1)}\right]^2 \prod_{i=1}^2 
\Big\{
 [\Gamma(\alpha + 3)\Gamma(\alpha)^{m-1}]^{
\mathbb{I}_{\mathsf{A}}(c_{0i},c_{1i},c_{2i})} \times  \\
    & & 
[\Gamma(\alpha+1)^3 \Gamma(\alpha)^{m-3}]^{ \mathbb{I}_{\mathsf{B}}(c_{0i},c_{1i},c_{2i}) }
[\Gamma(\alpha+2)\Gamma(\alpha+1)\Gamma(\alpha)^{m-2}]^{ I_{\mathsf{C}}(c_{0i},c_{1i},c_{2i}) }
\Big\}
\\
    & = & \left[\frac{\alpha\Gamma(\alpha)^m}{\alpha\Gamma(\alpha)^m(\alpha m + 2)(\alpha m + 1)}\right]^2  \  \prod_{i=1}^2 \Big\{ [(\alpha+2)(\alpha+1)]^{
\mathbb{I}_{\mathsf{A}}(c_{0i},c_{1i},c_{2i})} \times \\
& &
 \alpha^{2
\mathbb{I}_{\mathsf{B}}(c_{0i},c_{1i},c_{2i})}
 [(\alpha+1)\alpha]^{
I_{\mathsf{C}}(c_{0i},c_{1i},c_{2i})}\Big\} 
\\
    & = & \frac{[(\alpha+2)(\alpha+1)]^{n_{\mathsf{A}}}[\alpha^2]^{n_{\mathsf{B}}}
[(\alpha+1)\alpha]^{n_{\mathsf{C}}}
}{[(\alpha m + 2)(\alpha m + 1)]^2}
\end{eqnarray*}
where  $n_{\mathsf{A}},n_{\mathsf{B}}, n_{\mathsf{C}}$ are defined in \eqref{eq:na}-\eqref{eq:nc}; this concludes the proof.
\end{proof}

\subsection{Proof of Corollary \ref{cor:limiting_probabilities}}\label{app:prf_cor1}

\begin{proof}
To prove the limits as $\alpha \rightarrow 0$ it is enough to determine if it is possible for $n_{\mathsf{A}} = 2$.  
In such a scenario,  the numerator of the joint probability is not a factor of $\alpha$ and so hence will not
go to zero as $\alpha\rightarrow 0$.  To see this we provide closed form probabilities for 
$ \mathbb{P}(c_{11} = c_{21}, c_{12} = c_{22} |c_{01} = c_{02}, \alpha, m)$ and
$\mathbb{P}(c_{11} = c_{21}, c_{12} = c_{22} | c_{01} \ne c_{02}, \alpha, m)$.
 Without loss of generality: when $c_{01} = c_{02}$ we set $c_{01} = 1$, $c_{02}= 1$ and when $c_{01} \ne c_{02}$ we set $c_{01} = 1$, $c_{02}= 2$.  Then we have
$$
\mathbb{P}(c_{11} = c_{21}, c_{12} = c_{22} |c_{01} = c_{02}, \alpha, m)  
$$
\begin{eqnarray} \label{eq:prob_cond_equal}
 & = & \sum_{s= 1}^m \sum_{t = 1}^m\mathbb{P}(c_{11} = s, c_{21}=s, c_{12} = t, c_{22}=t | c_{01} = 1, c_{02}=1, \alpha, m)\label{eq:cor_prf1} \nonumber \\ 
    & = &  \frac{[(\alpha+2)(\alpha+1)]^2+3(m-1)[(\alpha+1)\alpha]^2 + (m-1)(m-2)[\alpha^2]^2}{[(\alpha m + 2) (\alpha m + 1)]^2}.
\end{eqnarray}
Thus, due to the first summand in \eqref{eq:cor_prf1} (which is only available as $n_{\mathsf{A}} =2$) we have
$$
\lim_{\alpha \rightarrow 0}\mathbb{P}(c_{11} = c_{21}, c_{12} = c_{22}|c_{01} = c_{02}, \alpha, m)  = 1.
$$
Since the last summand in the numerator of \eqref{eq:cor_prf1} is a factor of $m^2$ and the denominator is a factor of $m^4$, 
$$
\lim_{\alpha \rightarrow \infty}\mathbb{P}(c_{11} = c_{21}, c_{12} = c_{22} |c_{01} = c_{02}, \alpha, m)  = \frac{1}{m^2}.
$$

We also have
$$
\mathbb{P}(c_{11} = c_{21}, c_{12} = c_{22} |c_{01} \neq c_{02}, \alpha, m)  
$$
\begin{eqnarray}  \label{eq:prob_cond_notequal}
    & = &  
\sum_{s= 1}^m \sum_{t = 1}^m\mathbb{P}(c_{11} = s, c_{21}=s, c_{12} = t, c_{22}=t | c_{01} = 1, c_{02}=2, \alpha, m)
 \nonumber \\ 
    & = & \frac{2(\alpha+2)(\alpha+1)^2\alpha +2(m-2)[(\alpha+1)\alpha]^2 + (m-2)(m-3)[\alpha^2]^2}{[(\alpha m + 2) (\alpha m + 1)]^2}.\label{eq:2ndprob}
\end{eqnarray}
Note that when $c_{01} \ne c_{02}$ and both $c_{11} = c_{21}$ and/or $c_{12} = c_{22}$ then $n_{\mathsf{A}} = 2$ is not possible and  as such each summand in the numerator of \eqref{eq:2ndprob} is a factor of $\alpha$ and as such $\lim_{\alpha \rightarrow 0}\mathbb{P}(c_{11} = c_{21}, c_{12} = c_{22}|c_{01} \ne c_{02}, \alpha, m) = 0$.  
In addition,  similar to above arguments,  since the last summand of \eqref{eq:2ndprob} is a factor of $m^2$ and the denominator is a factor of $m^4$, $\lim_{\alpha \rightarrow \infty}\mathbb{P}(c_{11} = c_{21}, c_{12} = c_{22} | c_{01} \ne c_{02}, \alpha, m) = 1/m^2$.

The last four cases which consider the probabilities 
$\mathbb{P}(c_{11} \ne c_{21}, c_{12} = c_{22} | c_{01} = c_{02}, \alpha, m)$ and $\mathbb{P}(c_{11} \ne c_{21}, c_{12} \ne c_{22} | c_{01} \ne c_{02}, \alpha, m)$ are more challenging to derive.  We begin with considering the former.  It is straightforward to see that $n_{\mathsf{A}} = 2$ is not possible when $c_{01} = c_{02}$ along with $c_{11} \ne c_{21}$ and/or $c_{12} = c_{22}$.  As a result, $\lim_{\alpha \rightarrow 0}\mathbb{P}(c_{11} \ne c_{21}, c_{12} = c_{22} | c_{01} = c_{02}, \alpha, m) = 0$.  Additionally, since $\mathbb{P}(c_{11} \ne c_{21}, c_{12} = c_{22} | c_{01} = c_{02}, \alpha, m)$ is obtained via summing over $c_{11}$, $c_{21}$, and $c_{22}$ (as $c_{12}=c_{22}$)
one can show that the last summand of the probability calculation with be a factor of $m^3$.  As a result,  $\lim_{\alpha \rightarrow \infty}\mathbb{P}(c_{11} = c_{21}, c_{12} = c_{22} |c_{01} \ne c_{02}, \alpha, m) = 1/m$.

Finally, since $n_{\mathsf{A}}=2$ is a possibility when calculating $\mathbb{P}(c_{11} \ne c_{21}, c_{12} \ne c_{22} | c_{01} \ne c_{02}, \alpha, m)$, then $\lim_{\alpha \rightarrow 0}\mathbb{P}(c_{11} \ne c_{21}, c_{12} = c_{22} | c_{01} = c_{02}, \alpha, m) = 1$.  Since $\mathbb{P}(c_{11} \ne c_{21}, c_{12} \ne c_{22} | c_{01} \ne c_{02}, \alpha, m)$
is calculated by summing over values for all four cluster labels, the last summand in the probability calculation will be a factor of $m^4$ which results in $\lim_{\alpha \rightarrow \infty}\mathbb{P}(c_{11} = c_{21}, c_{12} = c_{22}| c_{01} \ne c_{02}, \alpha, m) = 1$. This completes the proof of the eight limits in the statement of the Corollary.
\end{proof}

\subsection{Proof of Theorem \ref{theo:part}}\label{app:theo_prf}

Recall \eqref{eq:cs1}-\eqref{eq:cs2} which is needed below.
In order to prove Theorem \ref{theo:part} we start with following calculations
\begin{eqnarray*}
p\left(\rho_1, \dots, \rho_J, \rho_0\right) & = & \sum_{m=1}^{\infty}p\left(\rho_1, \dots, \rho_J, \rho_0 \mid m\right)q_M(m) \\
& = & \sum_{m=1}^{\infty}
\sum_{\mathbf{c}_0\in \mathtt{C}_{k_0}(m)}\mathbb{E}\left[
\prod_{j=1}^J 
\pi\left(\rho_j \mid \mathbf{c}_0, m\right)
\prod_{i=1}^n \omega_{0c_{0i}}\Bigg|m
\right] q_M(m)
\end{eqnarray*}
where we use the letter $\pi$ to indicate the law of individual exchangeable partitions, i.e. the exchangeable partition probability function (EPPF) and the expectation operator is with respect to the prior structure on $\mathbf{c}_0$, $\mathbf{\omega}_{0}$ conditional on $M=m$.
Indeed, the partitions induced at each view $j$ are independent conditionally on $\mathbf{c}_0$ and the number of components $M = m$. This framework is analogous to that of conditionally partially exchangeable partitions proposed in \cite{franzolini2023conditional}, where the conditional measure at the base level indicates which random partition distribution is specified at the $j$-th level, given by the $M$-dimensional Dirichlet distributions with subject-specific shape parameters.

The proof now constitutes deriving %$ \pi\left(\rho_0\mid m\right)$ and 
$\pi\left(\rho_j \mid \mathbf{c}_0, m\right)$, computing the expectation
 and putting this together.  This will be achieved in the next two Lemmata with the final proof at the end.
For $\mathbf{c}_0\in \mathtt{C}_{k_0}(m)$, with $\mathtt{C}_{k_0}(m)$ indicating the set of partitions of $n$  we write $n_j$ are the number of units in cluster $j$.

\begin{lemma}\label{lem:1}
Let $(m,k_0)\in\mathbb{N}^2$ be given and $\varphi:[m]^n\rightarrow\mathbb{R}$ be bounded.
Under the model (1) we have
$$
\sum_{\mathbf{c}_0\in \mathtt{C}_{k_0}(m)} 
\varphi(\mathbf{c}_0)\mathbb{E}\left[
\prod_{i=1}^n \omega_{0c_{0i}}
\big|m\right] 
= 
\mathbb{I}_{[m]}(k_0)
\frac{1}{ \Gamma(\alpha_0)^{k_0}}
\frac{\Gamma(m\alpha_0)}{\Gamma(n+m\alpha_0)}
\sum_{\mathbf{c}_0\in \mathtt{C}_{k_0}(m)} 
\varphi(\mathbf{c}_0)
\prod_{j=1}^{k_0}\Gamma(\alpha_0 + n_j).
$$
\end{lemma}

\begin{proof}
We follow the work by \cite[][Therorem 1]{argiento2022infinity}:
\begin{equation*}
\mathbb{E}\left[
\prod_{i=1}^n \omega_{0c_{0i}}
\big|m\right]  = \mathbb{I}_{[m]}(k_0)\int_{0}^{\infty} \psi(u_0,\alpha_0)^{m-k_0} \frac{u_0^{n-1}}{\Gamma(n)}\prod_{j=1}^{k_0}\kappa\left(n_j,u_0,\alpha_0\right) du_0 
\end{equation*}
where $u_0$ is an auxiliary variable. Moreover, $\psi(u_0,\alpha_0)$ and $\kappa\left(n_j,u_0,\alpha_0\right)$ are the Laplace transform and cumulant of the density fo the unnormalized weights \citep[see][and references therein for details]{argiento2022infinity}. In this case, where the unnormalized weights are Gamma-distributed, we have:
\begin{align}
    & \psi(u_0,\alpha_0) = \int_0^{+\infty}e^{-u_0s}\frac{1}{\Gamma(\alpha_0)}s^{\alpha_0 - 1}e^{-s}ds = \frac{1}{\left(u_0 + 1\right)^{\alpha_0}} \label{eq:psi_kappa1}\\
    & \kappa(n_j,u_0,\alpha_0) = \int_0^{+\infty}s^{n_j}e^{-u_0s}\frac{1}{\Gamma(\alpha_0)}s^{\alpha_0 - 1}e^{-s}ds = \frac{\Gamma(\alpha_0 + n_j)}{\Gamma(\alpha_0)}\frac{1}{\left(u_0 + 1\right)^{\alpha_0 + n_j}}\label{eq:psi_kappa2}
\end{align}
yielding:
\begin{eqnarray*}
\mathbb{E}\left[
\prod_{i=1}^n \omega_{0c_{0i}}
\big|m\right]
 & = &  
\mathbb{I}_{[m]}(k_0)\int_{0}^{\infty} \frac{1}{\left(u_0 + 1\right)^{m \alpha_0 + n}} \frac{u_0^{n-1}}{\Gamma(n) \Gamma(\alpha_0)^{k_0}}\prod_{j=1}^{k_0}\Gamma(\alpha_0 + n_j) du_0 \\
& = & \mathbb{I}_{[m]}(k_0)
\frac{1}{ \Gamma(\alpha_0)^{k_0}}
\frac{\Gamma(m\alpha_0)}{\Gamma(n+m\alpha_0)}
\prod_{j=1}^{k_0}\Gamma(\alpha_0 + n_j).
\end{eqnarray*}
as the integral is a standard Beta function. This concludes the proof.
\end{proof}

\begin{lemma}\label{lem:2}
Under the model (1) we have
$$
\pi\left(\rho_j \mid \mathbf{c}_0, m\right) = 
\mathbb{I}_{[m]}(k_j) 
\sum_{\mathbf{c}_j\in\mathtt{C}_{k_j}(m)}
 \frac{\left(\alpha + 1\right)^{n_{j0}} \alpha^{n - n_{j0}}}{\left(m\alpha + 1\right)^n}
$$
\end{lemma}

\begin{proof}
We compute the probability distribution for the $j$-th view.
\begin{equation*}
\pi\left(\rho_j \mid \rho_0, m\right) = \mathbb{I}_{[m]}(k_j) \sum_{\mathbf{c}_j\in\mathtt{C}_{k_j}(m)} \mathbb{E}\left[\prod_{i=1}^n \omega_{ic_{ji}}\right]
\end{equation*}
The expectation operator is with respect to the joint distribution of the vectors of normalized weights $\bm \omega_i$, for $i = 1, \dots, n$ and conditional on $\mathbf{c}_{0},m$ (which are omitted from the notation for simplicity).

In the forthcoming calculations, we shall write $h_{il}(s_{il})$ to denote the Gamma prior
$S_{il}|c_{0i},\alpha\sim\text{Gamma}(\alpha+\mathbb{I}_{\{l\}}(c_{i0}),1)$. We also use the notation
\begin{eqnarray*}
\alpha_{ji}(c_{i0}) & = & \alpha + \mathbb{I}_{\{c_{ji}\}}(c_{i0})\\
\alpha_l(c_{i0}) & = & \alpha + \mathbb{I}_{\{l\}}(c_{i0}).
\end{eqnarray*}
We use the unnormalized weights construction of model~(2) to write for any $k_j\in[m]$:
\begin{eqnarray*}
\pi\left(\rho_j \mid \mathbf{c}_0, m\right) & = &  
\sum_{\mathbf{c}_j\in\mathtt{C}_{k_j}(m)}
 \mathbb{E}\left[\prod_{i=1}^n \omega_{ic_{ji}}\right]\\
& = & 
\sum_{\mathbf{c}_j\in\mathtt{C}_{k_j}(m)}
 \prod_{i=1}^n \int_{\left(\mathbb{R}^+\right)^m} \frac{s_{ic_{ji}}}{\sum_{l=1}^m s_{il}} \prod_{l = 1}^m h_{il}(s_{il})ds_{il} \\
& = &
\sum_{\mathbf{c}_j\in\mathtt{C}_{k_j}(m)}
 \prod_{i=1}^n \int_{\left(\mathbb{R}^+\right)^m} \int_0^{\infty}s_{ic_{ji}} e^{-u_i \sum_{l=1}^m s_{il}} du_i \prod_{l = 1}^m h_{il}(s_{il})ds_{il} \\
& = &
\sum_{\mathbf{c}_j\in\mathtt{C}_{k_j}(m)} \prod_{i=1}^n \int_0^{\infty}\kappa\left(1,u_i,\alpha_{ji}(c_{i0})\right) \prod_{\substack{l = 1 \\ l \neq c_{ji}}}^m \psi(u_i,\alpha_l(c_{i0})) du_i
\end{eqnarray*}
To go to the second line the product over $i$ can be swapped with the integrals due to the conditional independence of the $n$ units and the simple identity that for any $\alpha>0$ $\int_{0}^{\infty}e^{-\alpha x}dx=1/\alpha$ to go to the third line.  The last line is derived by using integration and the definitions \eqref{eq:psi_kappa1}-\eqref{eq:psi_kappa2}.

We have
for $k_j\in[m]$:
\begin{eqnarray}
\pi\left(\rho_j \mid \mathbf{c}_0, m\right) & = &  
\sum_{\mathbf{c}_j\in\mathtt{C}_{k_j}(m)}
 \prod_{i=1}^n \int_0^{\infty}\kappa\left(1,u_i,\alpha_{ji}(c_{i0})\right) \prod_{\substack{l = 1 \\ l \neq c_{ji}}}^m \psi(u_i,\alpha_l(c_{i0})) du_i
 \nonumber \\
& = & 
\sum_{\mathbf{c}_j\in\mathtt{C}_{k_j}(m)}
 \prod_{i=1}^n \int_0^{\infty}\frac{\alpha + \mathbb{I}_{\{c_{ji}\}}(c_{0i})}{\left(u_i + 1\right)^{\alpha + \mathbb{I}_{\{c_{ji}\}}(c_{0i}) + 1}} \prod_{\substack{l = 1 \\ l \neq c_{ji}}}^m \frac{1}{\left(u_i + 1\right)^{\alpha_l(c_{i0})}} du_i \nonumber\\
&  = & 
\sum_{\mathbf{c}_j\in\mathtt{C}_{k_j}(m)}
 \prod_{i=1}^n \int_0^{\infty} \frac{\alpha + \mathbb{I}_{\{c_{ji}\}}(c_{0i})}{\left(u_i + 1\right)^{m\alpha + 2}} du_i  \nonumber\\
& = & 
\sum_{\mathbf{c}_j\in\mathtt{C}_{k_j}(m)} \prod_{i=1}^n \frac{\alpha + \mathbb{I}_{\{c_{ji}\}}(c_{0i})}{m\alpha + 1} \nonumber\\
&   = & 
\sum_{\mathbf{c}_j\in\mathtt{C}_{k_j}(m)}
 \frac{\left(\alpha + 1\right)^{n_{j0}} \alpha^{n - n_{j0}}}{\left(m\alpha + 1\right)^n} \nonumber
\end{eqnarray}
where we recall that $n_{j0} = \text{Card}\left(\{i \in [n] : c_{ji} = c_{0i}\}\right)$ indicates the number of elements with labels in agreement between $\bm c_j$ and $\bm c_0$. 
\end{proof}

We now have the proof of Theorem \ref{theo:part}.

\begin{proof}
Now using Lemmata \ref{lem:1}-\ref{lem:2} we have
\begin{eqnarray*}
p\left(\rho_1, \dots, \rho_J, \rho_0\right) & = & 
\sum_{m=1}^{\infty}
\sum_{\mathbf{c}_0\in \mathtt{C}_{k_0}(m)}\mathbb{E}\left[
\prod_{j=1}^J 
\pi\left(\rho_j \mid \mathbf{c}_0, m\right)
\prod_{i=1}^n \omega_{0c_{0i}}\Bigg|m
\right] q_M(m)
 \\
& = & \sum_{m=1}^{\infty}
\mathbb{I}_{[m]}(k_0)
\frac{\Gamma(m\alpha_0)}{\Gamma(\alpha_0)^{k_0}\Gamma(n+m\alpha_0)}
\sum_{\mathbf{c}_0\in \mathtt{C}_{k_0}(m)}
\prod_{j=1}^J \mathbb{I}_{[m]}(k_j) 
\Big\{\\
& & 
\sum_{\mathbf{c}_j\in\mathtt{C}_{k_j}(m)}
 \frac{\left(\alpha + 1\right)^{n_{j0}} \alpha^{n - n_{j0}}}{\left(m\alpha + 1\right)^n}
\prod_{j=1}^{k_0}\Gamma(\alpha_0 + n_j)\Big\}
 q_M(m).
\end{eqnarray*}
\end{proof}

\section{MCMC algorithm}\label{app:MCMC}

\subsection{MCMC algorithm for unnormalised weights parameterisation}

We exploit the unnormalised weights construction to devise the following MCMC algorithm. Specifically, we update $\{\bm c_0, \bm c_1, \dots, \bm c_J, \bm \theta, u_0, u_1, \dots, u_n, \bm s_0, \bm s_1, \dots, \bm s_n\}$ by sampling from the corresponding full-conditional distributions.

\begin{itemize}
    \item Update $\bm c_0$. For each $i\in[n]$:
    $$
    \mathbb{P}\left(c_{0i} = m \mid \cdot\right) \propto s_{0m} s_{im}
    $$
    
    \item Update $u_0$.
    $$
    u_0 \mid \cdot \sim \text{Gamma}\left(n, t_0\right)
    $$
    
    \item Update $\bm s_0 = \left(s_{01}, \dots, s_{0M}\right)$. For each $m\in[M]$, let $n_{0m} = \text{Card}\left(\{i \in \{1, \dots, n\} : c_{0i} = m\}\right)$:
    $$
    s_{0m} \mid \cdot \sim \text{Gamma}\left(\alpha_0 + n_{0m}, u_0 + 1 \right)
    $$
    Note that $n_{0m}$ might be equal to 0 if no unit is allocated to the $m^{\text{th}}$ component in $\bm c_0$.
    
    \item Update $\bm c_j$. For each $(i,j)\in[n]\times[J]$:
    $$
    \mathbb{P}\left(c_{ji} = m \mid \cdot\right) \propto s_{im} f_j\left(y_{ji} \mid \bm \theta^*_{jm}\right)
    $$
    
    \item Update $u_i$. For $i\in[n]$:
    $$
    u_i \mid \cdot \sim \text{Gamma}\left(J, t_i\right)
    $$
    
    \item Update $\bm s_i = \left(s_{i1}, \dots, s_{iM}\right)$. For $i\in[n]$, let $n_{im} = \text{Card}\left(\{j \in \{1, \dots, J\} \mid c_{ji} = m\}\right)$:
    $$
    s_{im} \mid \cdot \sim \text{Gamma}\left(\alpha + \mathbb{I}_{\{m\}}(c_{0i}) + n_{im}, u_i + 1 \right)
    $$
    Note that $n_{im}$ might be equal to 0 if the $i$-unit is never allocated to the $m^{\text{th}}$ component across the $J$ views.

    \item Update $\bm \theta$. For each $j\in[J]$ and $m\in[M]$, the full conditional of the location parameters is proportional to:
    $$
    p(\bm \theta^*_{jm} \mid \cdot) \propto G_j(\bm \theta^*_{jm}) \prod_{i=1}^n
\left\{f_j\left(y_{ji} \mid \bm \theta^*_{jc_{ji}} \right)\right\}^{\mathbb{I}_{\{m\}}(c_{ji})}
    $$
    which in the proposed example is conjugate.

    \item Update $M$.

    In the following, for some of the updating step, it is convenient to split the component indices $\{1, \dots, M\}$ into two sets: those that are associated to clusters in $\bm c_0$, i.e. for which observations are allocated, and those that are associated to empty components, i.e. for which none of the observations are allocated. We indicate these set of indices as $\mathcal{M}^a$ and $\mathcal{M}^{na}$, respectively.

    Let $M^{na}$ be the number of non-allocated components, so that $M = K_n+ M^{na}$. The full-conditional distribution for $M^{na}$ is equal to:
    $$
    \mathbb{P}\left(M^{na} = m \mid \cdot\right) \propto \frac{(m+K_n)!}{m!} \psi(u_0)\prod_{i=1}^n\psi(u_i)^m q_M(m+K_n)
    $$
    When $M \sim \text{Poi}_1(\Lambda)$, then:
    $$
    \mathbb{P}\left(M^{na} = m \mid \cdot\right) = \frac{K_n}{K_n + \Lambda \psi(\bm u)}\text{Poi}_0(\Lambda \psi(\bm u)) + \frac{\Lambda \psi(\bm u)}{K_n + \Lambda \psi(\bm u)}\text{Poi}_1(\Lambda \psi(\bm u))
    $$
    where $\psi(\bm u) = (u_0 + 1)^{-\alpha_0}\prod_{i=1}^n(u_i+1)^{-\alpha}$.
\end{itemize}

\subsection{MCMC for GUSTO application}

\paragraph{BMI view}
We are interested in updating the parameters $\left(\bm \beta^*_1, \dots, \bm \beta^*_M, \bm \Sigma_Z,  \bm \eta^Z\right)$. These are conjugate updates.

\begin{itemize}
    \item update $\bm \beta^*_m$ from the following full-conditional

$$
\bm \beta^*_m \mid \cdot \sim \text{N}\left( \bm \mu^\star_{\bm \beta}, \bm \Sigma^\star_{\bm \beta} \right)
$$
where
$$
\bm \Sigma^\star_{\bm \beta} = \left( \sum_{i=1}^n \mathbb{I}_{\{m\}}(c_{1i})\bm B_i^\top \bm \Sigma^{-1}_{Z,i} \bm B_i + \bm \Sigma_{\bm \beta}^{-1} \right)^{-1}
$$
and
$$
\bm \mu^\star_{\bm \beta} = \bm \Sigma^\star_{\bm \beta} \left( \bm \Sigma_{\bm \beta}^{-1} \bm \mu_{\bm \beta} + \sum_{i=1}^n \mathbb{I}_{\{m\}}(c_{1i}) \bm B_i^\top \bm \Sigma^{-1}_{Z,i} \left( \bm Y_i - \bm X_i^\top \bm \eta^Z \right) \right)
$$

\item update $\sigma^2_{Z,t}$ for $t\in[T^Z]$ from:
    $$
    \sigma^2_{Z,t} \mid \cdot \sim \text{Inv-Gamma}\left(
  \alpha_{\sigma^2_Z} + \frac{1}{2} \text{Card}\left(\left\{i: t \in \bm t_i\right\},\right) \beta_{\sigma^2_Z} + \frac{1}{2} \sum_{i: t \in \bm t_i} \left( Y_{it} - \bm B_{it} \bm \beta^*_{c_{1i}} - \bm X_i^\top \bm \eta^Z \right)^2 \right) 
$$
then set $\bm \Sigma_Z = \text{diag}\left(\sigma^2_{Z,1}, \dots, \sigma^2_{Z,T^Z} \right)$.

\item update $\bm \eta^Z$ from:

$$
\bm \eta^Z \mid \cdot \sim \text{N}(\bm \mu^\star_{\bm \eta^Z}, \bm \Sigma^\star_{\bm \eta^Z})
$$

$$
\bm \Sigma^\star_{\bm \eta^Z} = \left( \sum_{i=1}^{n} \sum_{t \in \bm t_i} \bm X_i \bm X_i^\top /\sigma^2_{Z,t} + \bm \Sigma_{\bm \eta^Z}^{-1} \right)^{-1}
$$

$$
\bm \mu^\star_{\bm \eta^Z} = \bm \Sigma^\star_{\bm \eta^Z} \left( \bm \Sigma_{\bm \eta^Z}^{-1} \bm \mu_{\bm \eta^Z} + \sum_{i=1}^{n} \bm X_i \bm \Sigma^{-1}_{Z,i} \cdot \sum_{t=1}^{T^Z_i} \left( \bm Y_{it} - \bm B_{it}^\top \bm \beta^*_{c^Z_i} \right) \right)
$$
\end{itemize}

\paragraph{Hypertension view}
The parameters to update are $\left(\bm \lambda^*_1, \dots, \bm \lambda^*_M,  \bm \eta^H_{rs} \right)$, for $r \in \{0,1\}$ and $s=1-r$. We employ Metropolis-Hastings steps to update such parameters.

\begin{itemize}
    \item update $\bm \lambda^*_m$ from the following full-conditional
Propose a new value of $\bm \lambda^*_m$ from a random walk proposal with adaptive covariance matrix:
$$
\bm \lambda^{*,\text{new}}_m \sim \text{N}\left(\bm \lambda^*_m, \bm S_{\bm \lambda}\right)
$$
and accept $\bm \lambda^{*,\text{new}}_m$ with probability:
$$
\min\left\{1, \frac{\prod\limits_{i : c^H_i = m}p\left(\bm Y_i \mid \bm \lambda^{*,\text{new}}_m\right)}{\prod\limits_{i : c^H_i = m}p\left(\bm Y_i \mid \bm \lambda^*_m\right)} \frac{\text{log-Normal}\left(\bm \lambda^{*,\text{new}}_m \mid \bm \mu_{\bm \lambda}, \text{diag}\left(\bm \sigma^2_H \right) \right)}{\text{log-Normal}\left(\bm \lambda^*_m \mid \bm \mu_{\bm \lambda}, \text{diag}\left(\bm \sigma^2_H \right) \right)} \right\}
$$

\item update $\sigma^2_{H,rs}$ from:
    $$
    \sigma^2_{H,rs} \mid \cdot \sim \text{Inv-Gamma}\left(
  \alpha_{\sigma^2_H} + \frac{M}{2}, \beta_{\sigma^2_H} + \frac{1}{2} \sum_{m=1}^M \left( \log \lambda^*_{m,rs}  - \mu_{\bm \lambda, rs} \right)^2 \right) 
$$

\item update $\bm \eta^Z$ from:
Propose a new value of $\bm \eta^H_{rs}$ from a random walk proposal with adaptive covariance matrix:
$$
\bm \eta^{H,\text{new}}_{rs} \sim \text{N}\left(\bm \eta^H_{rs}, \bm S_{\bm \eta^H_{rs}}\right)
$$
and accept $\bm \eta^{H,\text{new}}_{rs}$ with probability:
$$
\min\left\{1, \frac{p\left(\bm Y \mid \bm \eta^{H,\text{new}}_{rs}, \bm \eta^H_{sr}\right)}{p\left(\bm Y \mid \bm \eta^H_{rs}, \bm \eta^H_{sr}\right)} \frac{p\left(\bm \eta^{H,\text{new}}_{rs} \mid \bm \mu_{\bm \eta^H_{rs}}, \bm \Sigma_{\bm \eta^H_{rs}} \right)}{p\left(\bm \eta^H_{rs} \mid \bm \mu_{\bm \eta^H_{rs}}, \bm \Sigma_{\bm \eta^H_{rs}}\right)} \right\}
$$

\end{itemize}

\paragraph{Wheezing view}

In this part of the model, the parameters of interest to be updated are $\left(b_{it}, W_{itl}, \zeta^*_m, p^*_m, r_{il}, \bm \eta^W\bm \eta^W\right)$, for $(m,i,l)\in[M]\times[n]\times[L]$.

\begin{itemize}
    \item for observations $W_{it}=0$, update $b_{it}$:
        $$b_{it} \mid \cdot \sim \text{Bern}\left(\frac{(1-p_i)\mathbb{P}\left(W_{it}=0 \mid \mu_{it}\right)}{p_i + (1-p_i)\mathbb{P}\left(W_{it}=0 \mid \mu_{it}\right)} \right)
        $$

    \item for $t \in B_i \cup C_i$, update $W_{it1},\dots, W_{itL}$:
        \begin{align*}
            p\left(W_{it1},\dots, W_{itL} \mid \cdot \right) \propto& 
            \prod_{l=1}^L \exp\left(-\mu_{itl}\right) \mu_{itl}^{W_{itl}} \frac{1}{W_{itl}!}
        \end{align*}
        where $\mu_{itl}=r_{il} \left(I_l(t_{it}) - I_l(t_{it-1})\right)\exp\left(\bm \eta^W \bm X_i\right)$. Thus,
        $$
            W_{it1},\dots, W_{itL} \mid \cdot \sim \text{Mult}\left(W_{it}; \mu_{it1}, \dots, \mu_{itL}\right)
        $$

    \item update $\zeta^*_m$, for $m\in[M]$:
        \begin{align*}
            \zeta^*_m \mid \cdot \sim \text{Gamma}\left(\alpha_{\zeta} + L n^W_m, \beta_{\zeta}+\sum_{l=1}^L \sum_{i=1}^n
\mathbb{I}_{\{m\}}(c_{3i})
r_{il}\right)
        \end{align*}
    where $n^W_m = \text{Card}\left(\{ i:c_{3i}=m\}\right)$.
          
    \item update $r_{il}$:
        \begin{align*}
            f(r_{il}\mid \cdot) &\propto \prod_{t \in \bm t^W_i} f(W_{itl}\mid r_{il}) f(r_{il} \mid \zeta_i)\\
            &\propto \prod_{t \in \bm t^W_i} \frac{e^{-\mu_{itl}} \mu_{itl}^{W_{itl}}}{W_{itl}!}  e^{-\zeta_i r_{il}} \\
            &\propto e^{-r_{il} \sum\limits_{t \in \bm t^W_i} \left(I_l(t_{it})-I_l(t_{it-1})\right)\exp\left(\bm \eta^W \bm X_i\right) } r_{il}^{\sum\limits_{t \in \bm t^W_i} W_{itl}} e^{-\zeta_i r_{il}} 
        \end{align*}
        Thus:
        \begin{align*}
            r_{il}\mid \cdot \sim \text{Gamma}\left(\sum\limits_{t \in \bm t^W_i} W_{itl}+1, \zeta_i + \sum\limits_{t \in \bm t^W_i} \left(I_l(t_{it})-I_l(t_{it-1})\right)\exp\left(\bm \eta^W \bm X_i\right)\right)
        \end{align*}

    \item update $p^*_m$, for $m\in[M]$ from the following full-conditional:
        \begin{align*}
            p(p^*_m\mid \cdot) &\propto \prod_{i=1}^n 
\left\{p_i^{n_{Ai}} (1-p_i)^{n_{Bi}+n_{Ci}} p_i^{\alpha_p-1} (1-p_i)^{\beta_p-1}\right\}^{\mathbb{I}_{\{m\}}(c_{3i})}
        \end{align*}
        Thus,
        \begin{align*}
            p^*_m\mid \cdot \sim \text{Beta}\left(\sum_{i=1}^n
\mathbb{I}_{\{m\}}(c_{3i})
n_{Ai}+\alpha_p, \sum_{i=1}^n
\mathbb{I}_{\{m\}}(c_{3i})
\{n_{Bi}+n_{Ci}\} + \beta_p\right)
        \end{align*}

    \item update $\bm \eta^W$ with an adaptive Metropolis-Hastings from the following full-conditional:
    \begin{align*}
    p\left(\bm \eta^W \mid \cdot \right) &\propto \exp -\frac{1}{2}\left(\bm \eta^W - \bm \mu_{\bm \eta^W} \right)^\top \Sigma_{\bm \eta^W}^{-1} \left(\bm \eta^W - \bm \mu_{\bm \eta^W} \right) \\
    &\prod_{i=1}^{n} \exp\left(-\sum_{t \in \bm t^W_i} \sum_{l=1}^L\mu_{itl}\right) \prod_{t \in C_i} \prod_{l=1}^L \mu_{itl}^{W_{itl}} \\
    &\propto \exp -\frac{1}{2}\left(\bm \eta^W - \bm \mu_{\bm \eta^W} \right)^\top \Sigma_{\bm \eta^W}^{-1} \left(\bm \eta^W - \bm \mu_{\bm \eta^W} \right) \\
    &\prod_{i=1}^{n} \exp\left(-\sum_{t \in \bm t^W_i}   \sum_{l=1}^L r_{il} \left(I_l(t_{it}) - I_l(t_{it-1})\right)\right)^{\exp(\bm \eta^W \bm X_i)} \\
    & \exp\left(\bm \eta^W \bm X_i\right)^{\sum_{t \in C_i}\sum_{l=1}^L W_{itl}}
    \end{align*}
    
\end{itemize}

\clearpage
\section{Simulation Study (Sensitivity Analysis)}\label{app:sensitivity}
We present in this section the analysis fo simulated data, focusing on: the recovery of both baseline partition $\bm c_0$ and view-level partitions $\bm c_j$, for $j \in [J]$ when the data are simulated from the proposed model. This study is run for several combinations of the hyperparameters $\alpha_0$ and $\alpha$, yielding a sensitivity analysis framework to assess the ability of the model to recover the true underlying partitions. We simulate data from two views ($n=150$, $J=2$) from mixtures of univariate Normals, such that for $i \in [n]$ and $j \in [J]$:
\begin{align*}
    Y_{ji} \mid \bm \mu_j, \bm \sigma_j^2, \bm c_j &\sim \text{N}\left(y_{ji} \mid \mu_{jc_{ji}}, \sigma^2_{jc_{ji}}\right) \nonumber\\
    \mu_{jm}, \sigma_{jm}^2 \mid M &\sim \text{N}\left(\mu_{jm} \mid m_0, \sigma_{jm}^2/k_0\right) \text{Inv-Gamma}\left(\sigma_{jm}^2 \mid a_0, b_0\right) , \quad m=1, \dots, M \nonumber
\end{align*}

\begin{figure}
\centering
\includegraphics[scale=0.75]{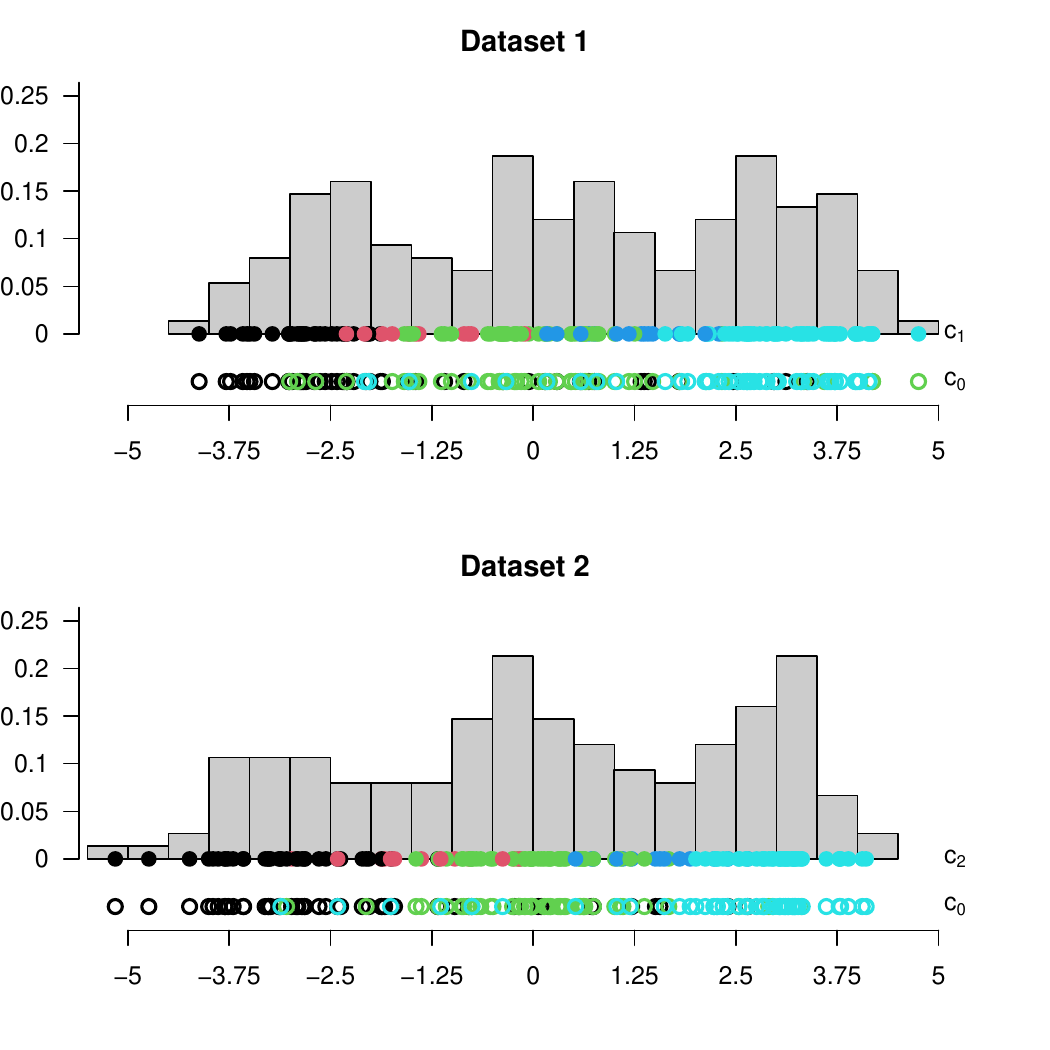}
\caption{Sensitivity Analysis. Histograms of the simulated data in the two views. Dots indicate the individual samples, while colours represent their true clustering allocation. In each panel, the top row of point indicates the corresponding view's partition, while the one below indicates the baseline partition $\bm c0$.}
\label{fig:Simul1_hist}
\end{figure}

We simulate the allocation probabilities from model (1) by setting $M = 5$ and $\bm c_0$ a partition with three clusters of equal size. The location parameters shared by the $J$ mixtures are the pairs $\bm \theta_m = \left(\mu_m, \sigma^2_m\right)$, for $m \in [M]$ and $j \in [J]$, such that $\mu_1, \dots, \mu_M$ are equally-spaced mean parameters between -3 and 3, and $\sigma^2_1, \dots, \sigma^2_M$ are variances equal to 0.5. Of the five components used in the simulation, the three that are present in $\bm c_0$ are $\bm \theta_1$, $\bm \theta_3$, and $\bm \theta_5$. The generated data sets are shown in Figure \ref{fig:Simul1_hist}.

We fit model (1) to this dataset for several combinations of the hyperparameters $\alpha_0, \alpha$ $\in \{0.05, 0.1, 0.5, 1, 5\}$. We fix $q_M \sim \text{Poi}_1\left(\Lambda\right)$, indicating the Poisson distribution shifted by one unit, with $\Lambda = 5$. We run the MCMC algorithm described in Appendix~\ref{app:MCMC} for 3500 iterations, of which the first 1000 are discarded as burn-in and the remaining 2500 are used for posterior inference.

We report measures of goodness of cluster recovery. Namely, we compute the posterior mode of the number of components $M$ and of the number of clusters $K_n$, the Rand Index \citep{rand1971objective} between the estimated partition and the truth, and the co-clustering error \citep{bassetti2018bayesian}. The estimated partitions are computed by minimizing either the Binder \citep{binder1981approximations} or the Variation of Information \citep{meilua2003comparing} loss functions.

Table \ref{tab:Simul1_Clust_all} shows the measures of interest for the partition estimated considering all the clustering labels together, i.e. neglecting the separation between baseline and views in $\bm c_0$, $\bm c_1, \dots, \bm c_J$. The number of components and clusters is correctly recovered for all combinations of $\left(\alpha_0, \alpha\right)$, while better estimates of the partition are produced for small values of $\left(\alpha_0, \alpha\right)$.

Similar measures are shown in Tables \ref{tab:Simul1_Clust_c0}, \ref{tab:Simul1_Clust_c1} and \ref{tab:Simul1_Clust_c2} of the individual partitions (baseline, and views). In particular, the partition of $\bm c_0$ is recovered well for values of $\alpha_0, \alpha \leq 0.1$.

\begin{table}[h]
    \centering
    \caption{Sensitivity Analysis. Comparison of different goodness-of-clustering measures for various values of $\alpha_0$ and $\alpha$. The quantities are computed by consider the whole dataset together, thus ignoring the views.}
    \label{tab:Simul1_Clust_all}
    \renewcommand{\arraystretch}{1}
\begin{tabular}{c|ccccc}
    \cline{2-6}
    & \multicolumn{5}{c}{\textbf{$\text{mode}(M \mid \mathbf{Y})$ / $\text{mode}(K_n \mid \mathbf{Y})$}} \\\hline
    \diagbox{$\alpha_0$}{$\alpha$} & 0.05 & 0.1 & 0.5 & 1 & 5 \\
    \hline
    0.05 & 4/4 & 5/5 & 5/5 & 5/5 & 5/5 \\
    0.1  & 5/5 & 5/5 & 5/5 & 5/5 & 5/5 \\
    0.5  & 5/5 & 5/5 & 5/5 & 5/5 & 5/5 \\
    1    & 5/5 & 5/5 & 5/5 & 5/5 & 5/5 \\
    5    & 5/5 & 5/5 & 5/5 & 5/5 & 5/5 \\
    \cline{2-6}

    & \multicolumn{5}{c}{\textbf{RI (Binder) / RI (VI)}} \\\hline
    \diagbox{$\alpha_0$}{$\alpha$} & 0.05 & 0.1 & 0.5 & 1 & 5 \\
    \hline
    0.05 & 0.83/0.82 & 0.85/0.82 & 0.82/0.68 & 0.74/0.69 & 0.74/0.68 \\
    0.1  & 0.83/0.82 & 0.85/0.83 & 0.82/0.69 & 0.70/0.68 & 0.70/0.68 \\
    0.5  & 0.85/0.82 & 0.84/0.82 & 0.82/0.68 & 0.81/0.53 & 0.74/0.26 \\
    1    & 0.85/0.82 & 0.84/0.82 & 0.82/0.48 & 0.80/0.51 & 0.76/0.26 \\
    5    & 0.84/0.82 & 0.82/0.82 & 0.83/0.48 & 0.80/0.51 & 0.77/0.26 \\
    \cline{2-6}

    & \multicolumn{5}{c}{\textbf{co-Clustering Error}} \\\hline
    \diagbox{$\alpha_0$}{$\alpha$} & 0.05 & 0.1 & 0.5 & 1 & 5 \\
    \hline
    0.05 & 0.20 & 0.21 & 0.26 & 0.29 & 0.33 \\
    0.1  & 0.20 & 0.21 & 0.26 & 0.29 & 0.32 \\
    0.5  & 0.20 & 0.21 & 0.26 & 0.28 & 0.30 \\
    1    & 0.21 & 0.22 & 0.26 & 0.28 & 0.30 \\
    5    & 0.20 & 0.21 & 0.26 & 0.27 & 0.30 \\
    \cline{2-6}
\end{tabular}
\end{table}

\begin{table}[h]
    \centering
    \caption{Sensitivity Analysis. Comparison of different goodness-of-clustering measures for various values of $\alpha_0$ and $\alpha$. The quantities are computed by considering only the baseline partition $\bm c_0$.}
    \label{tab:Simul1_Clust_c0}
    \renewcommand{\arraystretch}{1}
\begin{tabular}{c|ccccc}
    \cline{2-6}
    & \multicolumn{5}{c}{\textbf{$\text{mode}\left(K_n\mid \bm Y\right)$}} \\\hline
    \diagbox{$\alpha_0$}{$\alpha$} & 0.05 & 0.1 & 0.5 & 1 & 5 \\
    \hline
    0.05 & 3 & 3 & 3 & 2 & 2 \\
    0.1  & 3 & 4 & 3 & 2 & 2 \\
    0.5  & 5 & 5 & 4 & 5 & 5 \\
    1    & 5 & 5 & 5 & 5 & 5 \\
    5    & 5 & 5 & 5 & 5 & 5 \\
    \cline{2-6}

    & \multicolumn{5}{c}{\textbf{RI (Binder) / RI (VI)}} \\\hline
    \diagbox{$\alpha_0$}{$\alpha$} & 0.05 & 0.1 & 0.5 & 1 & 5 \\
    \hline
    0.05 & 0.75/0.75 & 0.76/0.73 & 0.70/0.33 & 0.33/0.33 & 0.33/0.33 \\
    0.1  & 0.74/0.73 & 0.76/0.75 & 0.72/0.33 & 0.33/0.33 & 0.33/0.33 \\
    0.5  & 0.76/0.73 & 0.73/0.73 & 0.72/0.33 & 0.62/0.33 & 0.33/0.33 \\
    1    & 0.75/0.72 & 0.74/0.73 & 0.70/0.33 & 0.71/0.33 & 0.52/0.33 \\
    5    & 0.75/0.73 & 0.72/0.73 & 0.72/0.33 & 0.71/0.33 & 0.57/0.33 \\
    \cline{2-6}

    & \multicolumn{5}{c}{\textbf{co-Clustering Error}} \\\hline
    \diagbox{$\alpha_0$}{$\alpha$} & 0.05 & 0.1 & 0.5 & 1 & 5 \\
    \hline
    0.05 & 0.30 & 0.33 & 0.45 & 0.52 & 0.61 \\
    0.1  & 0.30 & 0.32 & 0.44 & 0.52 & 0.58 \\
    0.5  & 0.30 & 0.32 & 0.43 & 0.45 & 0.47 \\
    1    & 0.30 & 0.33 & 0.41 & 0.43 & 0.44 \\
    5    & 0.30 & 0.32 & 0.39 & 0.40 & 0.40 \\
    \cline{2-6}
\end{tabular}

\end{table}

\begin{table}[h]
    \centering
    \caption{Sensitivity Analysis. Comparison of different goodness-of-clustering measures for various values of $\alpha_0$ and $\alpha$. The quantities are computed by considering only the partition for view $\bm c_1$.}
    \label{tab:Simul1_Clust_c1}
    \renewcommand{\arraystretch}{1}
\begin{tabular}{c|ccccc}
    \cline{2-6}
    & \multicolumn{5}{c}{\textbf{$\text{mode}\left(K_n\mid \bm Y\right)$}} \\\hline
    \diagbox{$\alpha_0$}{$\alpha$} & 0.05 & 0.1 & 0.5 & 1 & 5 \\
    \hline
    0.05 & 4 & 5 & 5 & 5 & 5 \\
    0.1  & 5 & 5 & 5 & 5 & 5 \\
    0.5  & 5 & 5 & 5 & 5 & 5 \\
    1    & 5 & 5 & 5 & 5 & 5 \\
    5    & 5 & 5 & 5 & 5 & 5 \\
    \cline{2-6}

    & \multicolumn{5}{c}{\textbf{RI (Binder) / RI (VI)}} \\\hline
    \diagbox{$\alpha_0$}{$\alpha$} & 0.05 & 0.1 & 0.5 & 1 & 5 \\
    \hline
    0.05 & 0.87/0.86 & 0.90/0.87 & 0.87/0.86 & 0.87/0.86 & 0.87/0.86 \\
    0.1  & 0.88/0.87 & 0.90/0.87 & 0.88/0.86 & 0.88/0.85 & 0.87/0.86 \\
    0.5  & 0.89/0.87 & 0.90/0.87 & 0.87/0.86 & 0.87/0.64 & 0.87/0.24 \\
    1    & 0.90/0.86 & 0.90/0.86 & 0.87/0.61 & 0.88/0.67 & 0.86/0.24 \\
    5    & 0.87/0.87 & 0.88/0.87 & 0.89/0.61 & 0.88/0.67 & 0.87/0.24 \\
    \cline{2-6}

    & \multicolumn{5}{c}{\textbf{co-Clustering Error}} \\\hline
    \diagbox{$\alpha_0$}{$\alpha$} & 0.05 & 0.1 & 0.5 & 1 & 5 \\
    \hline
    0.05 & 0.16 & 0.15 & 0.16 & 0.18 & 0.20 \\
    0.1  & 0.15 & 0.15 & 0.16 & 0.17 & 0.19 \\
    0.5  & 0.15 & 0.15 & 0.16 & 0.18 & 0.20 \\
    1    & 0.16 & 0.16 & 0.16 & 0.18 & 0.20 \\
    5    & 0.15 & 0.15 & 0.17 & 0.19 & 0.20 \\
    \cline{2-6}
\end{tabular}

\end{table}

\begin{table}[h]
    \centering
    \caption{Sensitivity Analysis. Comparison of different goodness-of-clustering measures for various values of $\alpha_0$ and $\alpha$. The quantities are computed by considering only the partition for view $\bm c_1$.}
    \label{tab:Simul1_Clust_c2}
    \renewcommand{\arraystretch}{1}
\begin{tabular}{c|ccccc}
    \cline{2-6}
    & \multicolumn{5}{c}{\textbf{$\text{mode}\left(K_n\mid \bm Y\right)$}} \\\hline
    \diagbox{$\alpha_0$}{$\alpha$} & 0.05 & 0.1 & 0.5 & 1 & 5 \\
    \hline
    0.05 & 4 & 5 & 5 & 5 & 5 \\
    0.1  & 5 & 5 & 5 & 5 & 5 \\
    0.5  & 5 & 5 & 5 & 5 & 5 \\
    1    & 5 & 5 & 5 & 5 & 5 \\
    5    & 5 & 5 & 5 & 5 & 5 \\
    \cline{2-6}

    & \multicolumn{5}{c}{\textbf{RI (Binder) / RI (VI)}} \\\hline
    \diagbox{$\alpha_0$}{$\alpha$} & 0.05 & 0.1 & 0.5 & 1 & 5 \\
    \hline
    0.05 & 0.87/0.87 & 0.90/0.88 & 0.86/0.87 & 0.86/0.86 & 0.87/0.87 \\
    0.1  & 0.88/0.87 & 0.89/0.88 & 0.87/0.87 & 0.88/0.87 & 0.87/0.86 \\
    0.5  & 0.89/0.87 & 0.88/0.88 & 0.87/0.87 & 0.87/0.68 & 0.87/0.24 \\
    1    & 0.90/0.87 & 0.89/0.88 & 0.86/0.61 & 0.88/0.66 & 0.87/0.24 \\
    5    & 0.88/0.87 & 0.87/0.88 & 0.89/0.61 & 0.87/0.66 & 0.86/0.24 \\
    \cline{2-6}

    & \multicolumn{5}{c}{\textbf{co-Clustering Error}} \\\hline
    \diagbox{$\alpha_0$}{$\alpha$} & 0.05 & 0.1 & 0.5 & 1 & 5 \\
    \hline
    0.05 & 0.14 & 0.15 & 0.16 & 0.17 & 0.19 \\
    0.1  & 0.14 & 0.14 & 0.16 & 0.17 & 0.18 \\
    0.5  & 0.15 & 0.15 & 0.16 & 0.17 & 0.19 \\
    1    & 0.15 & 0.15 & 0.16 & 0.18 & 0.19 \\
    5    & 0.15 & 0.15 & 0.17 & 0.18 & 0.19 \\
    \cline{2-6}
\end{tabular}
\end{table}

\clearpage
\section{Additional Figures}\label{app:figures}

\begin{figure}[ht]
\centering
\subfloat[Baseline - Z-BMI]{\includegraphics[width=0.35\linewidth]{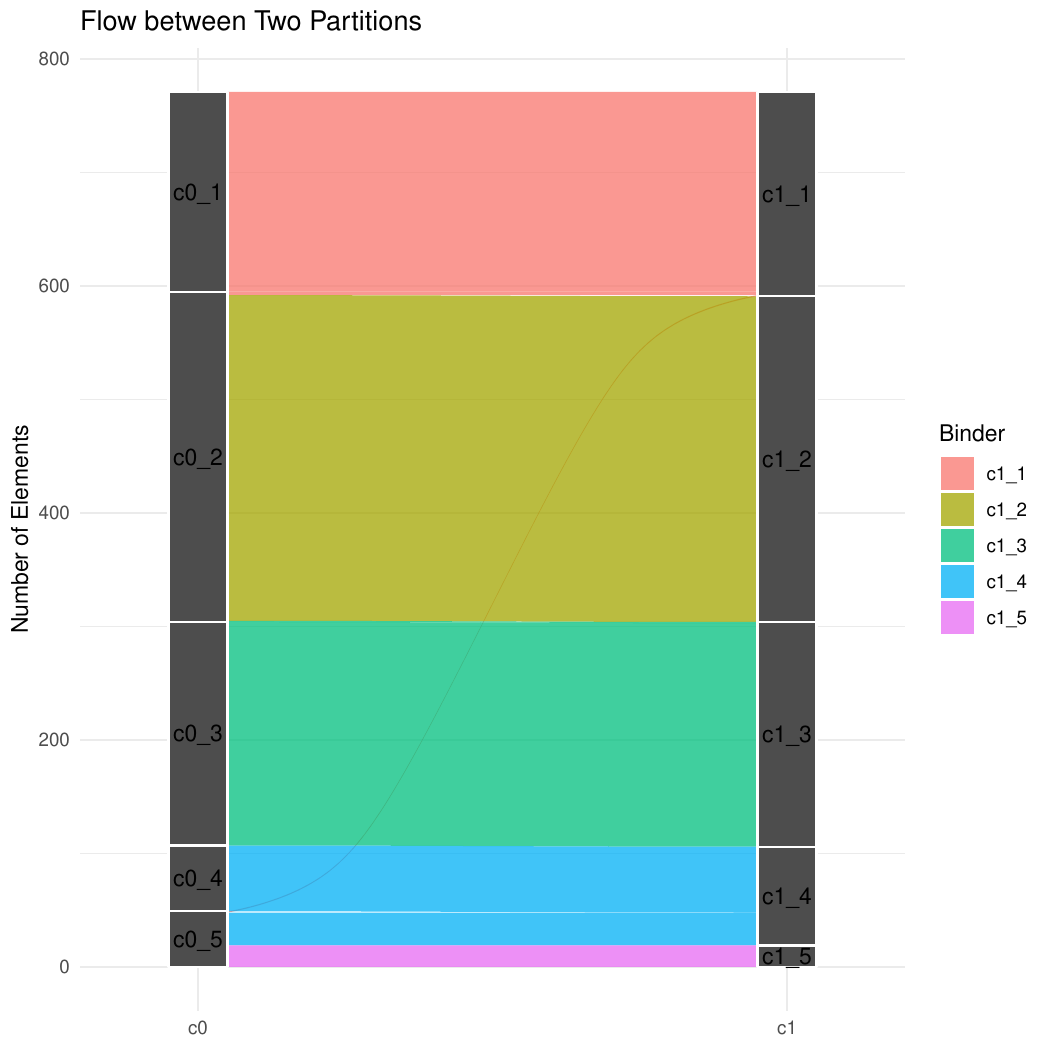}}
\subfloat[Baseline - Hypertension]{\includegraphics[width=0.35\linewidth]{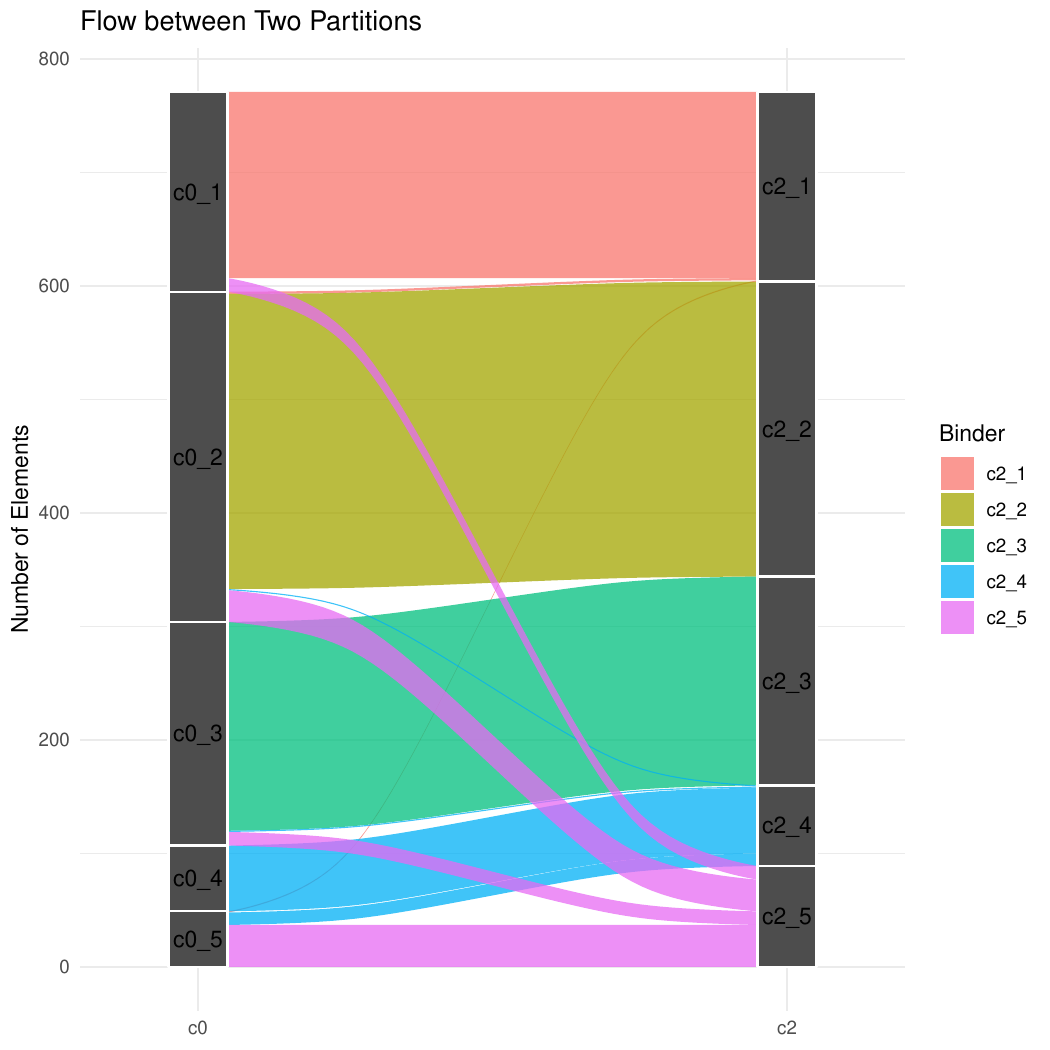}}
\subfloat[Baseline - Wheezing]{\includegraphics[width=0.35\linewidth]{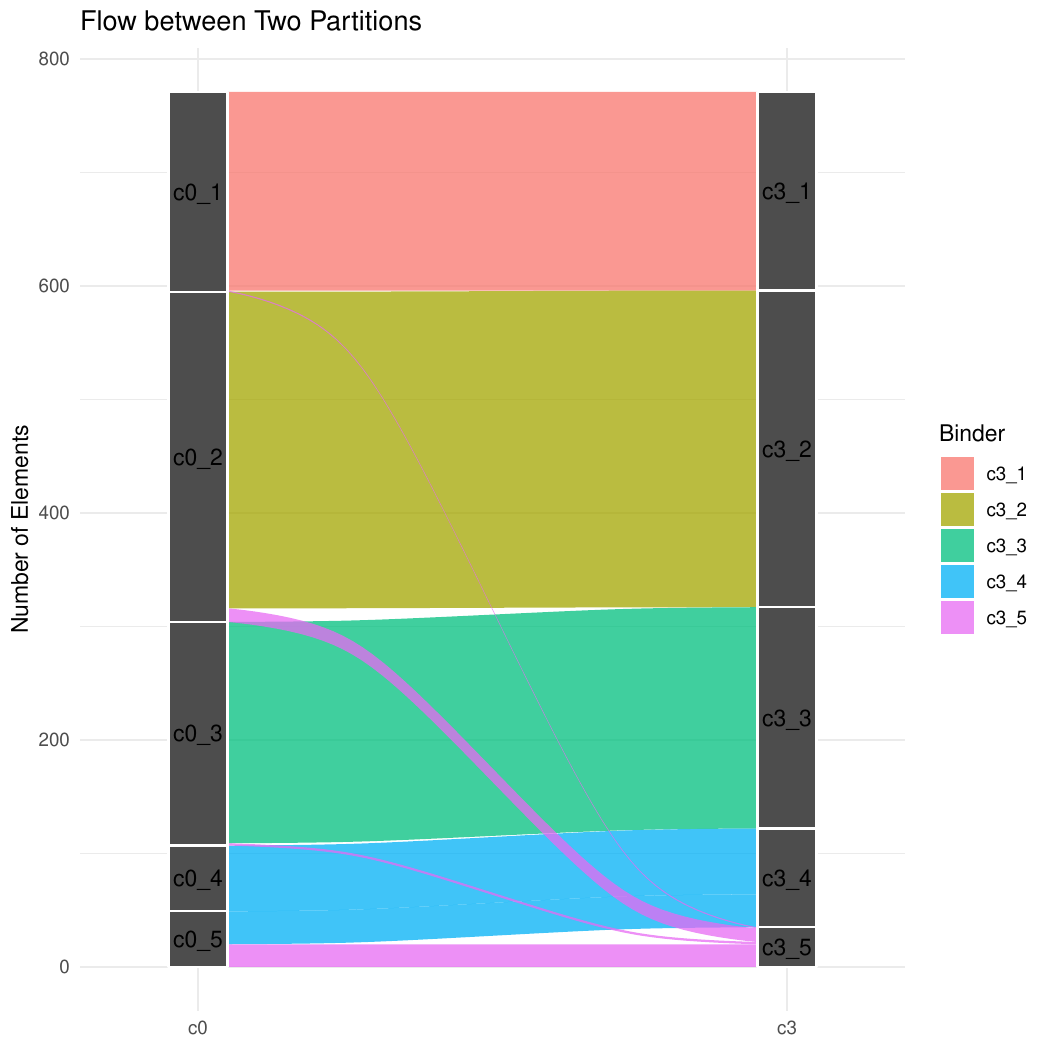}}
\caption{GUSTO data. Waterfall plots depicting how the estimated Binder partition of the units changes between the baseline partition $\bm c_0$ and the three views.}
\label{fig:GUSTO_flow_c0}
\end{figure}

\begin{figure}[ht]
\centering
\subfloat[Z-BMI - Hypertension]{\includegraphics[width=0.35\linewidth]{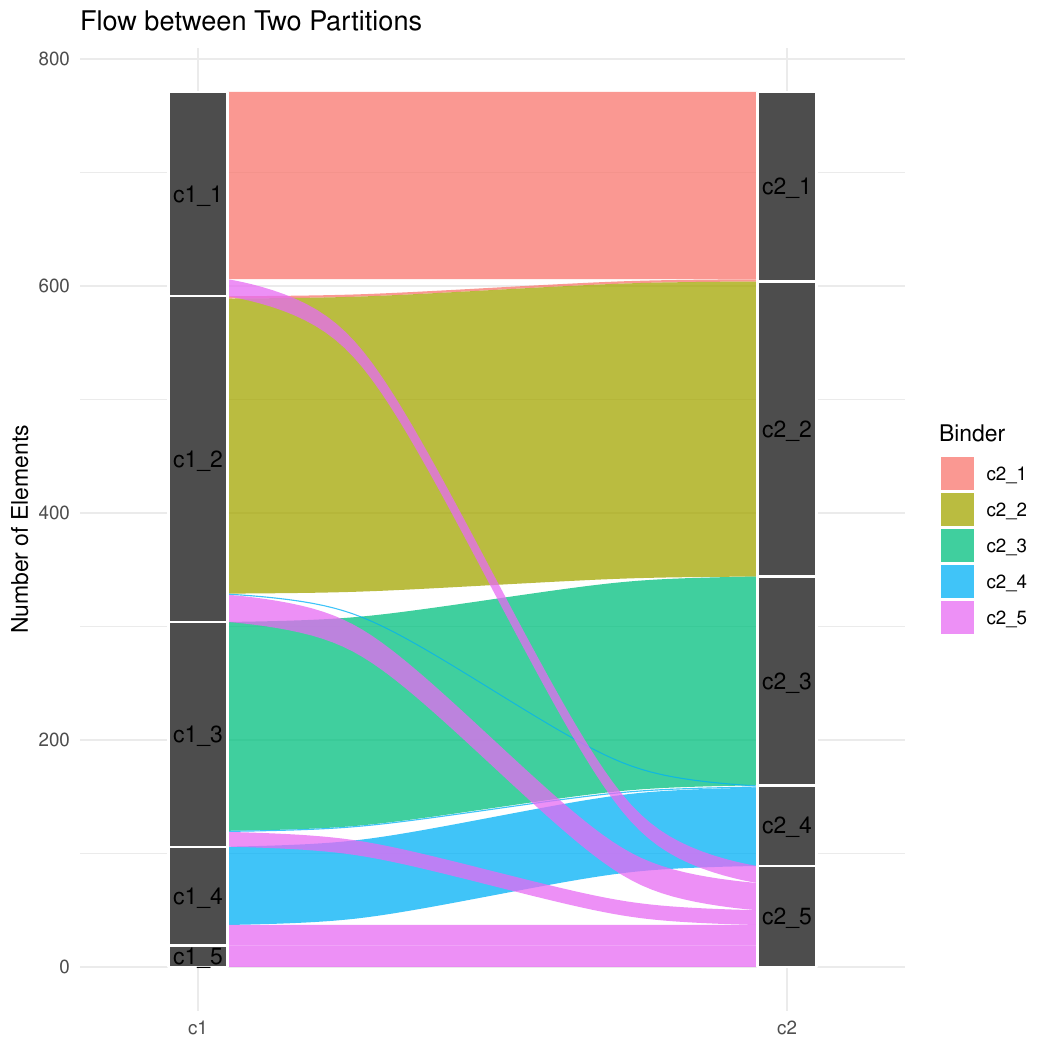}}
\subfloat[Z-BMI - Wheezing]{\includegraphics[width=0.35\linewidth]{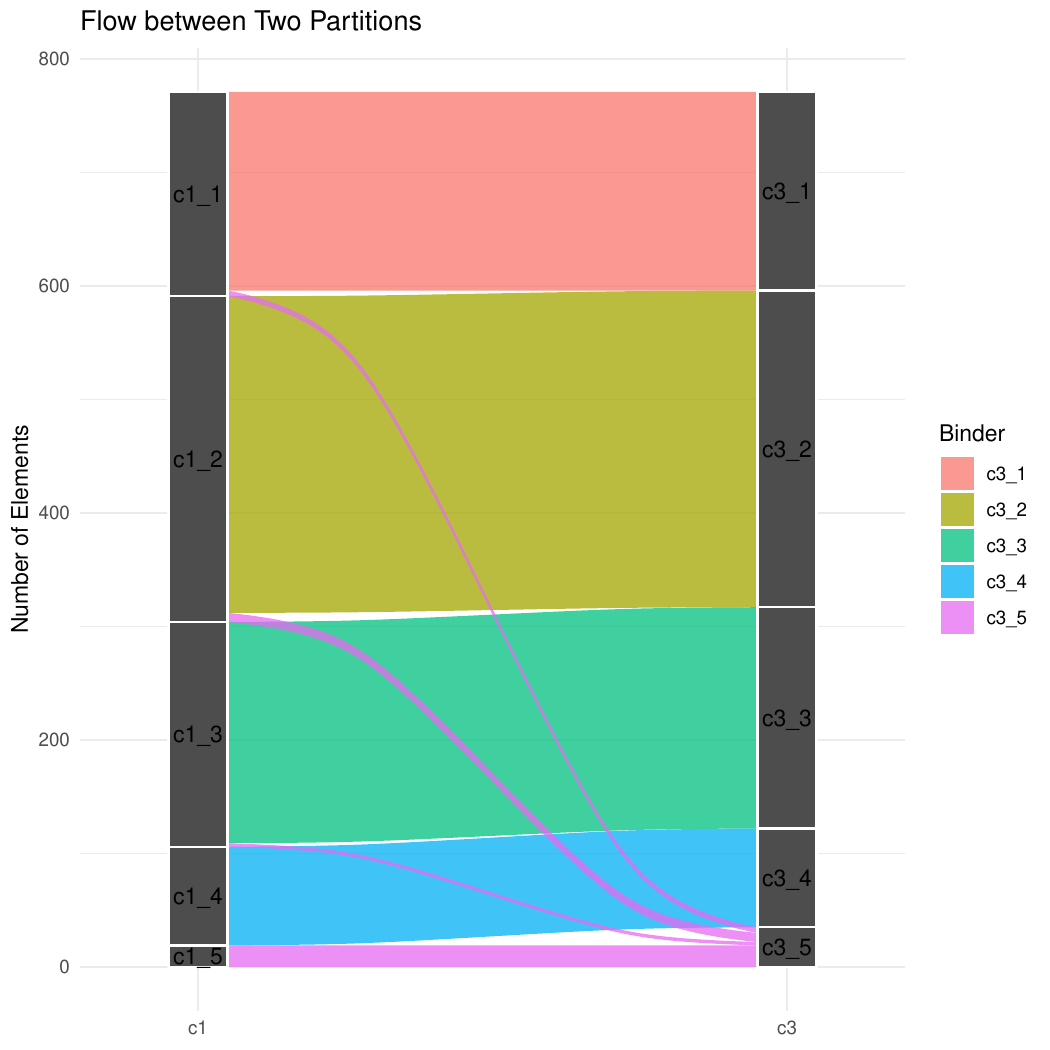}}
\subfloat[Hypertension - Wheezing]{\includegraphics[width=0.35\linewidth]{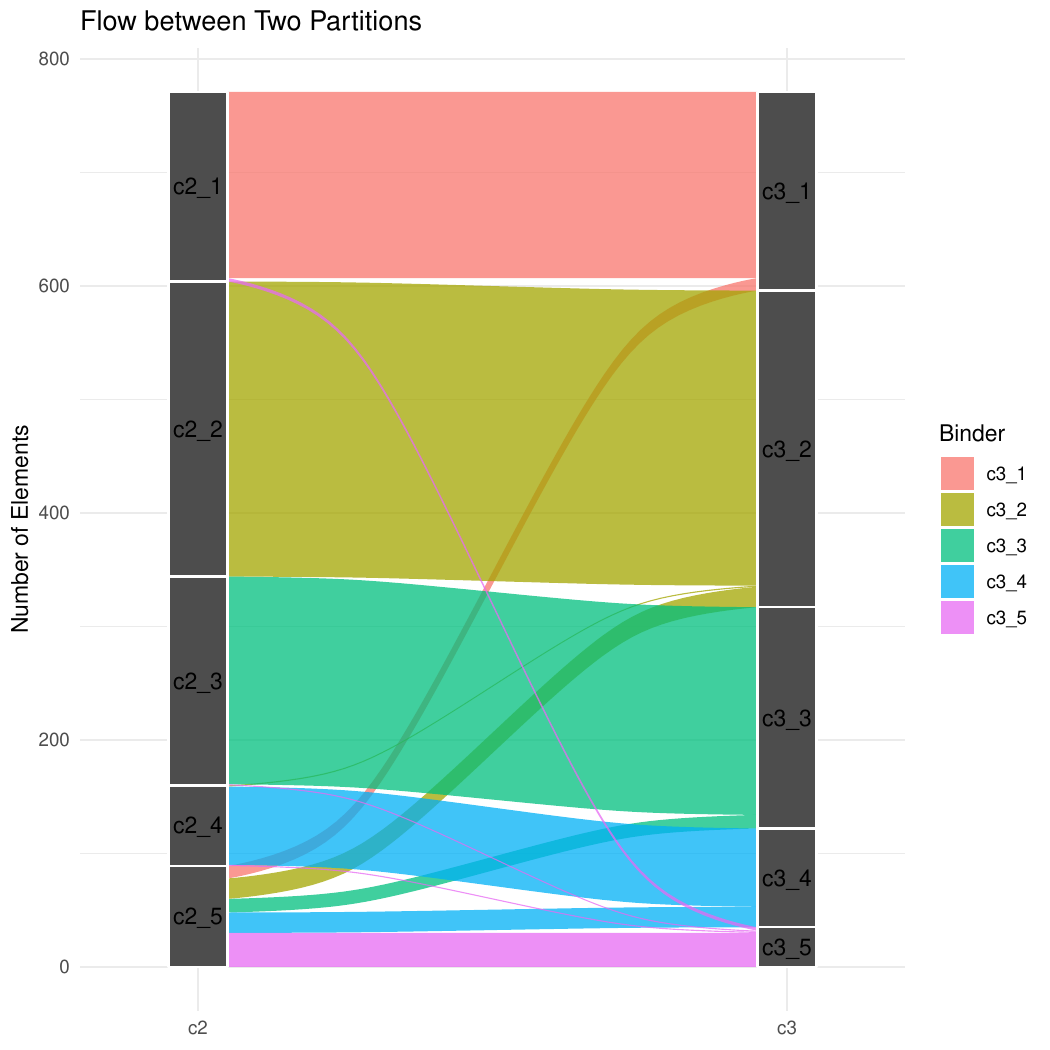}}
\caption{GUSTO data. Waterfall plots depicting how the estimated Binder partition of the units changes between the different views.}
\label{fig:GUSTO_flow_cj}
\end{figure}

\end{document}